\documentclass{aa}  
\usepackage{needspace}
\usepackage{graphicx}
\usepackage{caption}
\usepackage{enumitem}
\usepackage{url}
\usepackage{amsmath}
\usepackage{wrapfig} 
\usepackage{pdfpages}
\usepackage[rightcaption]{sidecap}
\usepackage{txfonts}
\usepackage{lipsum}
\usepackage{subcaption}         
\usepackage{lscape}             
\usepackage{placeins}           

\defcitealias{Smolcic2017}{S17}
\defcitealias{Rudnick2014}{RO14}
\newcommand{\jybeam}{$\mathrm{Jy}\,\mathrm{beam}^{-1}\,$}
\newcommand{\radmsq}{$\mathrm{rad}\,\mathrm{m}^{-2}\,$}
                                
\usepackage{hyperref}

\begin{document}

   \title{The VLA-COSMOS 3~GHz Large Project: \\Polarised source counts and catalogue}



   \author{S. Ranchod\inst{1}\thanks{E-mail: sranchod@mpifr-bonn.mpg.de}
        \and S. A. Mao\inst{1} \and R. P. Deane\inst{2, 3, 4} \and V. Smol{\v{c}}i{\'c}\inst{5} \and J. D. Wagenveld\inst{1} \and M. Bondi\inst{6} \and K. Mooley\inst{7, 8} \and \\  E. Schinnerer\inst{9}
        }

   \institute{Max-Planck Instit\"{u}t f\"{u}r Radioastronomie, Auf dem H\"{u}gel 69, 53121 Bonn, Germany 
            \and Inter-University Institute for Data Intensive Astronomy, Department of Astronomy, University of Cape Town, Cape Town, South Africa
            \and Department of Physics, University of Pretoria, Hatfield, Pretoria, 0028, South Africa 
            \and Wits Centre for Astrophysics, University of the Witwatersrand, 1 Jan Smuts Avenue, 2000, Johannesburg, South Africa
            \and Department of Physics, University of Zagreb, Bijeni\v{c}ka cesta 32, Zagreb, Croatia
            \and Istituto di Radioastronomia, Via Gobetti 101, 40129, Bologna, Italy
            \and Division of Physics, Math and Astronomy, California Institute of Technology, Pasadena CA 91106, USA
            \and Indian Institute of Technology Kanpur, Kanpur, UP 208016, India 
            \and Max-Planck-Institute f\"ur Astronomie, K\"onigstuhl 17, 69117 Heidelberg, Germany \\ }

   \date{Received September 30, 20XX}

 
  \abstract
   {The exploration of the faint polarised radio source population is essential for interpreting the nature and evolution of magnetic fields in galaxies. While recent studies have provided insight into source counts for the $\mu$Jy polarised source population at ~1.4 GHz, higher frequency surveys may be more sensitive to new populations that are depolarised at lower frequencies (i.e. due to internal or external depolarisation effects).}
   {We present the deepest polarised source counts at 3~GHz to date, at an angular resolution of $1.5''$. With these relatively higher frequency observations, we aim to probe the faint polarised star-forming galaxy (SFG) population. Furthermore, through spectral modelling, we aim to provide further insight into the frequency evolution of polarised source counts.} 
   {We processed the polarisation data of the VLA-COSMOS 3~GHz Large Project, one of the deepest high-resolution radio continuum surveys. We produced Stokes Q and U mosaicked channel maps. After selecting known sources in total intensity, we performed 3D rotation measure synthesis and searched for polarised emission using an empirically determined threshold.}
   {With a sensitivity of 2.6~$\mu$\jybeam in Faraday depth, we detect 65 polarised sources (51~deg$^{-2}$) above our threshold. We find that our cumulative and Euclidean-normalised source counts at 3~GHz are consistent with those in the literature at 1.4~GHz, which we attribute to the combined effect of spectral index and depolarisation in the detected sources. 
   We detect no SFGs in our sample and derive a 2$\sigma$ upper limit on the density of polarised SFGs of $<2.0~\mathrm{deg}^{-2}$. This implies that significantly deeper observations will be required to readily detect this population in the SKA-era.}
   {}


   \maketitle
   \nolinenumbers

\section{Introduction}

Magnetic fields have an active role in galaxy evolution, influencing star formation and the feedback processes of active galactic nuclei (AGNs). The strength and ordering of magnetic fields in synchrotron-emitting sources can be directly observed through the linear polarisation of this emission. To fully understand the role of magnetic fields in galaxy evolution, including how this varies with source characteristics and develops over cosmic time, we require statistically large samples of polarised extragalactic sources \citep[e.g.][]{Johnston-Hollitt2015,Heald2020}. In particular, source detections over a range of polarised intensities can probe various sub-populations, such as low-luminosity AGNs and star-forming galaxies (SFGs).

Historically, blind extragalactic polarisation surveys have been limited to the bright source population ($P_\mathrm{1.4\,GHz}>0.5$~mJy). Investigations into the number counts of polarised sources have found that the dominant source of polarised emission above this threshold is from AGNs \citep{Taylor2007}, likely from the extended lobes of Fanaroff-Riley \citep[FR;][]{Fanaroff1974} radio galaxies \citep{Grant2010}.

Over the past decade, modern surveys have begun to probe the fainter polarised population, with \citet[][hereafter, RO14]{Rudnick2014} pioneering the investigation into the $\mu$Jy regime. Their deep survey of the GOODS-N field with the Karl G. Jansky Very Large Array (VLA) down to $P_\mathrm{1.4\,GHz}>30~\mu$\jybeam revealed flattening in the cumulative polarised source counts for $P_\mathrm{1.4\,GHz}<1$~mJy, with respect to previous work using brighter sources \citep{Subrahmanyan2010,Grant2010}. This slope transition has also been observed by \citet{Eyles2020} and \citet{Berger2025}. While \citetalias{Rudnick2014} attribute this flattening to a change in population between FRI (core-dominated) and FRII (lobe-dominated) sources \citep[e.g.][]{OSullivan2008}, no modern surveys have had both the angular resolution for morphological source classification as well as the depth and area for a statistically large number of detections to confirm this trend. More recently, modern radio interferometers, including the Square Kilometre Array (SKA) precursor and pathfinder telescopes have given rise to sensitive polarisation surveys that readily detect the polarised population at $P_\mathrm{1.4\,GHz}<1$~mJy, producing unprecedented rotation measure (RM) densities for RM grid experiments. Rotation measure catalogues can be used to map the foreground Galactic magnetic field \citep[e.g.][]{Brown2007,Schnitzeler2019,Ma2020}, nearby galaxies \citep[e.g.][]{Mao2008,Livingston2022}, and galaxy clusters \citep[e.g.][]{Anderson2021,Loi2025}. The early science results from the Polarisation Sky Survey of the Universe's Magnetism \citep[POSSUM;][]{Gaensler2025} with the Australian SKA Pathfinder (ASKAP) have produced RM densities of $>30$~deg$^{-2}$ \citep{Vanderwoude2024} and will eventually cover 20~000 deg$^2$ of the southern sky to a sensitivity of ${\sim}25~\mu$\jybeam. The MPIfR MeerKAT Galactic Plane Survey \citep[MMGPS;][]{Padmanabh2023} is set to produce an RM density of $\sim40$~deg$^{-2}$ in the southern Galactic plane to a sensitivity of $\sim9~\mu$\jybeam. While these surveys are actively ongoing, there remains much to be understood about observational biases that may affect the polarised source density of these RM grid experiments. This can be addressed through deep extragalactic polarisation surveys. To date, the deepest are the MeerKAT International GHz Tiered Extragalactic Exploration \citep[MIGHTEE;][]{Taylor2024} with $P_\mathrm{1.4\,GHz}>16.5~\mu$\jybeam and the MeerKAT Fornax Survey \citep{Loi2025} with $P_\mathrm{1.4\,GHz}>9~\mu$\jybeam and an angular resolution of $13''$. Despite the depths of these surveys, all polarised sources are classified as AGNs, and the depth at which we readily detect the polarised SFG population is not well constrained \citep[e.g.][]{OSullivan2008}.

While the majority of the blind extragalactic surveys discussed here were conducted at 1.4~GHz (L band), higher frequencies can offer a new perspective. In some cases, the resulting higher angular resolution can better mitigate beam depolarisation and enable the morphological classification of radio sources. Given the $\lambda^2$-dependence of depolarisation (due to turbulent screens or internal effects), at 3~GHz, we are less affected by depolarisation and can possibly detect sources that are otherwise depolarised at lower frequencies. Through simulations, \citet{Stil2009} show that unresolved SFGs are expected to have an increased rest-frame fractional polarisation up to 20\% at 4.8~GHz as opposed to only 10\% (in extreme cases) at 1.4~GHz. This simulation assumes that all SFGs have the same polarisation properties as local galaxies. In a simulation of Milky Way-like SFGs, \citet{Sun2012} found a similar frequency dependence, with the distribution of integrated fractional polarisation peaking at 4.2\% at a rest-frame frequency of 4.8~GHz and dropping to a peak of 0.8\% at 1.4~GHz. In a review of available polarisation surveys at the time, \citet{Tucci2012} showed an increase in the median fractional polarisation from 1.5\% at 1.4~GHz to ${\sim}$3\% at $>10$~GHz for steep- and flat-spectrum AGNs. However, the frequency-dependence of polarised source counts is relatively unexplored in the literature. Detailed modelling of the frequency dependence of observed source counts for various source classifications can characterise the spectral index and depolarisation behaviour \citep[e.g.][]{Lamee_2016} of the polarised source population as a whole. Moreover, this is a necessary step in the planning of SKA-era surveys to maximise the number of detected sources for RM grid experiments \citep[e.g.][]{Heald2020}.

To study the faint polarised sky at 3~GHz, we used the VLA-COSMOS 3~GHz Large Project \citep[][hereafter, S17]{Smolcic2017}, which is a deep (${\sim}2.3~\mu$\jybeam), high-resolution (0.75''), broad bandwidth (2--4~GHz) survey with the VLA, over 2~deg$^2$ in the COSMOS field. This region has broad multi-wavelength coverage, which combined with the radio catalogue has produced new insights into the far-infrared-radio correlation \citep{Delhaize2017}, the redshift evolution of the AGN dichotomy \citep{Delvecchio2017}, and the cosmic star-formation rate history \citep{Novak2017,Leslie2020}. The original data release of VLA-COSMOS 3~GHz included only total intensity data products, and in this paper, we reprocess this survey to include polarisation calibration and Stokes Q and U imaging to investigate the polarised source population in the COSMOS field. The combined depth, resolution, bandwidth, and multi-wavelength coverage of this survey have allowed us to characterise the faint polarised source population over an unexplored parameter space and produce the deepest samples of polarised sources at 3~GHz to date. We aimed to leverage this to probe the polarised SFG population. In addition, we computed the polarised source counts at 3~GHz to investigate their frequency dependence with respect to legacy L-band surveys (e.g. \citetalias{Rudnick2014}). 

This paper is organised as follows. In Sect.~\ref{ch4:sec:vla-cosmos}, we detail the data processing, including calibration, imaging, source finding, and spectral analysis. We present the polarised source counts in Sect.~\ref{ch4:sec:counts-general} and the detection of SFGs in Sect.~\ref{sec:sfg-detection}. In Sect.~\ref{ch4:sec:spec-depol}, we present the results of the spectral index and depolarisation analysis. In Sect.~\ref{ch4:sec:discussion}, we discuss the frequency evolution of the polarised source counts and the biases regarding source detectability at 3~GHz. We summarise and conclude in Sect.~\ref{ch4:sec:conclusion}. Further analysis regarding the properties of the detected polarised sources will be presented in a companion paper (Ranchod et al., Paper II). Throughout this paper, we follow the spectral index $\alpha$ convention of $S_\nu \propto \nu^\alpha$, where $S_\nu$ is the radio flux density at frequency $\nu$.

\section{VLA-COSMOS 3 GHz Large Project}\label{ch4:sec:vla-cosmos}
The VLA-COSMOS 3 GHz large project is a radio continuum survey of the 2 deg$^2$ COSMOS field (centred at RA=10:00:28.6, Dec=+02:12:21.0) with the VLA. This survey observes the COSMOS field in the S band (2--4 GHz), and the total intensity (Stokes I) data release (\citetalias{Smolcic2017}) produced a total intensity mosaic image with a median root mean square (rms) of $\sigma = 2.3\,\mu $Jy beam$^{-1}$, and an angular resolution of 0.75". 

A total of 384 hours was observed with the VLA A-array (324 hours) and C-array (60 hours) configurations from November 2012 to May 2014. The survey is comprised of 192 pointings, arranged in three grids of 64 (8$\times$8), with a $10'$ separation in Right Ascension (RA) and Declination (Dec) within a given grid and an offset of $5'$ between grids. 
In each observation, J1331+3030 (3C 286) was observed as a flux, bandpass, and polarisation calibrator with a 3--5 min scan and J0713+4349 for 5 min as a polarisation leakage calibrator. The complex gain calibrator J1024--0052 was observed every 30~min over each observation. Further information on the pointing configuration, calibrator duty-cycle and observing conditions is available in \citetalias{Smolcic2017}.

\subsection{Data processing and imaging}\label{ch4:sec:calibration}
In the original total intensity data release (\citetalias{Smolcic2017}), the data were calibrated using the Astronomical Image Processing System (AIPS) and AIPSLite \citep{Greisen1998,Bourke2014,Mooley2016}. In this work, we recalibrated the data to enable a robust verification of the polarisation calibration, which was not previously prioritised. 
The uncalibrated data were retrieved from the VLA archive, under the Legacy ID AS1163, as 113 separate schedule blocks and processed using the VLA CASA pipeline \citep{McMullin_2007} version 2022.2.0.64 in \textsc{casa} version 6.4.1. We note that \citetalias{Smolcic2017} used an early version of the VLA CASA pipeline as a comparison to the tailored AIPS pipeline and find no significant difference in the output. We modified the standard version of the pipeline to include polarisation calibration, using the polarisation leakage calibrator J0713+4349 and the polarisation angle calibrator J1331+3030 \citep[3C~286;][]{Perley2017}. This was achieved by enabling \texttt{hifv\_circfeedpolcal}, which performs cross-hand delay, leakage, and polarisation angle calibration before applying solutions to the target and the final round of radio frequency interference (RFI) flagging.

We assessed the quality of the calibration through the diagnostic plots that are produced by the pipeline. Of the 113 observations, six were discarded due to corruption of the data in the archiving process. Inspecting the phase calibrator for the remaining 107 observations, we find an average flagged percentage of 64\%. As described in \citetalias{Smolcic2017}, spectral windows (SPWs) 2 and 3 have high levels of RFI corruption. They are 100\% flagged in the \citetalias{Smolcic2017} calibration and are similarly flagged above 70\% in this work.  

To assess the success of polarisation calibration, we inspected the corrected amplitude and phase of the polarisation leakage calibrator and the polarisation angle calibrator as a function of frequency. We find a consistent amplitude and phase with the expected values for both calibrator sources across all observations. As a further check, we concatenated one-third of the observations of the leakage calibrator and the polarisation angle calibrator into single measurement sets (MSs) for joint imaging. We then imaged both calibrator sources as full-Stokes spectral cubes using WSClean \citep{Offringa2014} with a spectral resolution of 16~MHz (128 channels). To assess the polarisation angle calibration, we performed RM synthesis (see also Sect.~\ref{ch4:sec:rmsynth}) on the polarisation angle calibrator using the \texttt{RMtools} package \citep[Version 1.1.1,][]{Vaneck2026} from the Canadian Initiative for Radio Astronomy Data Analysis (CIRADA) tools. We obtain an electric vector position angle (EVPA) of $\chi = +31.92 \pm 0.02$~deg and a Faraday depth of $\phi = +0.36 \pm 0.06$ rad m$^{-2}$, which is consistent with what we expect from the literature \citep{Perley2013}. In this section, we only discuss on-axis leakage quantification. An in-depth off-axis leakage quantification can be found in Appendix~\ref{ch4:sec:leakage}, for which we find a maximum off-axis leakage of 0.5\%. To quantify the residual on-axis leakage, we measured the linear $p_L$ and circular $p_V$ polarisation fraction of the leakage calibrator J0713+4349, which is unpolarised. We find $p_L$ = 0.027\% and $p_V$ = 0.35\% and take these to be the upper limit for on-axis leakage in the respective polarisations. 

Following calibration, the data were prepared for imaging and mosaicking. The calibrated data were split from each observation and combined into separate MSs according to their pointing positions, using \textsc{casa} tasks \texttt{split} and \texttt{mstransform}. Subsequently, each pointing was processed separately. 

\begin{figure}
    \centering
    \includegraphics[width=0.9\linewidth]{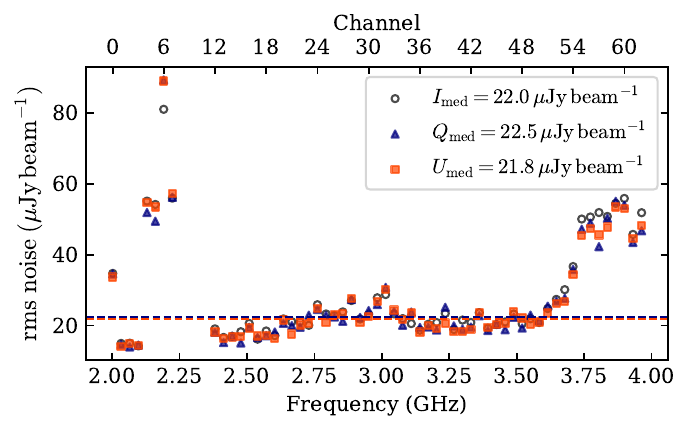}
    \caption[Mean rms noise as a function of frequency for Stokes I, Q, and U]{Mean rms noise per frequency channel for Stokes I (open black circles), Q (blue-filled triangles) and U (orange-filled squares). The dashed lines indicate the median per-channel rms across the band, as displayed in the legend.}
    \label{ch4:fig:rms_freq}
\end{figure}
\begin{figure*}
\centering
\includegraphics[width=0.9\textwidth]{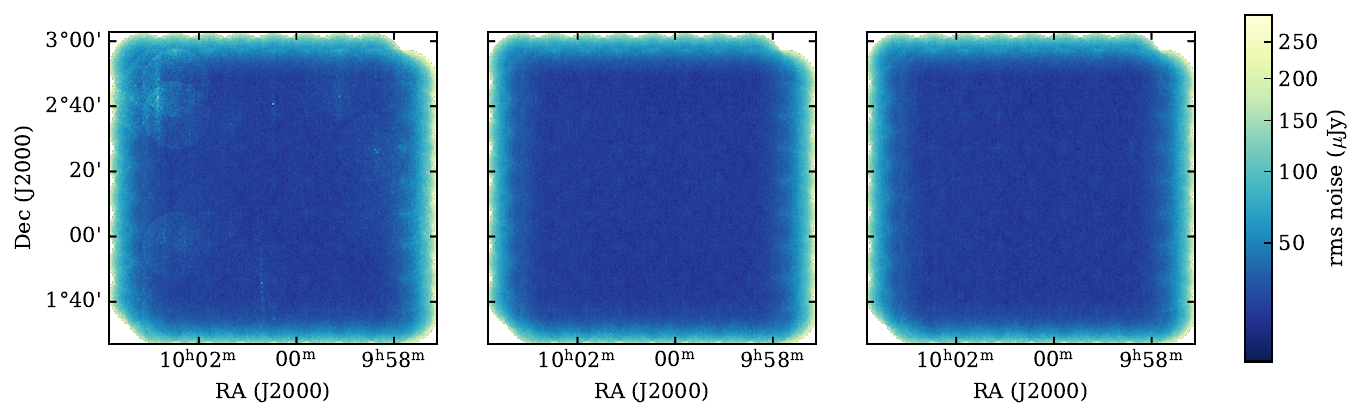}
\caption[rms noise maps for Stokes I, Q, and U mosaicked images]{rms noise maps for the central frequency ($\nu = 3002$~MHz) of the mosaicked images. Stokes~I is shown on the left, Stokes~Q in the centre, and Stokes~U on the right. The colour stretch of the maps uses a square-root scale to highlight outliers.}
\label{ch4:fig:noise_maps}
\end{figure*}
We imaged each pointing as multi-channel spectral cubes for Stokes I, Q, and U. For spectral cube imaging, we optimally selected the number of frequency channels and the channel width $\Delta \nu$,accounting for both the expected per-channel sensitivity and bandwidth depolarisation effects. We also consider that we are limited by computational time and disk capacity for the large volumes of data. Following \citet{Gaensler2001}, we determined the degree of bandwidth depolarisation $p_b$ for a range of potential $\Delta \nu$. Assuming a range of $+5<|\phi|<+400$~\radmsq, typical for extragalactic point sources at approximate arc second resolutions at high Galactic latitudes \citep{OSullivan2017,Taylor2024}, we calculated $p_b$ for $0 < \Delta\nu < 100 $~MHz. We assumed a reasonable bandwidth depolarisation limit of $p_b > 95\%$, and find that we can achieve this for $\Delta\nu < 70$~MHz over the given range of $\phi$. We opted for 64 channels with $\Delta \nu = 32$~MHz. This allows for an expected per-channel sensitivity of 18~$\mu$\jybeam for mosaicked channels, assuming natural weighting and uniform flagging across the band, and results in a sufficient number of channels for RM synthesis.

Imaging was carried out using \textsc{WSClean} for 8000$\times$8000 pixel images and a pixel size of $0.2''$. A deconvolution mask was produced using the Stokes I mosaicked image from \citetalias{Smolcic2017} by selecting pixels above 8$\sigma$, with respect to the local rms noise and performing two rounds of binary dilation. This mask was then cropped and centred to create individual masks for each pointing. Since the lower frequency channels have a field of view larger than their counterparts in the \citetalias{Smolcic2017} mosaic, we manually created masks for pointings at the edge of the mosaic by producing an initial shallow image with 100 iterations and masking below $6\sigma$. This mask was then combined with the mask from the \citetalias{Smolcic2017} image. We selected a Briggs \citep{Briggs1995} weighting of 0.7 (\textsc{casa} convention) and cleaned to a threshold of 0.3$\sigma$, based on the local rms. We also used the \texttt{join-channels} parameter. Using this method, we imaged 64 frequency channels for Stokes I, Q, and U for all 192 pointings.

To achieve a uniform angular resolution across the band, we smoothed each channel map to a target angular resolution of $\theta = 1.5''$,  a factor of two lower than the angular resolution of the multi-frequency synthesis (MFS) image in \citetalias{Smolcic2017} ($\theta = 0.75''$). This was followed by primary beam correction. We created primary beam models for each frequency channel and for Stokes I, Q, and U using the \textsc{casa} task \texttt{tclean}. The primary beam correction was applied to all channel maps using the task \texttt{pbcor}. We find that 20\% of pointings have large restoring beams (i.e. reduced angular resolution) with extreme ellipticity for one or more channels between 2129~MHz $<\nu<$ 2349~MHz (see e.g. Fig.~\ref{ch4:fig:rms_freq}) due to the large flagged percentage at these frequencies (SPWs 2 and 3). For such cases, we exclude the channel map in the mosaicked image. These areas in the mosaic are still fully covered by adjacent pointings. Excluding these channels from RM synthesis would have a minimal impact on derived parameters and uncertainties, as the affected channels are localised over a small frequency range.

We created a mosaic for each channel, and in Stokes I, Q, and U using \textsc{Montage}\footnote{\url{http://montage.ipac.caltech.edu/}}, where each image is weighted by the square of the primary beam model. On average, each channel mosaic has an rms of 22~$\mu$\jybeam in Stokes I,  Q and U. We produced rms maps for each mosaic using \textsc{PyBDSF} \citep{Mohan_2015} to map the spatial variation of noise per channel and across the mosaicked area. \textsc{PyBDSF} computes these images using a sliding box of 100 pix $\times$ 100 pix. The resultant rms maps (Stokes I, Q, and U) for the central frequency, ($\nu=3002$)~MHz are shown in Fig.~\ref{ch4:fig:noise_maps}. We note that there are regions with increased rms in the Stokes I noise map. This is attributed to increased artefacts surrounding bright sources, and is typically confined to individual pointings containing those sources. The increased noise is not present in the corresponding pointings in the Stokes Q and U maps, as these sources are significantly fainter or undetected in polarisation. The median rms noise as a function of frequency is shown in Fig.~\ref{ch4:fig:rms_freq}. Noise increases at $\nu < 2.4$~GHz and $\nu >3.6$~GHz due a larger flagged percentage at those frequencies.

\subsection{Source detection}
Due to the positive bias introduced when using traditional source finding algorithms on polarised intensity images caused by the Ricean noise distribution \citep[e.g.][]{Hales2012}, we elected not to use such algorithms for source detection in this field. Instead, we adopted the empirical approach described in \citetalias{Rudnick2014}. 

\subsubsection{Candidate selection}
The radio source population of the COSMOS field has been well defined in total intensity down to 10~$\mu$Jy in \citetalias{Smolcic2017}, and we used this catalogue as a prior when searching for polarised sources. We selected candidate sources such that each source falls within the mosaic area in all frequency channels to ensure that all sources sample the same $\lambda^2$ coverage for RM synthesis. We achieved this by inspecting the rms map at channel 63 (3.96~GHz), as it has the smallest mosaic area. We limit the extent of the \citetalias{Smolcic2017} catalogue to the area that has the lowest 20\% of rms in the 3.96~GHz mosaic. This area has an angular size of $80.43' \times 77.86'$ and covers 1.74 deg$^2$, amounting to 87\% of the total intensity image. This candidate catalogue contains 9253 of the total 10830 sources in the original catalogue.

We created cutouts in all 64 channels in Stokes I, Q, and U for each source in preparation for RM synthesis. The cutout FITS files were then combined to generate spectral cubelets for each source. By default, each cutout is 100$\times$100 pixels ($20''\times20''$). Special cases were made for sources classified as `multi-component' or sources containing more than 100 pixels, as defined by \citetalias{Smolcic2017}. There were 55 sources that met these criteria, for which we manually selected regions from which to extract the cutouts based on the spatial extent of the sources in the \citetalias{Smolcic2017} MFS image.

\subsubsection{RM synthesis}\label{ch4:sec:rmsynth}
The complex polarised intensity can be expressed as a function of Faraday depth $\phi$,
\begin{equation}
    P(\lambda^2) = Q(\lambda^2) +iU(\lambda^2) = \int ^\infty _{-\infty} F(\phi)\exp(2i\phi\lambda^2)\mathrm{d}\phi.
\end{equation}
Here, $F(\phi)$ is the Faraday spectrum, or Faraday dispersion function, which is the complex polarised intensity at a given Faraday depth \citep{Burn1966,Brentjens2005}. With RM synthesis, we obtained the observed Faraday spectrum, which is convolved with a sampling function of the $\lambda^2$ domain, the Rotation Measure Spread Function (RMSF). Through deconvolution, we can approximate the true Faraday spectrum. We performed RM synthesis to obtain the polarised intensity and RM for each candidate using \texttt{RM-tools} (see Sect.~\ref{ch4:sec:calibration}). We used \texttt{RMsynth3D} to run RM synthesis on every pixel in a given cutout cubelet, and \texttt{RMclean3D} to deconvolve these spectra. The spectra were cleaned to an absolute threshold of 27~$\mu$\jybeam. Considering the sampled $\lambda^2$ range, the resulting Faraday depth resolution and maximum Faraday depth (for which we are sensitive up to 50\%) are as follows \citep{Brentjens2005}:
\begin{align}
    \delta\phi = \frac{2\sqrt{3}}{\Delta\lambda^2} = 206~\mathrm{rad\, m}^{-2} \label{ch4:eq:faradayres},\\ 
    |\phi_\mathrm{max}| = \frac{\sqrt{3}}{\delta\lambda^2} = 15576~\mathrm{rad\, m}^{-2}.
\end{align}
The largest Faraday scale to which we are sensitive above 50\% \citep{Rudnick2023} is given as 
\begin{align}
    W_\mathrm{max} = 0.67 \left( \frac{1}{\lambda^2_\mathrm{min}}+\frac{1}{\lambda^2_\mathrm{max}} \right) = 149~\mathrm{rad\, m}^{-2}.
\end{align}

\begin{figure}
    \centering
    \includegraphics[width=0.9\linewidth]{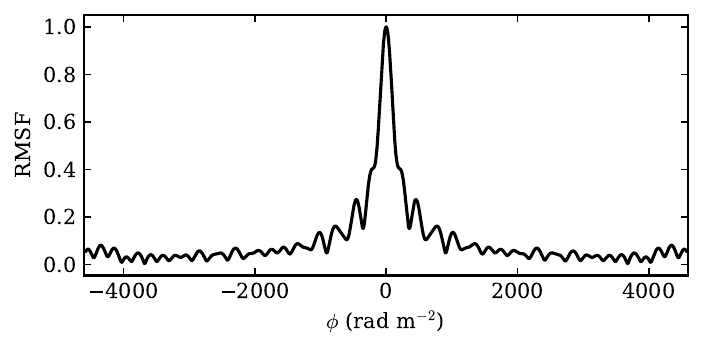}
    \caption[RMSF for VLA-COSMOS 3~GHz spectra]{Typical RMSF obtained through RM synthesis for the $\lambda^2$ range in these data.}
    \label{ch4:fig:rmsf}
\end{figure}

An example of the RMSF is shown in Fig.~\ref{ch4:fig:rmsf}. Each cleaned spectrum was then fitted with a parabola, resulting in a peak polarised intensity $P_\mathrm{peak}$ map, $\phi$ map, and a mean absolute deviation (MAD) noise map for each candidate source. In addition, the $P_\mathrm{peak}$ maps were corrected for polarisation bias, using the method described in \citet{George2012}, as implemented in the \texttt{RM-tools} software.

\subsubsection{Detection threshold}\label{ch4:sec:detection-thresh}
\begin{figure}
    \centering
    \includegraphics[width=0.9\linewidth]{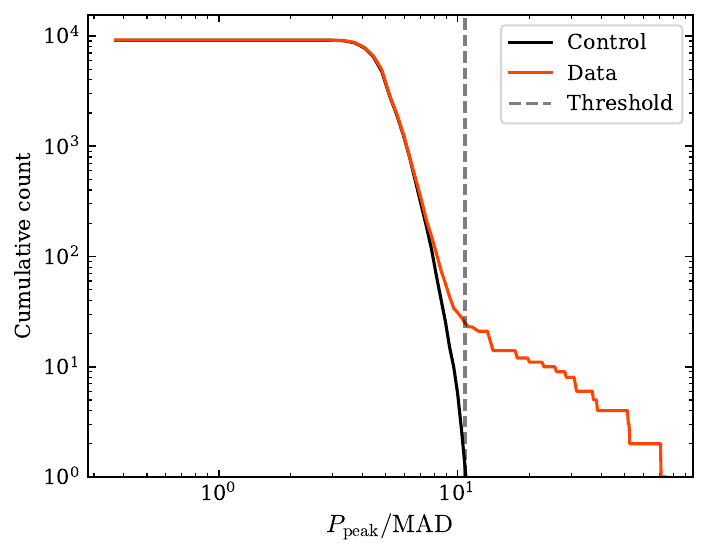}
    \caption[Cumulative distributions of the $P_\mathrm{peak}$/MAD at the peak of total intensity and for the control sample]{Cumulative distribution of the $P_\mathrm{peak}$/MAD ratios at the brightest pixel for all total intensity detections (red), and the cumulative distribution of the control (off-source) sample of $P_\mathrm{peak}$/MAD (black). The selected source-finding threshold is indicated by the dashed line at 10.7$\sigma$.}
    \label{ch4:fig:detection-thresh}
\end{figure}

As indicated above, we only searched for polarised emission at the location of sources detected in total intensity. We adapted the empirical method from \citetalias{Rudnick2014}, using a control sample. First, for all candidates classified as point sources in the \citetalias{Smolcic2017} sample (${\sim}9000$), we recorded the $P_\mathrm{peak}$ at the position of the brightest total intensity emission and determined the signal-to-noise ratio (S/N) within the Faraday spectrum at that point by dividing by the MAD. In addition, we recorded the $P_\mathrm{peak}$/MAD at four off-source positions at 6" distance (four beams) from the source as the control sample. Here, we considered $P_\mathrm{peak}$ across the full Faraday depth range $-2280<\phi<2280$~\radmsq. In Fig.~\ref{ch4:fig:detection-thresh}, we plot the cumulative distributions for both the point sources and the control sample. We define our detection threshold, at the $P_\mathrm{peak}$/MAD value for which we can expect 5\% of sources to be false detections, i.e. beyond which the 5\% of detected sources would be from the control sample. This results in a detection threshold of 10.7$\sigma$ in the Faraday depth domain. 

In our source-finding routine, we searched the full spatial extent of the total intensity maps. A source was detected if any pixel within this region exhibited a peak in the Faraday depth spectrum above 10.7$\sigma$. Of the 9253 candidate sources, we detect 65 polarised sources above this threshold, with 89 separate components. Sources typically have one or two components, with the exception being three distinct components (two lobes and a core) for particularly large sources. This yields an RM grid density of 51~deg$^{-2}$ for source components. The median MAD across all candidate sources is 2.6~$\mu$\jybeam, and the faintest detected source has a peak polarised intensity of 23.4~$\mu$\jybeam. Of the 65 sources, 55 are extended sources and ten are point sources. This source finding routine was followed by a manual inspection of both images and Faraday cubes, after which we did not find spurious detections caused by noise peaks, for example, extreme Faraday depths $|\phi|>1000$~\radmsq, typically characteristic of false detections (\citetalias{Rudnick2014}). We computed and discuss the completeness of the sample in Appendix~\ref{ch4:sec:completeness}.

\subsubsection{Total polarised intensity}
The total polarised intensity of an extended source is non-trivial to measure, as it is difficult to define the extent of the polarised region due to the positive noise bias in polarised intensity maps. Modern, large-scale polarisation surveys typically only report the peak polarised intensity -- a more practical quantity -- or do not have the angular resolution to detect large samples of extended polarised sources \citep[e.g.][]{OSullivan2023,Vanderwoude2024,Taylor2024}. Here, however, we needed the total polarised intensity to compute the Euclidean-normalised source counts (Sect.~\ref{ch4:sec:counts-euclidean}). \citetalias{Rudnick2014} identified high angular resolution polarised source detections at $1.6''$ and smoothed sources to $10''$ to measure their total polarised intensity. As many of our sources have angular size much larger than $10''$, we did not employ this method. To measure the integrated polarised intensity, we considered the extent of the source (for which we can expect to detect polarised emission) to encompass all flux density above 12$\sigma$ in the Stokes I MFS image, as determined through iterative testing and comparison. We then masked all pixels with peaks lower than $7\sigma$ in their Faraday depth spectra. To mitigate the inclusion of noise peaks, we applied this mask to the RM map and further masked all pixels with RMs more than five times the standard deviation from the mean RM (i.e. significant outliers), where the standard deviation is determined by a Gaussian fit to the unmasked RMs. Finally, we integrated over the unmasked region. Here, we assumed that polarised components of all sources are unresolved in the Faraday depth domain, such that the total polarised intensity for a given pixel has units of \jybeam, rather than \jybeam~RMSF$^{-1}$.
This is a fair assumption, given the low resolution of our Faraday spectra, which are typically consistent with the width of the RMSF. Image cutouts of Stokes I and polarised intensity maps for all sources are presented in Appendix~\ref{app:maps}.

\subsection{Spectro-polarimetric analysis}\label{ch4:sec:spectral-analysis}
\begin{figure}
    \centering
    \includegraphics[width=0.95\linewidth]{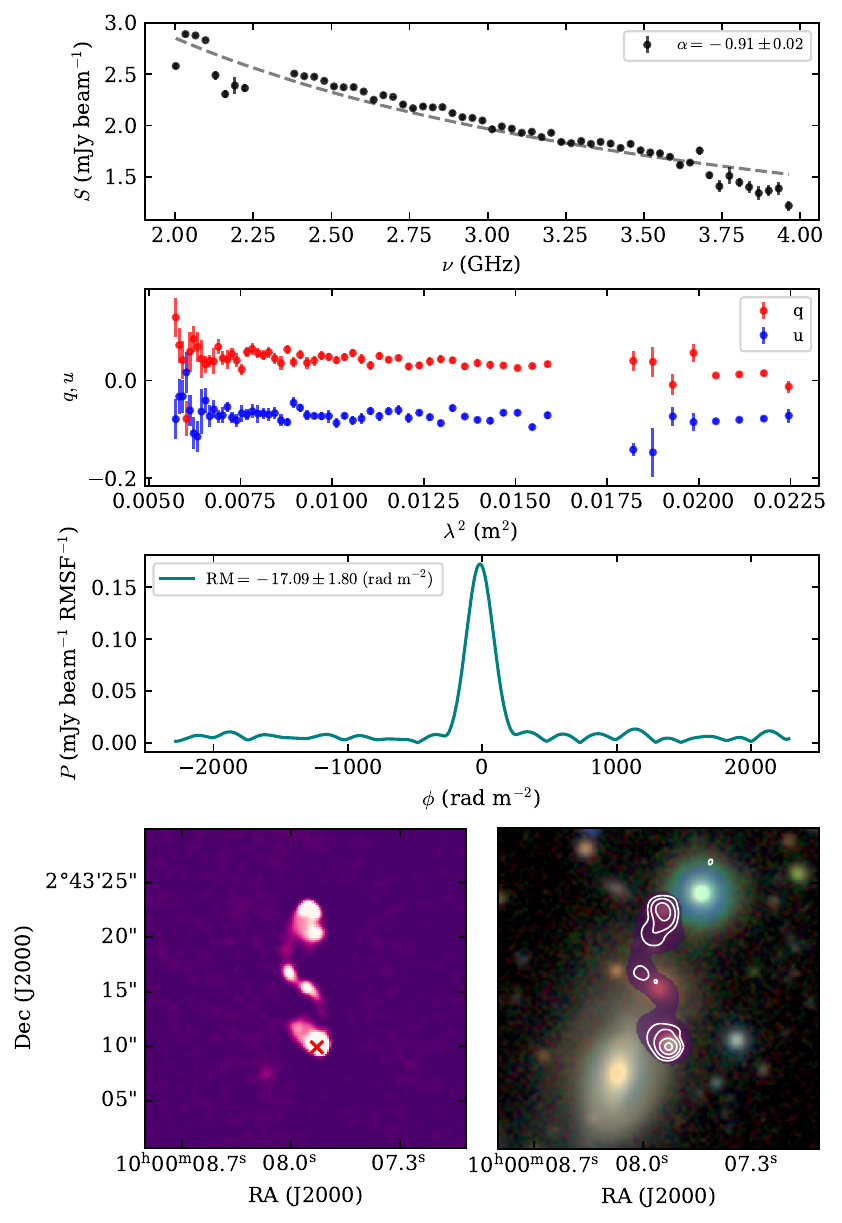}
    \caption[Example of extracted spectra and images for a source in the VLA-COSMOS 3~GHz sample]{Example of extracted spectra and images for source ID 10909. \textit{Top:} Total intensity spectrum, with the spectral index $\alpha$ indicated on the plot. \textit{Second row:} Fractional Stokes $q$ and $u$ spectra.
    \textit{Third row:} Clean Faraday spectrum, with the resultant RM indicated on the plot. \textit{Bottom left:} Total intensity image from \citetalias{Smolcic2017} ($\theta=0.75''$). The red cross shows the position at which the spectra are extracted, corresponding to the position of brightest polarised intensity. \textit{Bottom right:} Total intensity MFS image smoothed to $\theta = 1.5''$  (purple) and the polarised intensity contours plotted on an RGB optical image (UltraVISTA \citep{ultravista} $Ks$, $J$, and $H$ bands). The polarised intensity contour levels correspond to [22, 45, 90, 150]~$\mu$\jybeam. The host is an elliptical galaxy, centred on the image, while the other bright sources are likely not associated.}
    \label{ch4:fig:example_spec}
\end{figure}

For each detected source, we extracted Stokes I, Q, and U spectra at the pixel with peak polarised intensity. An example of the extracted spectra and the resulting Faraday spectrum are shown in Fig.~\ref{ch4:fig:example_spec}.

We calculated the in-band spectral index $\alpha$ by fitting a power law to the total intensity spectrum as $I(\nu) \propto \nu^{\alpha}$. In general, the total intensity spectra are well fit by a single power law, with no spectral turnovers or complex structures present in the sample. Similarly, we fitted the fractional polarisation as a function of $\lambda$ to measure the depolarisation index $\beta$, introduced by \citet{Farnes2014} as $p(\lambda)\propto\lambda^{\beta}$, where the fractional linear polarisation is given as 
\begin{align}
    p = \frac{\sqrt{Q^2+U^2}}{I}.
\end{align}
Sources with $\beta < 0$ are depolarised, likely due to internal or external Faraday dispersion, media with disordered magnetic fields, or magneto-ionic turbulence within the beam volume \citep[e.g.][]{Ranchod2024}. Ahead of fitting, we flagged the band edges and all channels with extreme uncertainties, i.e. greater than 100\%. We classified sources that show significant depolarisation as Faraday complex. We defined this limit as $\beta<0$ at the $2\sigma$ level, where $\sigma$ describes the uncertainties on the $\beta$ fit. While the $\beta$ parameter is model independent, we note its limitations in describing the physical depolarisation mechanism. The resulting $\alpha$ and $\beta$ distributions as well as the Faraday complexity are discussed in Sect.~\ref{ch4:sec:spec-depol}.

\section{Polarised source counts}\label{ch4:sec:counts-general}

\subsection{Cumulative polarised source counts}\label{ch4:sec:cumulative-counts}
\begin{figure}
    \centering
    \includegraphics[width=0.95\linewidth]{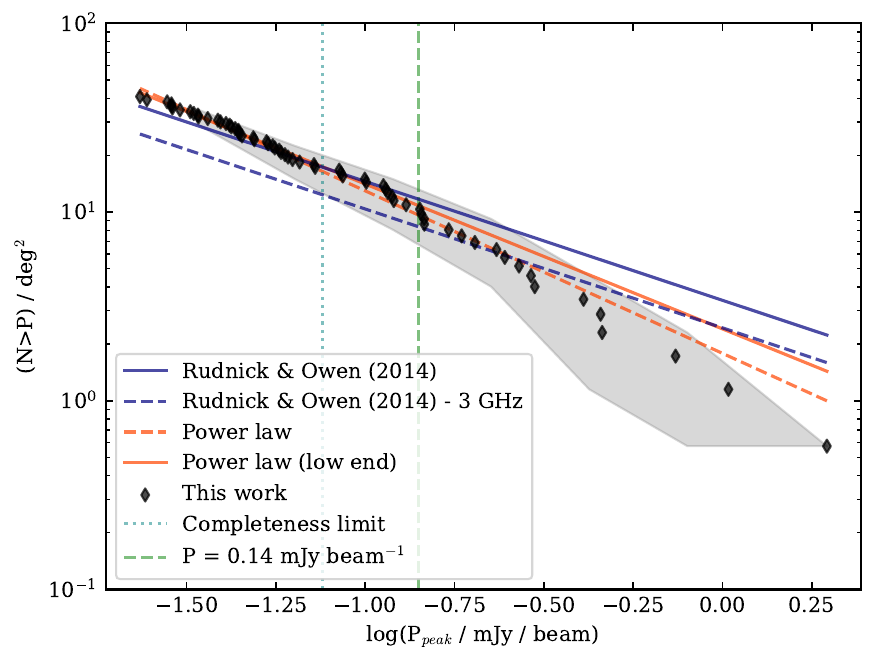}
    \caption[Cumulative source counts]{Cumulative source counts based on the peak polarised intensity of detected sources (black). These counts are corrected for completeness. The uncertainties are indicated in grey and are estimated through bootstrap resampling. The solid blue line shows the counts from \citetalias{Rudnick2014} at 1.4~GHz and extrapolated to 3~GHz (dashed blue line). The orange line represents a power-law fit (Eq.~\ref{ch4:eq:cum-counts}) to the full sample (solid line) and to only the low flux densities (dashed line). The limit above which we are 100\% complete is indicated by the dotted cyan line, and the perceived break at $P =  0.14~\mathrm{mJy~beam}^{-1}$ is shown by the dashed green line.}
    \label{ch4:fig:cumulative-counts}
\end{figure}

The cumulative counts of peak polarised intensity are particularly useful for characterising the polarised flux density distribution for deep surveys and for predicting the number of polarised sources in future observations. In Fig.~\ref{ch4:fig:cumulative-counts}, we present the cumulative counts for peak polarised intensity. They were corrected for completeness (Appendix~\ref{ch4:sec:completeness}), and the uncertainties were estimated through bootstrap resampling, as detailed in this section. We note that the bootstrap uncertainties used here are consistent with Poisson uncertainties on the source counts. To investigate the frequency dependence of the cumulative counts, we compare them to those presented in \citetalias{Rudnick2014}, cumulative counts at 1.4 GHz at a comparable angular resolution ($\theta = 1.4''$), also using the VLA but over a smaller area of 0.3~deg$^2$ in the GOODS-N field. While they only detected 11 polarised sources in this small region with good certainty, their work produced well-constrained analytical expressions for cumulative source counts due to their correction of systematic effects and is most widely cited in the literature. Here, we applied this method with a larger sample size for the first time, at 3~GHz. \citetalias{Rudnick2014} found a flattening in the cumulative distribution below 1 m\jybeam, which they attribute to the change in source population at fainter polarised intensities. This flattening has subsequently been reported for shallower surveys using the Australia Telescope Compact Array \citep[ATCA;][]{Eyles2020} and the Westerbork Synthesis Radio Telescope \citep[WSRT;][]{Berger2025}. We find that our results are consistent with those of \citetalias{Rudnick2014} at low flux density $P<0.14~$m\jybeam within the bootstrap uncertainties but appear to have a steeper slope at higher flux densities.  
 
To understand if the change in slope is statistically significant, we performed Bayesian model selection between a power law and two different broken power laws, assuming a Poisson likelihood function. We used nested sampling to evaluate the evidence $\log(Z)$ with the \textsc{python} package \texttt{bilby} and the \texttt{dynesty} sampler. For all parameters, we defined uniform priors. We assessed the Bayesian evidence ratio between a power law (model A) given by
\begin{align}\label{ch4:eq:cum-counts}
    N (>P) = N_0 \left( \frac{P}{P_0}\right) ^{\alpha_1},
\end{align}
and a broken power law
\begin{align}
N( >P) =
\begin{cases}
N_0 \left( \dfrac{P}{P_0} \right)^{\alpha_2} & \text{if } P \leq {P_{\text{break}}} \\
N_0 \left( \dfrac{{P_{\text{break}}}}{P_0} \right)^{\alpha_2 - \alpha_1} \left( \dfrac{P}{P_0} \right)^{\alpha_1} & \text{if } P > {P_{\text{break}}},
\end{cases}
\end{align}
considering options where $P_\mathrm{break}$ is either fixed at 0.14~m\jybeam (model B) or is a free parameter (model C). As our motivation is to determine the statistical significance of our perceived slope break, model B is defined by the visual identification of the break in Fig.~\ref{ch4:fig:cumulative-counts}. Above, $N_0$ is a normalisation constant, while $\alpha_1$ and $\alpha_2$ are the distribution slopes below and above the break, respectively. $P_\mathrm{break}$ is the polarised intensity at which the break occurs, and $P_0$ is the polarised intensity of the faintest source in the sample. For the broken power law with free $P_\mathrm{break}$, we find a Bayes factor of $\log(Z) = 0.162$ with respect to the power law. With fixed $P_\mathrm{break} =  0.14$~m\jybeam, we find $\log(Z) = 0.517$, with respect to the power law. While this does not strongly favour the power-law model, we selected the power law, as it is the simplest model. Furthermore, the best fit for $P_\mathrm{break}$ is $P_\mathrm{break} = 0.66^{+3.61}_{-0.60}~\mathrm{mJy\,beam}^{-1}$, above which we only have one data point. A break here would be consistent with the change of slope found in the literature.

As an additional test for the perceived change in slope, we performed a bootstrapping experiment where we selected 65 sources (with replacement) from our sample and construct $10^4$ realisations of the cumulative source counts. The shaded grey region in Fig.~\ref{ch4:fig:cumulative-counts} shows the uncertainty envelope defined by the 16th and 84th percentiles of the bootstrapped cumulative distributions. From this we can conclude that the perceived change of slope is not statistically significant and is likely an effect of small number statistics, particularly for higher polarised flux densities, where only 25\% of sources are above 0.14~m\jybeam. This is not unexpected as it is well established in total intensity that few bright sources are detected in the COSMOS field, an effect of the small sky area due to cosmic variance \citep[e.g.][]{Heywood2013}. We therefore limited the power-law fit to the lower flux densities with $P< 0.14$~m\jybeam. The resulting power-law fit is
\begin{align}\label{ch4:eq:my-cum-counts}
    N (>P) = 42.89\pm0.37 \left(\frac{P}{23\,\mu \mathrm{Jy~beam}^{-1}}\right)^{-0.77\pm0.01}.
\end{align}
From Fig.~\ref{ch4:fig:cumulative-counts}, we find that at 3~GHz we obtain consistent cumulative source counts with \citetalias{Rudnick2014} at 1.4~GHz. This consistency is interesting, as we expect sources to be fainter at higher frequencies due to the effect of spectral indices (typically $\alpha=-0.7$ for radio galaxies). In Fig.~\ref{ch4:fig:cumulative-counts} we plot the \citetalias{Rudnick2014} cumulative counts extrapolated to 3~GHz, assuming $\alpha=-0.7$. The inconsistency of our data with this relation at 3~GHz suggests the spectral index does not fully describe the frequency dependence of the cumulative counts. Through an analysis of the polarised sources in the S-PASS survey (2.3~GHz) in comparison to the NVSS polarisation catalogue (1.4~GHz), \citet{Lamee_2016} suggest that number counts may be consistent between 2.3~GHz and 1.4~GHz for bright sources ($P_{2.3~\mathrm{GHz}}>230$~mJy) at low angular resolution (${\sim}9$~arcmin). They attribute this to the increase in polarised intensity at 2.3~GHz in their sample, an effect of $\lambda^2$-dependent depolarisation. We present our depolarisation results in Sect.~\ref{ch4:sec:spec-depol} and further discuss the frequency-dependence of the polarised source counts in Sect.~\ref{ch4:sec:freq-depend}.

\subsection{Euclidean-normalised counts}\label{ch4:sec:counts-euclidean}
\sidecaptionvpos{figure}{c}
\begin{figure*}
    \centering
    \includegraphics[scale=0.6]{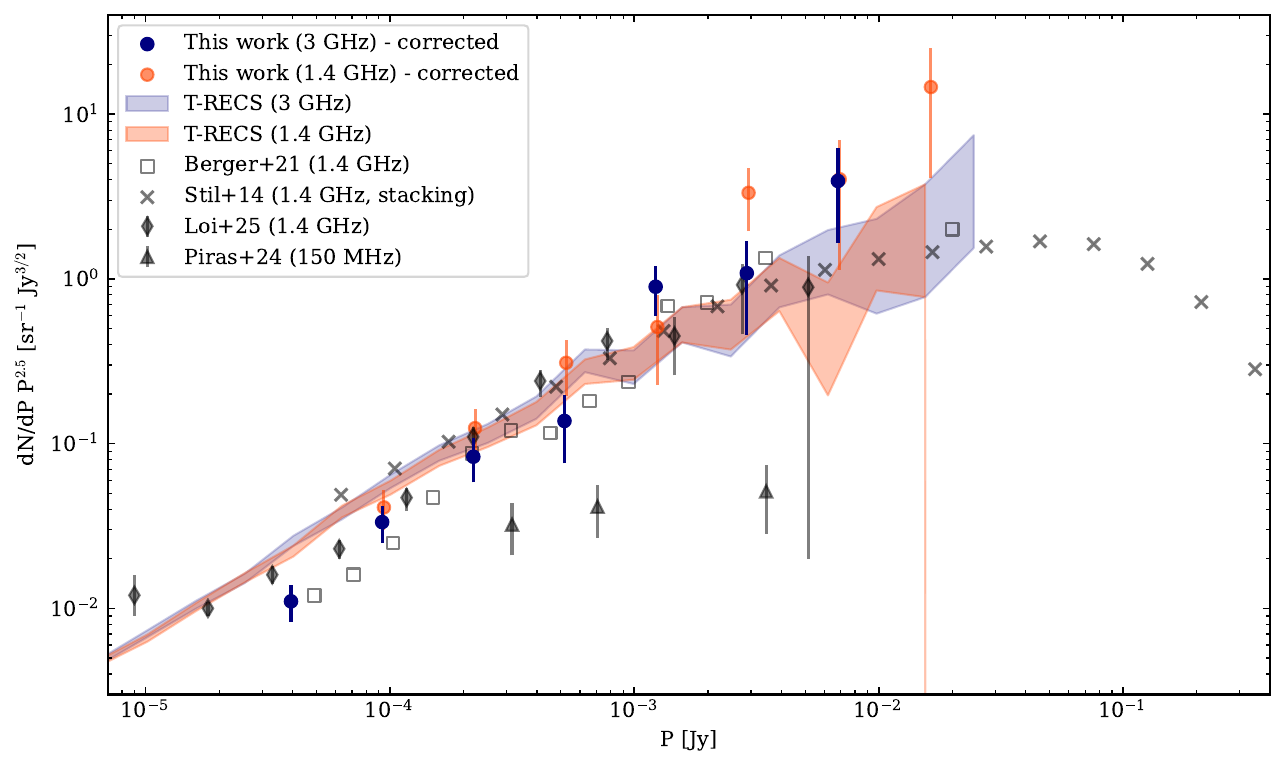}
    \caption[Euclidean-normalised differential source counts]{Euclidean-normalised differential source counts for the polarised source population in the VLA-COSMOS 3~GHz survey. The completeness-corrected counts are plotted as filled circles at 3~GHz (blue) and extrapolated to 1.4~GHz (orange). We plot the source counts from the following recent deep surveys (grey): Lockman Hole field \citep[][open squares]{Berger2021}, MeerKAT Fornax Survey \citep[][filled diamonds]{Loi2025}, ELAIS-N1 \citep[][triangles]{Piras2024}, and stacking of NVSS sources \citep[][crosses;]{Stil2014}. The polarised source counts computed from the T-RECS simulation \citep{Bonaldi2019} at 1.4~GHz (orange) and 3~GHz (blue) are also plotted as shaded regions.}
    \label{ch4:fig:euclidean-counts}
\end{figure*}

Euclidean-normalised differential source counts are a standard depiction of source counts in the literature \citep{Condon1984}, assuming a static, Euclidean Universe. This representation has historically been used for total intensity \citep[e.g.][]{Hales2014,Smolcic2017,Matthews2021} and polarised intensity \citep[e.g.][]{Hales2014,Berger2021} counts. Typically, any deviation from a monotonic decrease towards lower flux densities is indicative of a change in source population, (e.g. the emergence of the star-forming galaxy population as observed in total intensity; \citealt[e.g.][]{Matthews2021}). However, this regime has not yet been observed in polarised intensity, with the exception of \citet{Loi2025} who find an increase at $P_\mathrm{1.4~GHz} < 10~\mu$Jy. They suggest that these are sensitive to the radio-quiet quasar population but not SFGs. In Fig.~\ref{ch4:fig:euclidean-counts}, we present the Euclidean-normalised differential polarised source counts for the VLA-COSMOS 3~GHz survey. These source counts were computed using $P_\mathrm{tot}$, in contrast to $P_\mathrm{peak}$ used above. This is an important distinction as the majority of detected sources are resolved. We also display the source counts extrapolated to 1.4~GHz to compare with the literature, assuming $\alpha=-0.7$. We calculated Poisson uncertainties for the number of sources detected in each bin. Similar to Sect.~\ref{ch4:sec:cumulative-counts}, we corrected the source counts for completeness, as calculated in Appendix~\ref{ch4:sec:completeness} and plot the corrected counts with only one bin below our completeness limit of 75~$\mu$\jybeam.

In Fig.~\ref{ch4:fig:euclidean-counts}, we plot the source counts from other sub-mJy polarised source count studies: The Lockman Hole field at 1.4~GHz with the WSRT \citep{Berger2021}, ELAIS-N1 at 150~MHz observed with the LOw Frequency ARray \citep[LOFAR;][]{Piras2024}, the MeerKAT Fornax Survey at 1.4~GHz with MeerKAT \citep{Loi2025}, and the result of source stacking in NVSS at 1.4 GHz \citep{Stil2014}. The source counts presented in this work are the deepest polarised source counts at $\nu >2$~GHz and the second deepest overall \citep[cf.][ at 1.4~GHz as the deepest polarised counts]{Loi2025}. For the first time, we probe the sub-mJy regime at higher frequencies over a significant area. In Table~\ref{ch4:tab:source-counts}, we summarise the polarised intensity bins, the number of sources per bin (with and without completeness correction), as well at the Euclidean normalised counts (with completeness correction). 
\begin{table}[]
    \centering
    \caption[Euclidean-normalised differential polarised source counts]{Euclidean-normalised differential polarised source counts.}
    \begin{tabular}{ccccc}
    \hline
    $P_\mathrm{3\,GHz}$ & $\Delta P_\mathrm{3\,GHz}$ & $N$ & $N_\mathrm{corr}$ & $\mathrm{d}N/\mathrm{d}P\, P^{2.5}$ \\  
    {}[mJy] & {}[mJy] & & & $[\mathrm{Jy}^{1.5}\mathrm{sr}^{-1}]$ \\ \hline
       0.039  & 0.023--0.055 & 18 & 19 & 0.041$\pm$0.021\\
       0.093  & 0.055--0.131 & 16 & 16 & 0.124 $\pm$ 0.075\\
       0.219  & 0.131--0.308 & 11 & 11 & 0.31 $\pm$ 0.22\\
       0.517  & 0.308--0.728 & 5  & 5  & 0.51 $\pm$ 0.55\\
       1.222  & 0.728--1.717 & 9  & 9  & 3.3 $\pm$ 2.7\\
       2.885  & 1.717--4.054 & 3  & 3  & 4.0 $\pm$ 5.5\\
       6.812  & 4.054--9.570 & 3  & 3  & 14 $\pm$ 20\\ \hline
    \end{tabular}
    \label{ch4:tab:source-counts}
\end{table}

For our extrapolated counts at 1.4~GHz, we are consistent with \citet{Loi2025} within uncertainties. We note that we do not plot the three highest bins from \citet{Loi2025}, as they include bins with zero counts. An increase in bin size here would change the number counts in these bin, and they are therefore not bin-independent. A comparison of the counts in the three highest bins to our work is therefore non-trivial. In general, our source counts are compatible with \citet{Stil2014} at higher polarised intensities with $P>$~0.131~mJy, below which our counts are lower by a discrepancy of 0.5 dex. We also show general agreement with the source counts from \citet{Berger2021}; however, a subsequent paper from these authors \citep{Berger2025} acknowledge that certain corrections were not accounted for here, so we do not compare further. Finally, for completeness, we include the counts from \citet{Piras2024}, as these are the only sub-mJy counts at megahertz frequencies. However, a spectral extrapolation to 100~MHz is non-trivial due to the more pronounced $\lambda$-dependent depolarisation effects at $\nu<1$~GHz. 

The extrapolated 1.4 GHz counts have both higher flux densities and Euclidean-normalised values than at 3~GHz, as expected from Fig.~\ref{ch4:fig:cumulative-counts}, with the exception of the bin at $P = 1.22$~mJy. This bin in $P_\mathrm{1.4~GHz}$ corresponds to the bin in $P_\mathrm{3~GHz}$ that indicates the transition range between resolved and unresolved sources in our sample, and this dearth may be related to the limited surface brightness sensitivity of our observations, i.e. missing faint, diffuse emission in the flux integration across a given source, or an effect of the binning with low number statistics. Further investigations are beyond the scope of this work. Given the modest number of detections in our sample and the large per-bin uncertainties, this increase is not statistically significant, and we find that cumulative counts are a more appropriate measure for a frequency-dependent comparison.

Additionally, we compare the polarised source counts to those from the Tiered Radio Extragalactic Continuum Simulations \citep[T-RECS;][]{Bonaldi2019}. This is a set of simulated catalogues of the radio continuum sky (150~MHz $<\nu<$ 20~GHz), including both the AGN and SFG population, designed for predicting results of deep radio surveys with the SKA. This catalogue is unique in its inclusion of polarisation information for all sources, which were not included in previous iterations due to the lack of available data. In this work we used the T-RECS medium catalogue, which covers a 25~deg$^2$ area down to a depth of $S_\mathrm{1.4~GHz} < 10$~nJy. In Fig.~\ref{ch4:fig:euclidean-counts}, we plot the 1$\sigma$ uncertainty range of the Euclidean-normalised source counts from T-RECS at 1.4~GHz and 3~GHz from $5~\mu\mathrm{Jy} < P < 0.1~\mathrm{Jy}$. Here, we only include the AGN population from the T-RECS catalogue (see Sect.~\ref{sec:sfg-detection}). The source counts and their uncertainties are calculated as above. We find that there is no significant difference between the source counts at 1.4~GHz and 3~GHz and that these simulated counts show good agreement with our measurements and those from the literature for $P>500~\mu$Jy. At fainter polarised intensities, we find that the simulated source counts are significantly overestimated. We note that \citet{Bonaldi2019} assumed the fractional polarisation distribution from \citet{Hales2014}, which was derived from source counts with $P_\mathrm{1.4~GHz}>200~\mu$Jy, and at the time, source counts below this limit were not readily available. Moreover, the fractional polarisation for flat-spectrum sources was determined from a sample selected at 20~GHz with $P_\mathrm{20~GHz}>200~\mu$Jy \citep{Galluzzi2018}, which may introduce a bias in the extrapolation down to 3~GHz. In general, this highlights the necessity for deep polarised source counts, as presented in this work, over a large range of frequencies to inform and update simulations of the faint source population ahead of the SKA era.

\section{Detection of star-forming galaxies}\label{sec:sfg-detection}
The role of magnetic fields in SFGs can be investigated through the detection of a statistically significant sample of polarised SFGs. However, due to the faint nature of this source population and beam depolarisation effects, the polarised SFG population has not been readily detected in deep extragalactic surveys \citep[e.g.][]{Berger2021,Taylor2024,Loi2025}. At 3~GHz, we expect a higher integrated fractional polarisation for unresolved SFGs \citep{Stil2009,Sun2012}. This, combined with the high angular resolution and sensitivity of the VLA-COSMOS survey provides an advantageous perspective for the detection of the polarised SFG population. 

\citet{Smolcic2017b} and \citet{Delvecchio2017} provide a detailed classification of 93\% of the total intensity detections into SFG and AGN populations, based on their multi-wavelength counterparts. To summarise, SFGs were identified based on their dust-extinction-corrected rest-frame colour and the absence of radio excess with respect to the radio luminosity--star-formation rate relation. They classified sources as `Clean SFG' where they show no evidence of AGN activity. We cross-matched our detected polarised sources to the source classification catalogue from \citet{Smolcic2017b} and confirm that all polarised sources fall under the AGN classification. We do not detect polarised SFGs, as defined by the `Clean SFG' tag. This is despite the relatively higher frequency and depth of these observations. Furthermore, in Fig.~\ref{ch4:fig:euclidean-counts}, our source counts show the expected monotonic decrease towards low flux densities. This is another indication that we do not have the sensitivity to detect a change in population.  We discuss the detection of polarised SFGs in Sect.~\ref{ch4:sec:detect-disc}.

\section{Spectral index and depolarisation}\label{ch4:sec:spec-depol}
In Sect.~\ref{ch4:sec:spectral-analysis}, we determined the spectral index $\alpha$ in total intensity, as well as the depolarisation index $\beta$. The broadband characterisation of polarised sources is currently an active field of research for both the interpretation of the Galactic foreground \citep[e.g.][]{Anderson2015,Ma2020,Ranchod2024} as well as the physical properties of radio galaxies \citep[e.g.][]{Farnes2014,OSullivan2017,Pasetto2018}. However, to date this has been limited to the bright polarised population, typically with $P>1$~mJy. While we do not present the results of QU-fitting with detailed models in this work, we quantified and report the depolarisation using $\beta$ as a first step in characterising the properties of these faint polarised sources. 
Additionally, the $\lambda^2$ range, which we sampled at 3~GHz, is less susceptible to $\lambda$-dependent depolarisation, allowing us to better understand which source populations would be undetected at lower frequencies due to strong depolarisation effects.
 
\begin{figure}
    \centering
    \includegraphics[width=0.9\linewidth]{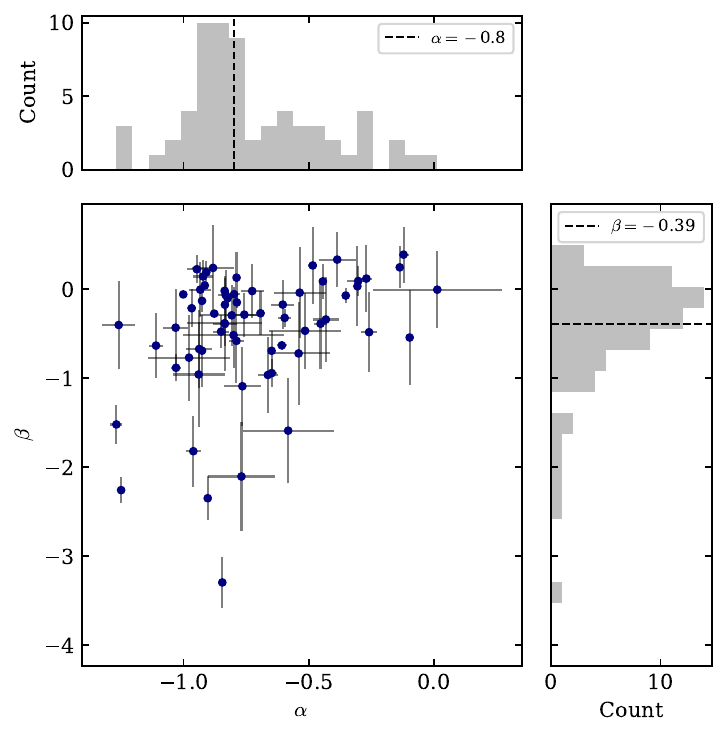}
    \caption[Spectral index ($\alpha$) vs depolarisation index ($\beta$) for the $P_\mathrm{peak}$ of all sources]{Spectral index ($\alpha$) against depolarisation index ($\beta$) for the $P_\mathrm{peak}$ of all sources.
    Histograms of the $\alpha$ and $\beta$ distributions are shown at the top and on the right, respectively. The solid black lines indicate their median values, as provided in the legend.}
    \label{ch4:fig:alpha-beta}
\end{figure}

In Fig.~\ref{ch4:fig:alpha-beta} we present the distribution of the total intensity spectral index $\alpha$ and depolarisation index $\beta$, as defined in Sect.~\ref{ch4:sec:spec-depol}. These values and their uncertainties are presented in the catalogue associated with this work (Appendix~\ref{sec:catalogue}). The $\alpha$ measurements exhibit typical errors of ${\sim}10$\%, and the $\beta$ fits typically exhibit larger errors of ${\sim}50$\% due to the low S/N and the complex nature of the spectra. The $\alpha$ distribution is as expected for the S-band frequencies \citep[e.g.][]{Smolcic2017b,Ranchod2025}: a bimodal distribution peaking at $\alpha_\mathrm{med} = -0.8$, with the secondary peak for flat-spectrum sources centred at $\alpha\approx-0.4$. In terms of radio-AGNs, flat spectra ($\alpha>-0.5$) usually indicate core-dominated sources, i.e. the optically thick region at the base of the jets, while steep spectra are dominated by the optically thin jet or lobe. Since the majority of sources are spatially resolved in total intensity, we can visually identify from which part of the source the peak spectrum is extracted. In the case of the flat-spectrum sources in this sample, all except one (ID 145) are compact sources, or the $P_\mathrm{peak}$ is at the AGN core. In the special case of ID 145 (see Appendix~\ref{app:maps}), a resolved FRII galaxy, the $P_\mathrm{peak}$ pixel is not associated with the core or the total intensity hotspots on the lobes but is rather located towards the edge of the source. This is characteristic of the classical increased polarisation fraction at the edge of radio lobes due to magnetic field alignment after interaction with the intergalactic medium, as determined through observations \citep[e.g.][]{Baidoo2023} and numerical simulations \citep[e.g.][]{Stimpson2025}. Therefore, the total intensity spectrum for this position has an insufficient S/N for a reliable measurement of $\alpha$.
 
The relationship of $\alpha-\beta$ seen in this work is consistent with that from \citet{Farnes2014}, where the authors characterised the broadband polarisation properties of ${\sim}10^3$ bright sources ($P_{1.4~\mathrm{GHz}}>3$~mJy) from 400~MHz to 100~GHz using narrowband observations from a range of different surveys. This distribution of sources signifies that depolarisation is an intrinsic property of the source, rather than an external effect. As in \citet{Farnes2014}, there is a clear separation between flat- and steep-spectrum sources. Steep-spectrum sources show a moderate ($\beta>-1$) or high depolarisation ($\beta < -1$), while flat-spectrum sources exhibit a $\beta$ distribution symmetric about $\beta=0$. This includes measurements of $\beta>0$, or `repolarisation', which is caused by the more complex total intensity spectral behaviour of an optically thick, core-dominated source (e.g. synchrotron self-absorption), or by fractional polarisation models with sinc-like variation as a function of $\lambda^2$ (e.g. Burn slab). Our results show that repolarisation is observed for flat-spectrum sources, even with a limited frequency range, consistent with \citet{Taylor2024}. This consistency with \citet{Farnes2014} and \citet{Taylor2024} despite the smaller $\lambda^2$ coverage, which suggests that the ${\sim}1-4$~GHz range may be the most relevant when characterising the broadband behaviour of polarised extragalactic sources. The majority of sources occupy the parameter space where, within uncertainties, $\alpha\approx\beta$, which may contribute to the absence of frequency dependence in the sources counts presented in Sect.~\ref{ch4:sec:counts-general}. This relation between depolarisation and source counts will be discussed in detail in Sect.~\ref{ch4:sec:freq-depend}.
 
We consider sources to be Faraday complex when they show significant depolarisation, i.e. $\beta < 0$ within $2\sigma$ of uncertainties. We find that 12 sources show Faraday complexity, which is 19\% of the sample. It is non-trivial to compare source Faraday complexity fractions with those in the literature, as this is highly dependent on the bandwidth, angular resolution, and complexity metric. Furthermore, the complexity can be based on depolarisation \citep[e.g.][]{Lamee_2016}, QU-fitting \citep[e.g.][]{Ranchod2024,Ma2025}, or resolution of the the Faraday spectra \citep[e.g.][]{Livingston2021}. The latter depends on $\delta\phi$ and max-scale of the observations. Despite the variety of methods for measuring Faraday complexity, we obtain a consistent complexity fraction compared to other extragalactic polarisation surveys in the literature. For observations with the same bandwidth of 2--4~GHz for a range of extragalactic fields, \citet{Pandhi2025} find 21\% complexity based on the Faraday spectra. Similarly, over a small frequency range in the POSSUM pilot observations, \citet{Vanderwoude2024} find a complexity fraction of 22\% at $b\sim50^\circ$.  

We note that the Faraday-simple sources, identified through their Faraday spectra, showing depolarisation with $\beta$, typically exhibit high uncertainties in polarised intensity, i.e. they are detected with relatively lower significance. This is likely due to a bias towards higher S/N sources being Faraday complex \citep[e.g.][]{Anderson2015}, as we are not sensitive to the Faraday complexities in the QU-spectra above the noise level for faint sources. This also explains why for $|b|>20^\circ$, \citet{OSullivan2017} report a comparatively high complexity fraction of 55\% for broadband ($1-3$~GHz) ATCA observations of 100 extragalactic sources with $I_\mathrm{2.1~GHz}>10$~mJy, as polarised sources are detected with higher significance.

\section{Discussion}\label{ch4:sec:discussion}
In the following section, we discuss the limitations of source detectability for deep, high-resolution polarisation surveys, as well as the drivers behind the consistent source counts between 1.4~GHz and 3~GHz.

\subsection{Nature of the faint polarised source population}\label{ch4:sec:detect-disc}
\begin{figure}
    \centering
    \includegraphics[width=0.95\linewidth]{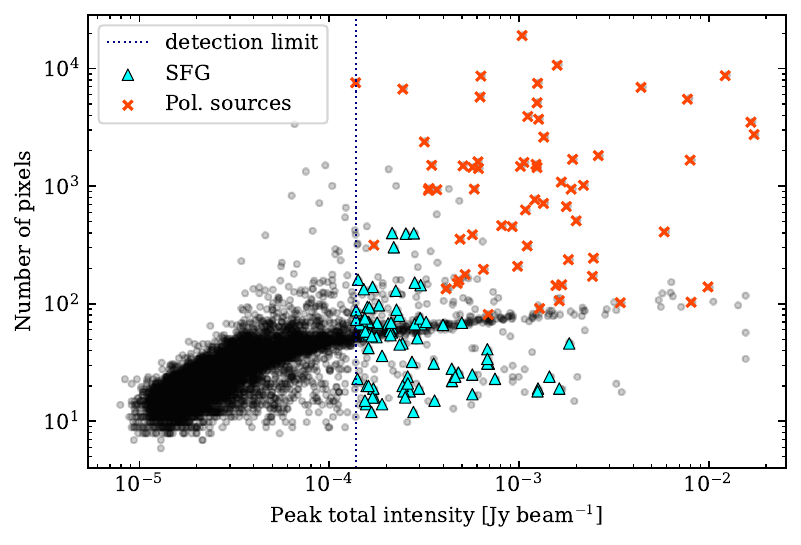}
    \caption[Peak total intensity plotted against the number of pixels, as detected in the total intensity catalogue from \citetalias{Smolcic2017}]{Peak total intensity plotted against the number of pixels, as detected in the total intensity catalogue from \citetalias{Smolcic2017} (black). The sources detected in polarisation are indicated by orange crosses, and the polarisation detection limit from this work is shown by the vertical dotted line (i.e. the minimum peak total intensity of sources detected in polarisation). The cyan triangles represent the SFGs above the detection threshold.}
    \label{ch4:fig:undetected}
\end{figure}

In Sect.~\ref{ch4:sec:counts-euclidean}, we show that all detected polarised sources are AGNs. To understand the limitations of this survey in terms of source detectability for various populations, we investigate the difference between the sources detected and not detected in polarisation. Figure~\ref{ch4:fig:undetected} shows the peak total intensity against the number of pixels (in total intensity) for each source in the \citetalias{Smolcic2017} catalogue. At the beam size of ${\sim}75$~pixels, there is a linear clustering of point sources, above which sources can be considered resolved. From here it is apparent that the detected sources are among the brightest and largest of the catalogue. We detect in polarisation almost all sources with more than ${\sim}1200$ pixels (17 beams), with one exception. Similarly, all extended sources with $S_\mathrm{peak} >1$~mJy are also detected in polarisation. These highly extended sources are in the minority compared to the full sample of mostly point-like and slightly resolved sources. This trend of preferentially detecting extended sources in polarisation is consistent with that found in \citetalias{Rudnick2014} for their sample of 11 sources also at arc second resolutions. 
The trend of preferentially detecting bright sources does not follow for point sources, where the polarised point sources are not necessarily the brightest and numerous bright point sources are not detected in polarisation. Our results show that these compact sources have lower $\beta$ values and are affected by depolarisation. While the detectability of radio sources is linked to their fractional polarisation, an observable for magnetic field ordering, this does not necessarily imply that larger, brighter sources have more ordered magnetic fields. It may also be an effect of beam depolarisation and due to the fact that small sources become depolarised below the detection threshold even at arc second resolutions. This can be further explored through high angular resolution observations of resolved polarised AGNs.

As noted in Sect.~\ref{sec:sfg-detection}, we do not detect any SFGs within the survey area. This is likely a combination of observational effects and the physical properties of SFGs. Various observations and simulations have shown that SFGs have significantly lower integrated fractional polarisations than AGNs \citep{Stil2009,Beck2015}. As a result, only the magnetic field properties of nearby, resolved SFGs have been studied in detail, and such sources are rare and absent from the limited area of the COSMOS survey.  
For example, \citet{Kierdorf2020} present broadband observations of the nearby, resolved face-on SFG M51, and report fractional polarisations averaging at $p\sim15\%$ at 3~GHz, and $p\sim25\%$ at 4.85~GHz (VLA C band).  For unresolved galaxies, \citet{Stil2009} and \citet{Sun2012} showed that the detection of SFGs is highly dependent on the inclination angle of the galaxy, and as such there is a low probability of detecting unresolved SFGs due to geometrical (beam) depolarisation and Faraday effects. SFGs are more susceptible to this effect in comparison to AGNs, as their large-scale magnetic fields vary azimuthally, roughly following the orientation of the spiral arms \citep{Beck2015}, resulting in a large range of EVPAs and thus higher beam depolarisation. \citet{Sun2012} investigated the frequency dependence of the inclination angle-dependence using 3D emission models of Milky Way-like galaxies, including the regular disk field, the regular halo field and the random magnetic field. Assuming all modelled galaxies were observed as point sources, their results showed a strong increase in fractional polarisation with inclination angle at frequencies $\nu>2.7$~GHz. They found that for $\nu=4.8$~GHz, this peaked at 70$^\circ$ with a fractional polarisation of 4.3\%, significantly lower than found in resolved nearby galaxies. 

In terms of observational effects, the array configuration selected for the VLA-COSMOS 3~GHz survey favours higher angular resolution over sensitivity to diffuse emission. As such, we may resolve out diffuse emission from extended SFGs. \citet{Delhaize2017} convolved the \citetalias{Smolcic2017} total intensity mosaic to various angular resolutions larger than the native $0.75''$ up to a maximum of $3''$ to investigate this effect. From this, they detect 455 additional SFGs based on prior positions determined from infrared data, which are unaccounted for in the \citetalias{Smolcic2017} catalogue. While this illustrates the resolution bias in these data, their newly detected sources do not exceed $S_\mathrm{3\,GHz} = 45~\mu$\jybeam and therefore would require extreme fractional polarisations of $p>30$\% to be detected in polarisation. Based on our non-detection of SFGs, we can determine an upper limit on the sky density of polarised SFGs at 3~GHz. Following \citet{Gehrels1986}, we find the $2\sigma$ upper limit of the density of polarised SFGs to be $<2.04~\mathrm{deg}^{-2}$ at $P_\mathrm{3\,GHz}> 15~\mu$\jybeam for a high Galactic latitude field. 

In Fig.~\ref{ch4:fig:undetected}, we indicate the SFGs with a peak total intensity above the minimum peak total intensity of the polarised sources. The majority of these galaxies are unresolved. For the resolved SFGs above the detection threshold, we calculate an upper limit on their polarisation fraction to be a maximum of $p < 8$\% based on the pixel of peak total intensity. Although we cannot directly compare it with that derived from the simulations of \citet{Sun2012} -- they modelled only unresolved galaxies -- this upper limit is consistent with that found for $\nu=4.8$~GHz, as mentioned above. This consistency supports the assumptions made in \citet{Sun2012} concerning the presence of field reversals in the regular disk field, and the contribution of the various random field components within the disk to increased depolarisation for disk galaxies. Our results substantiate the understanding that the star-forming source population does not contribute to the polarised source counts, even at sub-mJy flux densities \citep{OSullivan2008}. Future SKA-era surveys will require observations much deeper than $2.6~\mu$\jybeam to readily detect the SFG population in polarisation.

\subsection{Frequency evolution of polarised source counts}\label{ch4:sec:freq-depend}
Given the $\lambda$-dependence of depolarisation, one might expect an increased polarised source count at higher frequencies due to the detection of sources that would be depolarised at lower frequencies. In Sect.~\ref{ch4:sec:cumulative-counts}, we presented the cumulative source counts for the COSMOS field at 3~GHz, and find consistency with the source counts of \citet{Rudnick2014} at 1.4~GHz (making no correction for spectral index). With these observations, we confirm the prediction by \citet{Lamee_2016} that source counts with the VLA Sky Survey (i.e. at the same frequency range and angular resolution of this work) would be consistent with those at 1.4~GHz. Their prediction was based on the comparison between the polarised source detections in the S-PASS (2.3~GHz) and NVSS (1.4~GHz) surveys for $P>230$~mJy, where they investigated the spectral index and depolarisation ratio $P_\mathrm{2.3\,GHz}/P_\mathrm{1.4\,GHz}$. To further understand the interplay between spectral index and depolarisation for the sub-mJy population, we modelled the effect of these parameters on the polarised intensities at 3~GHz and the detected source counts below.
 
\begin{figure}
    \centering
    \includegraphics[width=0.95\linewidth]{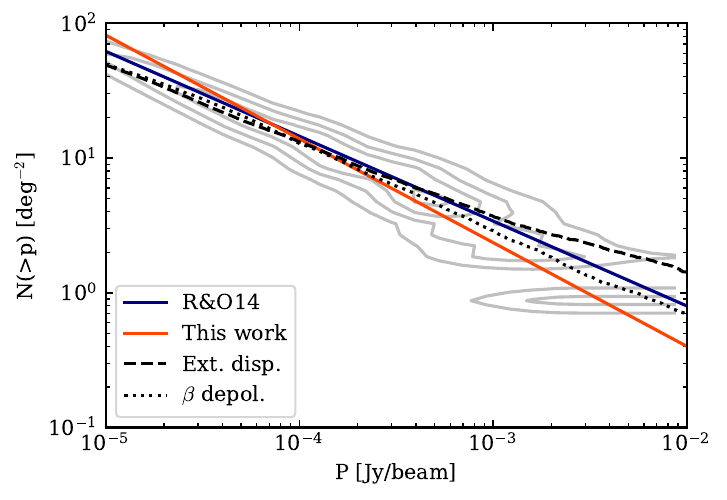}
    \caption[Modelled cumulative source counts]{Modelled cumulative source counts, as in Fig.~\ref{ch4:fig:cumulative-counts}. The solid blue and orange lines show the analytical expressions for the source counts from \citet{Rudnick2014} and this work (Eq.~\ref{ch4:eq:my-cum-counts}), respectively. The model assuming the $\beta$ distribution is indicated as a dotted line and the external dispersion model as a dashed line. The normalised spread determined from bootstrap resampling is shown as grey contours.}
    \label{ch4:fig:cum-counts-model}
\end{figure}

The polarised intensity, as a function of frequency $\nu$ can be expressed as $P(\nu) = S(\nu)p(\nu)$. For total intensity $S$, the $\nu$-dependence is the effect of spectral index $\alpha$, $S(\nu) \propto\nu^\alpha$. While there are multiple models that can describe the fractional polarisation $p (\nu)$, we quantified depolarisation using $\beta$. Simply assuming no polarised intensity dependence on the $\alpha$ and $\beta$ parameters, we sampled $10^4$ simulated sources from the \citetalias{Rudnick2014} distribution as $P_\mathrm{1.4\,GHz}$. We assigned $\alpha$ and $\beta$ values for each of the simulated sources, as randomly sampled from the spectral index and depolarisation index distributions in Fig.~\ref{ch4:fig:alpha-beta}. We included a limitation so that flat-spectrum sources would not have strong depolarisation, as shown in the figure. We then extrapolated to $P_\mathrm{3\,GHz}$ considering both the spectral index and depolarisation. The resulting cumulative counts are shown in Fig.~\ref{ch4:fig:cum-counts-model}. For a more robust comparison with \citetalias{Rudnick2014}, we computed the statistical uncertainty on this relation by performing bootstrap resampling, selecting $10^4$ samples of 65 sources, and generating their cumulative source count distribution. The resultant spread is plotted contours in Fig.~\ref{ch4:fig:cum-counts-model}. Both the modelled and measured source counts are consistent within the statistical uncertainties of \citetalias{Rudnick2014}. 

To demonstrate how the source counts might differ at S-band frequencies in a region with foreground magneto-ionic turbulent screen (e.g. the Galactic plane), we also modelled the 3~GHz source counts as above but assuming that the physical depolarisation mechanism is external Faraday dispersion \citep{Burn1966}:
\begin{align}\label{ch4:eq:qu-ext}
    \mathbf{p}(\lambda^2) \sim e^{-2\sigma_\mathrm{RM}^2 \lambda^4}.
\end{align}
Here, $\sigma_\mathrm{RM}$ is the Faraday dispersion. We randomly assigned a $\sigma_\mathrm{RM}$ value to each simulated sources, as above, sampling from the catalogue of \citet{Ma2025}, a Faraday complexity study of sources within $|b|<5^\circ$ with the VLA L band. \citet{Ma2025} find that 30\% of their sources show external Faraday dispersion though QU-fitting, with $5\leq\sigma_\mathrm{RM}\leq27$~\radmsq. Conserving this complexity fraction and $\sigma_\mathrm{RM}$ range for our simulated sample, we find that the resulting source count remains consistent with \citetalias{Rudnick2014} and our findings. Since we can detect higher Faraday dispersions at higher frequencies, we repeated this simulation, allowing 1\% of the simulated sources to have $27<\sigma_\mathrm{RM}<70$~\radmsq. The resultant modelled counts are plotted in Fig.~\ref{ch4:fig:cum-counts-model}. While these source counts remain within the \citetalias{Rudnick2014} statistical uncertainties, we find that just a small fraction of the more extreme Faraday dispersions shows an excess in source counts at $P_\mathrm{3~GHz} > 1$~m\jybeam with respect to our observations, suggesting an increased S-band source count in the Galactic plane in regions with increased Faraday dispersion. 
 
We briefly discuss the effect of observation bandwidth in the context of frequency dependence. Although we detect Faraday complexity in our sample, the low Faraday depth resolution from 2--4~GHz limits our sensitivity to model depolarisation within this $\lambda^2$ range \citep{Brentjens2005}. While with QU-fitting, we have the advantage of obtaining $p$ closer to the intrinsic value, the more severe depolarisation effects are more prevalent at $\nu<2$~GHz. Moreover, the observed frequency range is a limitation for accurately modelling the frequency dependence down to 1.4~GHz. A more detailed modelling of the frequency dependence of source counts will be addressed with future multi-band surveys with a more statistically significant number of detections, for example POSSUM combined with MeerKAT+ S-band Legacy Survey\footnote{\url{www.meerkatplus.tel}}. 
 
An important implication of this work is the assessment of whether S-band or L-band observations are more optimal for maximising the polarised source counts for future surveys. Both bands have their advantages, including higher angular resolutions in the S band and higher Faraday depth resolutions and larger fields of view in the L band. The band selection should be based on these properties, as the source counts are unlikely to be affected between the L and S bands. This is valid for extragalactic fields only, as we expect the fraction of depolarised sources to increase drastically towards the Galactic plane \citep[e.g.][]{Ranchod2024,Vanderwoude2024}. We emphasise that broadband observations are essential for fully characterising the depolarisation mechanisms affecting the polarised extragalactic population and, in turn, the magnetic field structures of AGNs, their various components, and their interactions with their surrounding environments.

\section{Conclusion}\label{ch4:sec:conclusion}
In this paper, we presented the polarisation component of the VLA-COSMOS 3~GHz Large Project, covering the 1.74 deg$^{2}$ central region of the continuum survey \citep{Smolcic2017}. By producing Stokes Q and U cubes and performing RM synthesis, we achieve a sensitivity of 2.6~$\mu$\jybeam in the Faraday depth domain. With a 10.7$\sigma$ detection threshold, we detect 65 polarised sources, all of which can be classified as AGNs, according to their multi-wavelength properties. Despite the depth and frequency band of this survey, we do not detect any SFG in polarisation. This work provides the deepest polarised source counts at 3~GHz to date.
 
At 3~GHz, we find polarised cumulative source counts consistent with those found by \citet{Rudnick2014} at 1.4~GHz, despite the effect of the spectral index. The source count consistency with 1.4 GHz can be described by a combination of this depolarisation effect and spectral index. We derived an analytical model that can be used to predict polarised source counts for future S-band surveys:
\begin{align}
    N (>P) = 42.89\pm0.37 \left(\frac{P}{23\,\mu \mathrm{Jy~beam}^{-1}}\right)^{-0.77\pm0.01}.
\end{align}
We also find that the Euclidean-normalised counts are consistent with those in the literature and that the T-RECS simulated catalogue of polarised sources returns a higher source count at low polarised intensity than observed in this work. This highlights the need for deep polarised source counts across a large frequency range to constrain population models ahead of the SKA era. 

As we detect no SFGs in polarisation in the VLA-COSMOS survey, we derive a $2\sigma$ upper limit on the density of polarised SFGs of $<2.04$~deg$^{-2}$ at $P_\mathrm{3\,GHz}>15~\mu$\jybeam and suggest that survey sensitivities better than $2.6~\mu$\jybeam will be required to readily detect these sources in the SKA era. These results imply that despite the advantages of high frequency (${\sim}3~$GHz) polarisation surveys and the consistent observed source counts at low polarised intensities, we require broad multi-band observations over, for example, the L and S bands to fully characterise the depolarisation properties of faint extragalactic AGNs. This will be possible by combining large-area pre-SKA polarisation surveys, such as POSSUM and the MeerKAT+ S-band Legacy survey, with common multi-band observations using SKA-mid and SKA-low.

In the companion paper, Ranchod et al., Paper II, we will present an analysis on the fractional polarisation of the detected polarised sources in the VLA-COSMOS survey and how this relates to their radio morphologies, luminosities, and host galaxy properties. Additionally, Paper II will include a data release of the Stokes I, Q, and U spectral cubes.

\begin{acknowledgements}
We would like to thank the anonymous referee for their comments which improved the manuscript. We thank Larry Rudnick and Rainer Beck for helpful discussions that improved this work. We thank Gianni Zamorani for helpful input on the SFG upper limit and Yik Ki (Jackie) Ma for comments on the manuscript. SR acknowledges the support from the International Max Planck Research School (IMPRS) for Astronomy and Astrophysics at the Universities of Bonn and Cologne. RPD acknowledges funding from the South African Radio Astronomy Observatory (SARAO), which is a facility of the National Research Foundation (NRF), an agency of the Department of Science, Technology and Innovation (DSTI). RPD acknowledges funding by the South African Research Chairs Initiative of the DSTI/NRF (Grant ID: 77948), as well as financial support from the Inter-University Institute for Data Intensive Astronomy (IDIA). IDIA is a partnership of the University of Cape Town, the University of Pretoria and the University of the Western Cape. VS acknowledges support from the project ``Implementation of cutting-edge research and its application as part of the Scientific Center of Excellence for Quantum and Complex Systems, and Representations of Lie Algebras'', Grant No. PK.1.1.10.0004, co-financed by the European Union through the European Regional Development Fund - Competitiveness and Cohesion Programme 2021-2027. This research made use of Montage. It is funded by the National Science Foundation under Grant Number ACI-1440620, and was previously funded by the National Aeronautics and Space Administration's Earth Science Technology Office, Computation Technologies Project, under Cooperative Agreement Number NCC5-626 between NASA and the California Institute of Technology.
\end{acknowledgements}

\bibliographystyle{bibtex/aa.bst}
\bibliography{bib}

\begin{appendix}
\section{Completeness}\label{ch4:sec:completeness}
To measure the completeness of the survey, and understand the efficacy of our source finding method, we produced mock image cubes with 300 modelled polarised point sources. The polarised flux density for each source was sampled from a uniform distribution ranging from $2 < P < 200~\mu$Jy across nine logarithmically spaced bins. The lower limit is selected as the MAD noise in Faraday depth, and the upper limit corresponds to the brightest detected source. We assumed the RM of all sources to be $\phi =0$~\radmsq, which is the mean expected Faraday depth for a high Galactic latitude extragalactic sample, and an EVPA of $\chi = 45$~deg. The sources are injected into the QU channel maps at random positions, with the exception that sources should have at minimum a three beam (4.5") separation from any known total intensity sources in the field.
\begin{figure}
    \centering
    \includegraphics[width=0.9\linewidth]{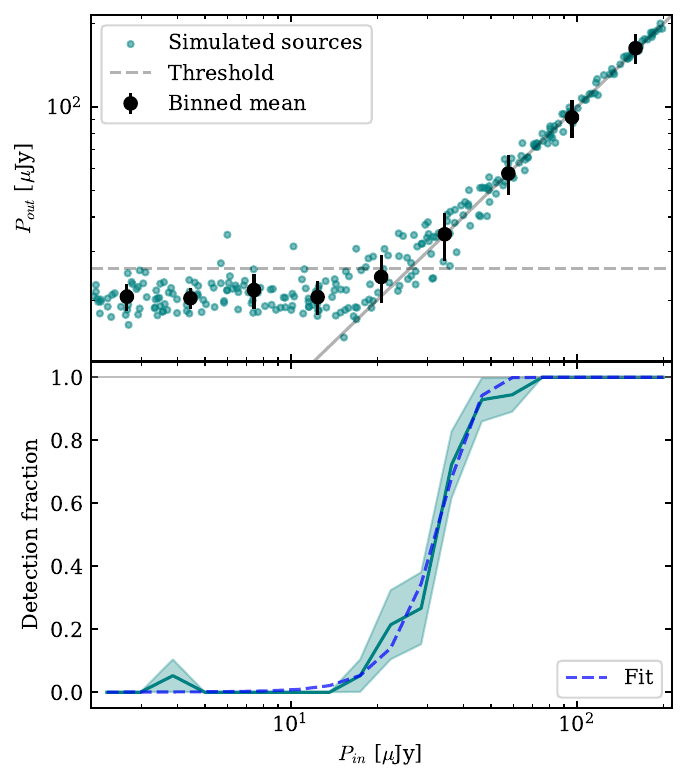}
    \caption[Injected vs recovered polarised intensity and completeness as a function of $P_\mathrm{in}$]{\textit{Top:} Injected $P_\mathrm{in}$ vs recovered $P_\mathrm{out}$ polarised intensity for the sources in the simulated cubes (cyan; Sect.~\ref{ch4:sec:completeness}). The binned data, in log space is shown in black. The solid grey line represents the 1-to-1 line, and the dashed grey line represents an indicative threshold of $27~\mu$\jybeam. \textit{Bottom:} Detection fraction of sources per $P_\mathrm{in}$ bin (teal), with their associated binomial uncertainties. The fitted normalised cumulative distribution is indicated by the dashed blue line. The horizontal grey line shows 100\% completeness.}
    \label{ch4:fig:pin_pout}
\end{figure}
 
For a given injected source, the Stokes $Q(\lambda^2)$ and $U(\lambda^2)$ spectra are first generated. We assume Faraday-simple sources, i.e. single component and without depolarisation, and no spectral index effect, i.e. $\alpha=0$.:
\begin{align}
    Q(\lambda^2) = P\cos[2(\chi_0 + \mathrm{RM} \lambda^2)] \\
    U(\lambda^2) = P\sin[2(\chi_0 + \mathrm{RM} \lambda^2)].
\end{align}
The spectrum is then convolved with the restoring beam in the image plane, and inserted into the Q and U mosaicked channel maps. We then extract Q and U spectral image cutout cubes for each source. These cubes are processed with \texttt{rmsynth3D}, as described in Sect.~\ref{ch4:sec:rmsynth}. We then recover the polarised intensity and MAD noise at the position of all mock sources. The injected polarised intensity is plotted against the recovered polarised intensity in Fig.~\ref{ch4:fig:pin_pout}. We find good agreement between these measurements down to 27~$\mu$\jybeam. This is consistent with a $10.7\sigma$ detection threshold for an average MAD of 2.6~$\mu$\jybeam, as found in Sect.~\ref{ch4:sec:detection-thresh}. In Fig.~\ref{ch4:fig:pin_pout}, we also show the fraction of recovered sources as a function of polarised intensity, which can be fit with a normalised cumulative distribution. We are 100\% complete at 75~$\mu$\jybeam and do not recover sources below 14~$\mu$\jybeam. The error bars shown here are binomial uncertainties.

\section{Off-axis leakage quantification}\label{ch4:sec:leakage}
 \begin{figure*}
    \centering
    \includegraphics[width=0.7\linewidth]{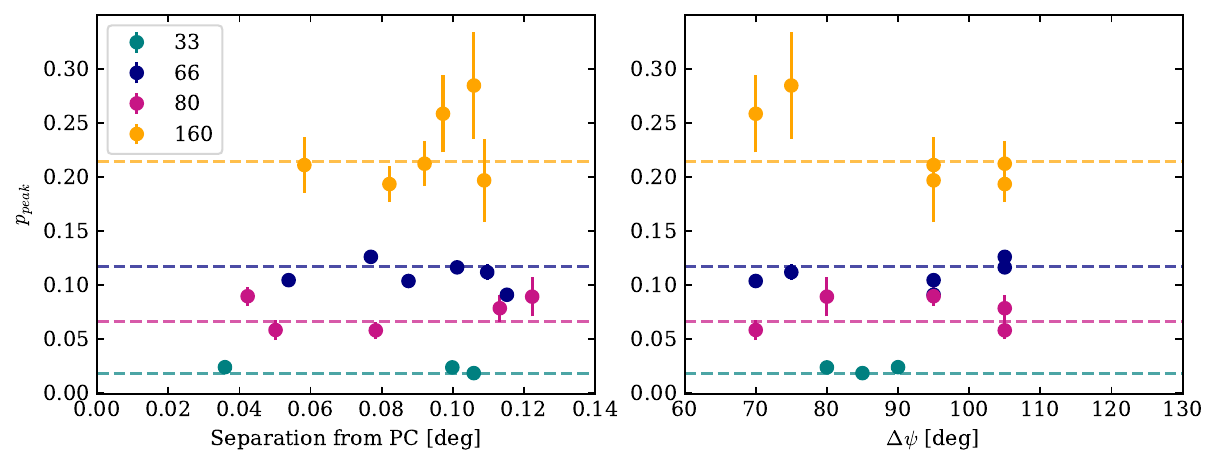}
    \caption[Fractional polarisation as a function of separation from pointing centre and range of parallactic angle for the leakage analysis]{Fractional polarisation of test sources (Table~\ref{ch4:tab:leakage_pointings}) as a function of separation from pointing centre (left) and range of parallactic angle (right), as observed in various pointings. The dashed lines show the fractional polarisation, as measured in the mosaic.}
    \label{ch4:fig:multi-point-test}
\end{figure*}
It is essential to quantify the effect of on-axis and off-axis leakage to identify spurious polarised detections caused by instrumental polarisation, i.e. the leakage of Stokes I emission into Stokes Q and U. This is typically a result of imperfections in the orthogonal feeds in the telescope receivers. We correct for on-axis leakage during polarisation calibration, which we discuss and quantify to be 0.027\% in Sect.~\ref{ch4:sec:calibration}. Off-axis leakage is much more difficult to correct for, as it depends on position with respect to the pointing centre, frequency, and parallactic angle coverage. 
\citet{Jagannathan2017} show how leakage can introduce errors in both polarised intensity and RM. However, they find that this effect is reduced through long observations sampling a significant range of parallactic angles, and becomes negligible for $\Delta\psi\sim 180$ deg. In this section, we conduct tests to estimate the degree to which off-axis leakage affects our observations.

Due to the dense pointing configuration of the survey, each source appears in at maximum nine pointings. We select four of the brightest polarised point sources to measure the variation in fractional polarisation $p_\mathrm{peak}$ as a function of position in a given pointing. The source IDs and the number of associated pointings are summarised in Table~\ref{ch4:tab:leakage_pointings}. We create Stokes I, Q, and U cubelets from each pointing and run \texttt{rmsynth3D} and \texttt{rmclean3d}, as described in Sect.~\ref{ch4:sec:rmsynth}. We then compare the Faraday spectra, Stokes Q and U spectra, as well as the $p_\mathrm{peak}$ between pointings. We find that across the various pointings, we obtain consistent measured RMs, and this is not affected by the position of the source with respect to the pointing centre. In the Faraday spectra, there is a polarised intensity offset from pointing to pointing and the measured $p_\mathrm{peak}$ from the mosaic is consistent with the mean $p_\mathrm{peak}$ across all individual pointings. We plot the $p_\mathrm{peak}$ as a function of separation from the pointing centre, and as a function of parallactic angle coverage over all observations for a given pointing (Fig.~\ref{ch4:fig:multi-point-test}). There are no clear trends for either of these parameters, but we find on average, the offset between various pointings is ${\sim}3$\% in $p_\mathrm{peak}$, within uncertainty of the mosaic measurement. For source ID 160, we find an instrumentally enhanced fractional polarisation for sources further than 0.1 degrees from the pointing centre and with $\Delta\psi<80$~deg, though it is unclear which effect dominates. Further investigation into this is beyond the scope of this work.
\begin{table}[]
    \caption[Summary of sources used in the leakage analysis]{Summary of sources used in the leakage analysis.}
    \centering
    \begin{tabular}{cc}
    \\ \hline
      Source ID   & Number of pointings \\ \hline
        33 & 3\\
        66 & 6\\
        80 & 5\\
        160 & 6\\ \hline
        \end{tabular}
        \tablefoot{The source ID is defined as in \citetalias{Smolcic2017}.}
    \label{ch4:tab:leakage_pointings}
\end{table}

To better quantify the off-axis leakage, and to estimate the expected percentage leakage for spurious detections, we create a leakage mosaicked map. Using \texttt{plumber} \citep{Sekhar2022}, we generate models for the off-diagonal elements of the Mueller matrix, representing the leakage between various Stokes parameters (e.g. $I\rightarrow Q$, $I\rightarrow U$). 

\begin{figure}
    \centering
    \includegraphics[width=0.9\linewidth]{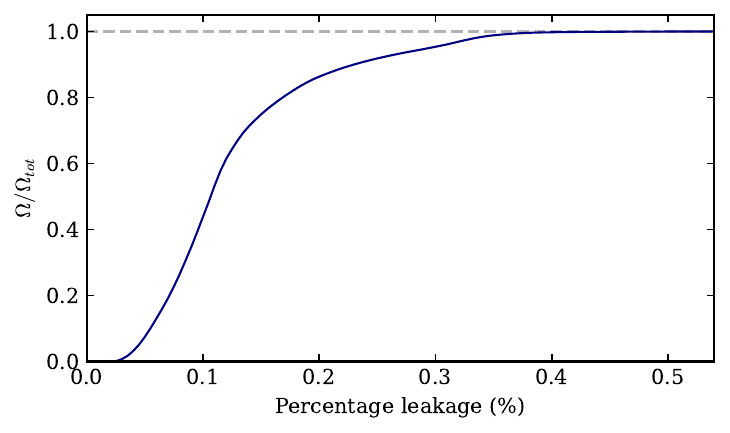}
    \caption[Cumulative distribution of the percentage leakage]{Cumulative distribution of the percentage leakage as determined from the leakage map. The distribution is normalised by the total survey area.}
    \label{ch4:fig:leakage_dist}
\end{figure}

Across all observations, we have a parallactic angle coverage of $\psi = [-60;+60]$~deg. We generate rotated beam responses over this range in increments of $\Delta \psi = 10$~deg for the $I\rightarrow Q$ and $I\rightarrow U$ elements of the Mueller matrix. We then create an averaged beam response for each pointing, weighted by its parallactic angle distribution. The $I\rightarrow Q$ and $I\rightarrow U$ beam responses are then mosaicked to produce leakage maps of the entire survey area for Stokes Q and U. We add the absolute on-axis leakage to the Stokes Q and U maps, respectively, and combine them in quadrature to produce a linear polarisation leakage map $P_\mathrm{leak}$ \citep[e.g.][]{Hales2014}. A cumulative distribution of the percentage leakage in linear polarisation, normalised by survey area, is plotted in Fig.~\ref{ch4:fig:leakage_dist}. The maximum off-axis leakage is 0.5\%, while half of the mosaic has leakage of $<0.1$\%.  The long `tail', where ${\sim}15\%$ of the area has leakage ranging from $0.2-0.4\%$ corresponds to the pointings around the edge of the mosaic, which do not have as many overlapping pointings as those in the centre of the mosaic, and therefore a higher leakage percentage. We did not consider the frequency dependence of leakage for this map, as it indicates a low leakage percentage in general, and we can exclude possible spurious sources with the upper limit of $p=0.5$\%.
 
As a final verification for the off-axis leakage limit, we attempt to detect spurious polarised emission from the brightest sources detected in total intensity within the area of the subgrid. Source ID~3 has a peak total intensity of 11 m\jybeam. For the corresponding position in the $P_\mathrm{peak}$ map produced from RM synthesis, we measure a fractional polarisation of 0.25\%, consistent with our model. This measurement has S/N = 4, and was not detected above our source finding threshold. From inspection, there are no apparent peaks at $\phi = 0$~\radmsq in the Faraday spectrum. It is therefore unlikely that we will detect any spurious sources caused by instrumental polarisation.

\clearpage
\Needspace{100\baselineskip}
\onecolumn
\section{Catalogue}\label{sec:catalogue}

\begin{longtable}{lcccccccc}
\caption{Catalogue of the 65 sources detected in polarisation.}
\label{tab:catalogue}\\

\hline
ID (1) & RA (2) & Dec (3) & $P_\mathrm{peak}$ (4) & $P_\mathrm{total}$ (5) & $p$ (6) & S/N (7) & $\alpha$ (8) & $\beta$ (9) \\
& (J2000) & (J2000) & [$\mu\mathrm{Jy\,beam^{-1}}$] & [$\mu\mathrm{Jy}$] & \% & & & \\
\hline
\endfirsthead

\caption[]{continued.}\\
\hline
ID (1) & RA (2) & Dec (3) & $P_\mathrm{peak}$ (4) & $P_\mathrm{total}$ (5) & $p$ (6) & S/N (7) & $\alpha$ (8) & $\beta$ (9) \\
& (J2000) & (J2000) & [$\mu\mathrm{Jy\,beam^{-1}}$] & [$\mu\mathrm{Jy}$] & \% & & & \\
\hline
\endhead

\hline
\endfoot

\hline
\\
\multicolumn{9}{p{\textwidth}}{\small \textbf{Notes.} Columns (1)--(3) list the IDs and source positions from the \citetalias{Smolcic2017} catalogue; columns (4)--(6) list the polarisation properties of the sources; column (7) lists the significance of the detection. Columns (8) and (9) list the spectral fit parameters discussed in Sect.~\ref{ch4:sec:spec-depol}.}
\endlastfoot
6	&	10:01:54	&	+2:49:54.0	&	202.71 $\pm$ 2.86	&	206.63 $\pm$ 4.67	&	2.62	&	53.8	&	$-$0.91 $\pm$ 0.01	&	0.19 $\pm$ 0.12 \\
16	&	09:58:03	&	+2:13:57.6	&	114.23 $\pm$ 4.48	&	120.23 $\pm$ 4.48	&	1.63	&	25.2	&	$-$0.90 $\pm$ 0.01	&	$-$2.35 $\pm$ 0.25 \\
26	&	10:01:53	&	+2:11:52.6	&	112.45 $\pm$ 2.88	&	109.20 $\pm$ 2.88	&	4.69	&	38.7	&	$-$0.14 $\pm$ 0.01	&	0.25 $\pm$ 0.24 \\
29	&	10:01:47	&	+2:03:14.2	&	100.68 $\pm$ 3.03	&	76.02 $\pm$ 3.03	&	3.25	&	33.4	&	$-$0.85 $\pm$ 0.01	&	$-$3.29 $\pm$ 0.28 \\
33	&	10:01:31	&	+2:29:24.7	&	39.37 $\pm$ 3.84	&	39.37 $\pm$ 3.84	&	1.56	&	13.7	&	$-$0.31 $\pm$ 0.02	&	0.03 $\pm$ 0.45 \\
39	&	09:58:29	&	+1:54:58.8	&	28.87 $\pm$ 2.35	&	28.87 $\pm$ 2.35	&	2.34	&	11.2	&	$-$0.93 $\pm$ 0.02	&	$-$0.69 $\pm$ 0.41 \\
41	&	09:58:14	&	+1:37:51.8	&	62.61 $\pm$ 2.02	&	45.60 $\pm$ 2.02	&	3.44	&	29.8	&	$-$0.27 $\pm$ 0.02	&	0.12 $\pm$ 0.38 \\
45	&	09:59:37	&	+2:23:47.2	&	38.71 $\pm$ 2.89	&	38.71 $\pm$ 2.89	&	2.81	&	13.5	&	$-$0.10 $\pm$ 0.02	&	$-$0.54 $\pm$ 0.53 \\
64	&	10:02:00	&	+2:39:04.8	&	57.45 $\pm$ 2.47	&	63.07 $\pm$ 2.47	&	3.48	&	27.1	&	$-$0.79 $\pm$ 0.01	&	0.13 $\pm$ 0.29 \\
66	&	10:00:05	&	+2:30:29.0	&	146.37 $\pm$ 3.03	&	152.81 $\pm$ 4.71	&	8.35	&	52.1	&	$-$0.84 $\pm$ 0.01	&	$-$0.02 $\pm$ 0.14 \\
80	&	10:00:00	&	+2:46:08.9	&	41.75 $\pm$ 2.53	&	87.06 $\pm$ 4.55	&	5.72	&	16.3	&	$-$0.26 $\pm$ 0.03	&	$-$0.48 $\pm$ 0.45 \\
83	&	09:59:58	&	+2:18:09.7	&	84.58 $\pm$ 1.73	&	206.63 $\pm$ 4.02	&	6.12	&	53.5	&	$-$0.69 $\pm$ 0.02	&	$-$0.27 $\pm$ 0.25 \\
89	&	10:01:09	&	+2:17:21.7	&	142.26 $\pm$ 2.55	&	145.28 $\pm$ 4.21	&	13.78	&	49.3	&	$-$1.03 $\pm$ 0.02	&	$-$0.88 $\pm$ 0.15 \\
101	&	10:01:43	&	+2:21:34.6	&	58.13 $\pm$ 2.86	&	43.84 $\pm$ 2.86	&	8.18	&	16	&	$-$0.96 $\pm$ 0.03	&	$-$1.82 $\pm$ 0.40 \\
112	&	10:03:05	&	+2:49:17.4	&	55.11 $\pm$ 3.52	&	47.48 $\pm$ 3.52	&	5.21	&	19.5	&	$-$0.79 $\pm$ 0.03	&	$-$0.58 $\pm$ 0.48 \\
123	&	10:01:22	&	+2:01:11.8	&	130.16 $\pm$ 2.65	&	897.00 $\pm$ 10.00	&	7.13	&	50.3	&	$-$0.65 $\pm$ 0.02	&	$-$0.94 $\pm$ 0.15 \\
137	&	10:01:31	&	+2:26:39.2	&	460.36 $\pm$ 4.30	&	1263.32 $\pm$ 14.15	&	24.23	&	119.7	&	$-$0.79 $\pm$ 0.02	&	$-$0.15 $\pm$ 0.06 \\
138	&	09:59:09	&	+2:48:13.0	&	291.02 $\pm$ 4.72	&	1516.91 $\pm$ 22.11	&	14.19	&	100.5	&	$-$0.82 $\pm$ 0.01	&	$-$0.10 $\pm$ 0.09 \\
145	&	10:00:48	&	+1:59:11.5	&	53.49 $\pm$ 2.59	&	569.49 $\pm$ 10.95	&	--	&	--	&	$-$0.39 $\pm$ 0.07	&	0.33 $\pm$ 0.31 \\
160	&	10:00:07	&	+2:40:49.9	&	48.64 $\pm$ 2.63	&	56.33 $\pm$ 2.63	&	5.79	&	21.9	&	$-$1.11 $\pm$ 0.03	&	$-$0.63 $\pm$ 0.37 \\
166	&	10:01:24	&	+2:17:06.4	&	52.96 $\pm$ 3.09	&	52.96 $\pm$ 3.09	&	7.83	&	18.7	&	$-$0.66 $\pm$ 0.04	&	$-$0.96 $\pm$ 0.42 \\
177	&	10:00:58	&	+1:51:33.4	&	146.17 $\pm$ 3.48	&	397.81 $\pm$ 10.13	&	15.05	&	51.2	&	$-$0.97 $\pm$ 0.02	&	$-$0.21 $\pm$ 0.22 \\
187	&	09:59:46	&	+2:36:02.2	&	245.85 $\pm$ 3.59	&	1520.29 $\pm$ 15.14	&	27.81	&	67.7	&	$-$0.60 $\pm$ 0.03	&	$-$0.32 $\pm$ 0.10 \\
208	&	09:58:46	&	+2:16:02.4	&	118.90 $\pm$ 3.89	&	194.13 $\pm$ 6.98	&	12.63	&	41	&	$-$0.85 $\pm$ 0.03	&	$-$0.48 $\pm$ 0.18 \\
213	&	10:01:12	&	+1:41:23.7	&	72.44 $\pm$ 4.36	&	61.37 $\pm$ 4.36	&	13.18	&	17.5	&	$-$0.60 $\pm$ 0.05	&	$-$0.17 $\pm$ 0.27 \\
229	&	09:58:37	&	+2:35:48.9	&	48.97 $\pm$ 3.40	&	48.97 $\pm$ 3.40	&	10.35	&	15	&	$-$1.03 $\pm$ 0.05	&	$-$0.43 $\pm$ 0.43 \\
236	&	09:58:29	&	+2:05:31.1	&	40.78 $\pm$ 2.28	&	40.78 $\pm$ 2.28	&	7.63	&	13.7	&	$-$0.94 $\pm$ 0.05	&	$-$0.67 $\pm$ 0.44 \\
246	&	10:03:00	&	+2:12:00.0	&	34.04 $\pm$ 2.54	&	34.04 $\pm$ 2.54	&	17.78	&	9.3	&	$-$0.80 $\pm$ 0.20	&	$-$0.52 $\pm$ 0.37 \\
247	&	09:59:01	&	+2:47:42.5	&	36.26 $\pm$ 2.65	&	128.34 $\pm$ 5.77	&	--	&	--	&	$-$0.58 $\pm$ 0.18	&	$-$1.59 $\pm$ 0.59 \\
319	&	10:00:15	&	+2:52:31.4	&	44.25 $\pm$ 4.32	&	44.25 $\pm$ 4.32	&	8.65	&	13.2	&	$-$1.26 $\pm$ 0.07	&	$-$0.40 $\pm$ 0.49 \\
437	&	09:58:23	&	+2:08:59.5	&	42.20 $\pm$ 2.74	&	91.45 $\pm$ 4.86	&	11.74	&	9.6	&	$-$0.77 $\pm$ 0.08	&	$-$1.09 $\pm$ 0.44 \\
444	&	10:02:30	&	+2:42:12.7	&	30.33 $\pm$ 2.69	&	62.33 $\pm$ 4.39	&	--	&	--	&	0.01 $\pm$ 0.26	&	$-$0.01 $\pm$ 0.43 \\
976	&	09:59:59	&	+2:07:15.1	&	28.70 $\pm$ 2.23	&	28.70 $\pm$ 2.23	&	15.67	&	10.4	&	$-$0.84 $\pm$ 0.15	&	$-$0.38 $\pm$ 0.53 \\
10900	&	09:59:08	&	+2:43:09.6	&	408.28 $\pm$ 3.91	&	1352.59 $\pm$ 18.68	&	7.94	&	98.1	&	$-$0.83 $\pm$ 0.01	&	$-$0.17 $\pm$ 0.06 \\
10901	&	09:57:58	&	+1:58:25.1	&	233.10 $\pm$ 3.42	&	551.42 $\pm$ 10.88	&	5.45	&	42.2	&	$-$1.25 $\pm$ 0.01	&	$-$2.26 $\pm$ 0.15 \\
10902	&	09:58:23	&	+2:26:28.5	&	1038.49 $\pm$ 4.91	&	5454.30 $\pm$ 41.03	&	9.72	&	168.9	&	$-$1.00 $\pm$ 0.01	&	$-$0.06 $\pm$ 0.04 \\
10903	&	10:02:09	&	+2:41:03.3	&	65.55 $\pm$ 2.41	&	64.91 $\pm$ 2.41	&	1.17	&	25.2	&	$-$0.65 $\pm$ 0.01	&	$-$0.69 $\pm$ 0.22 \\
10904	&	10:02:43	&	+1:59:43.4	&	738.82 $\pm$ 7.06	&	3663.45 $\pm$ 52.24	&	10.76	&	94.8	&	$-$0.92 $\pm$ 0.01	&	0.04 $\pm$ 0.04 \\
10905	&	10:02:30	&	+2:32:25.1	&	45.21 $\pm$ 2.02	&	78.26 $\pm$ 3.29	&	23.46	&	11.6	&	$-$0.52 $\pm$ 0.14	&	$-$0.47 $\pm$ 0.43 \\
10906	&	10:02:12	&	+2:31:35.0	&	298.18 $\pm$ 3.47	&	840.70 $\pm$ 13.20	&	14.8	&	98.6	&	$-$0.80 $\pm$ 0.02	&	$-$0.06 $\pm$ 0.08 \\
10907	&	10:03:09	&	+2:27:14.1	&	34.16 $\pm$ 2.29	&	34.16 $\pm$ 2.29	&	1.72	&	13.4	&	$-$0.48 $\pm$ 0.02	&	0.27 $\pm$ 0.43 \\
10909	&	10:00:08	&	+2:43:15.4	&	171.35 $\pm$ 2.61	&	480.42 $\pm$ 7.62	&	8.16	&	70.5	&	$-$0.93 $\pm$ 0.01	&	$-$0.13 $\pm$ 0.12 \\
10910	&	10:00:50	&	+1:49:23.7	&	44.59 $\pm$ 3.46	&	133.74 $\pm$ 7.61	&	14.5	&	7.5	&	$-$0.81 $\pm$ 0.13	&	$-$0.29 $\pm$ 0.34 \\
10911	&	10:01:15	&	+2:02:08.7	&	120.25 $\pm$ 2.44	&	168.30 $\pm$ 5.04	&	10.74	&	53.1	&	$-$0.44 $\pm$ 0.02	&	0.09 $\pm$ 0.20 \\
10913	&	10:00:28	&	+2:41:03.4	&	143.42 $\pm$ 1.82	&	4239.94 $\pm$ 17.11	&	17.68	&	81.5	&	$-$0.30 $\pm$ 0.04	&	0.09 $\pm$ 0.17 \\
10914	&	10:02:30	&	+2:09:13.3	&	85.58 $\pm$ 3.65	&	251.78 $\pm$ 9.28	&	22.09	&	25.9	&	$-$0.76 $\pm$ 0.06	&	$-$0.29 $\pm$ 0.25 \\
10915	&	10:00:00	&	+1:48:37.8	&	99.75 $\pm$ 3.18	&	214.76 $\pm$ 6.74	&	12.1	&	32.9	&	$-$1.27 $\pm$ 0.02	&	$-$1.52 $\pm$ 0.22 \\
10916	&	10:01:40	&	+1:51:29.7	&	71.91 $\pm$ 2.66	&	470.95 $\pm$ 9.25	&	9.29	&	24.5	&	$-$0.83 $\pm$ 0.03	&	$-$0.06 $\pm$ 0.28 \\
10917	&	10:01:52	&	+2:45:35.3	&	55.86 $\pm$ 2.93	&	69.62 $\pm$ 4.49	&	10.74	&	18.4	&	$-$0.73 $\pm$ 0.05	&	$-$0.02 $\pm$ 0.30 \\
10918	&	09:58:24	&	+2:49:16.2	&	455.75 $\pm$ 4.45	&	3907.63 $\pm$ 35.57	&	37.09	&	125.7	&	$-$0.61 $\pm$ 0.02	&	$-$0.63 $\pm$ 0.05 \\
10919	&	10:01:14	&	+1:54:44.2	&	59.64 $\pm$ 3.16	&	141.10 $\pm$ 6.54	&	10.18	&	19.6	&	$-$0.93 $\pm$ 0.04	&	0.00 $\pm$ 0.31 \\
10920	&	09:58:39	&	+1:35:57.7	&	115.72 $\pm$ 3.34	&	156.07 $\pm$ 6.43	&	19.56	&	37.4	&	$-$0.92 $\pm$ 0.04	&	0.14 $\pm$ 0.19 \\
10923	&	10:03:04	&	+1:47:36.0	&	186.19 $\pm$ 3.72	&	1377.35 $\pm$ 20.85	&	23.95	&	62.9	&	$-$0.95 $\pm$ 0.04	&	0.22 $\pm$ 0.16 \\
10927	&	10:01:02	&	+2:05:11.5	&	24.50 $\pm$ 2.61	&	24.50 $\pm$ 2.61	&	9.33	&	8.8	&	$-$0.94 $\pm$ 0.10	&	$-$0.96 $\pm$ 0.59 \\
10928	&	09:58:22	&	+2:47:22.2	&	86.53 $\pm$ 3.61	&	999.79 $\pm$ 19.14	&	12.93	&	21.7	&	$-$0.84 $\pm$ 0.05	&	$-$0.39 $\pm$ 0.25 \\
10933	&	10:00:43	&	+1:46:07.9	&	269.40 $\pm$ 5.09	&	3676.22 $\pm$ 41.57	&	6.33	&	53	&	$-$0.35 $\pm$ 0.01	&	$-$0.07 $\pm$ 0.09 \\
10935	&	09:59:27	&	+2:37:29.4	&	43.28 $\pm$ 3.66	&	95.21 $\pm$ 6.62	&	15.57	&	12.8	&	$-$0.88 $\pm$ 0.08	&	0.24 $\pm$ 0.49 \\
10936	&	10:00:28	&	+1:35:08.6	&	44.17 $\pm$ 2.00	&	877.33 $\pm$ 10.52	&	20.05	&	19	&	$-$0.54 $\pm$ 0.10	&	$-$0.04 $\pm$ 0.51 \\
10948	&	10:00:22	&	+2:00:00.0	&	60.84 $\pm$ 3.84	&	52.01 $\pm$ 3.84	&	6.03	&	20.3	&	$-$0.12 $\pm$ 0.02	&	0.39 $\pm$ 0.31 \\
10949	&	10:01:24	&	+2:49:36.7	&	32.40 $\pm$ 2.03	&	40.06 $\pm$ 2.03	&	3.79	&	13.2	&	$-$0.45 $\pm$ 0.03	&	$-$0.39 $\pm$ 0.51 \\
10952	&	10:02:39	&	+2:21:52.2	&	28.77 $\pm$ 2.34	&	28.77 $\pm$ 2.34	&	5.46	&	9.1	&	$-$0.43 $\pm$ 0.05	&	$-$0.34 $\pm$ 0.47 \\
10953	&	10:00:19	&	+2:32:56.3	&	23.43 $\pm$ 3.01	&	23.43 $\pm$ 3.01	&	9.38	&	5.8	&	$-$0.54 $\pm$ 0.12	&	$-$0.72 $\pm$ 0.58 \\
10956	&	10:00:27	&	+2:21:23.3	&	27.92 $\pm$ 2.62	&	87.56 $\pm$ 5.35	&	24.18	&	11.2	&	$-$0.98 $\pm$ 0.16	&	$-$0.77 $\pm$ 0.48 \\
10959	&	10:02:45	&	+2:45:16.1	&	33.13 $\pm$ 2.89	&	104.49 $\pm$ 6.05	&	14.73	&	9.8	&	$-$0.77 $\pm$ 0.13	&	$-$2.10 $\pm$ 0.61 \\
10962	&	10:02:51	&	+2:42:48.1	&	1958.71 $\pm$ 9.40	&	9570.27 $\pm$ 60.01	&	14.88	&	172.8	&	$-$0.88 $\pm$ 0.01	&	$-$0.27 $\pm$ 0.04 \\
\end{longtable}
\twocolumn

\onecolumn
\newpage
\FloatBarrier
\clearpage
\section{Maps of polarised sources}\label{app:maps}
\begin{figure*}[h]
    \centering
    \includegraphics[width=0.24\linewidth]{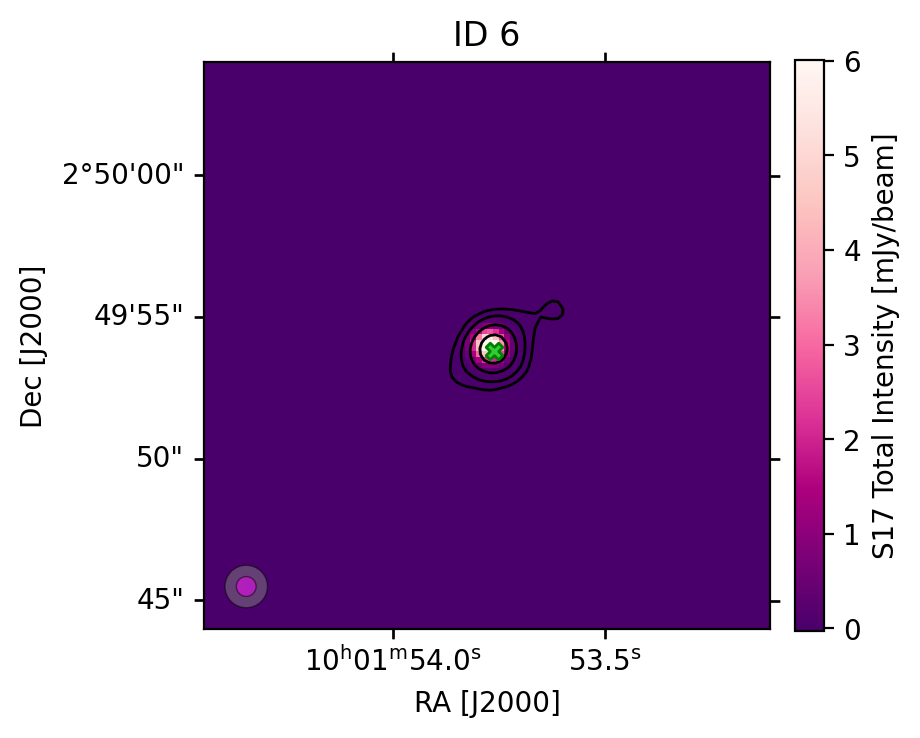}
    \includegraphics[width=0.24\linewidth]{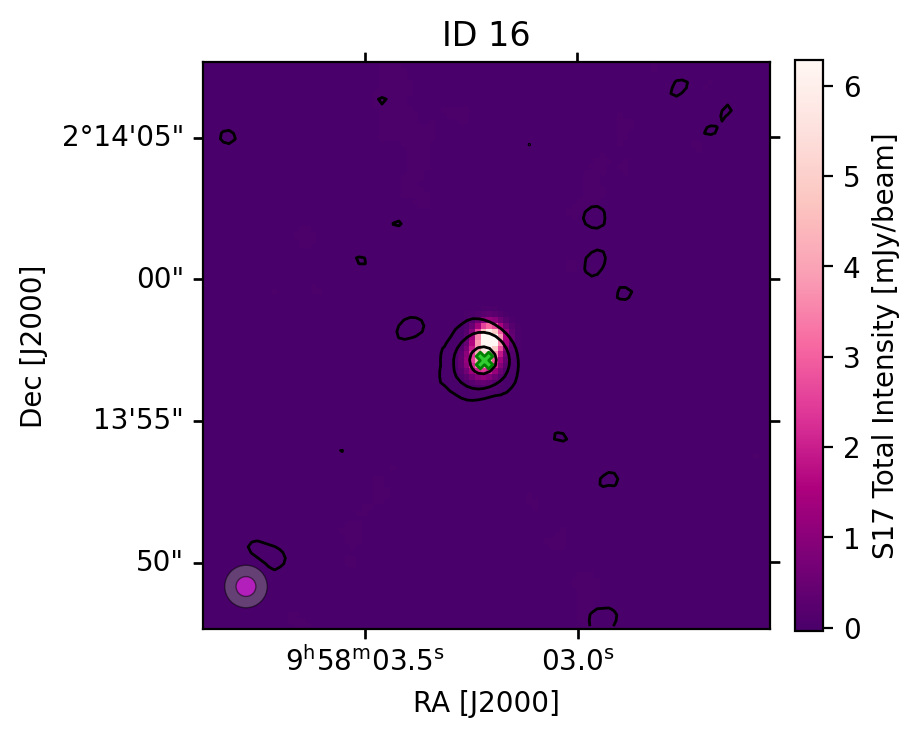}
    \includegraphics[width=0.24\linewidth]{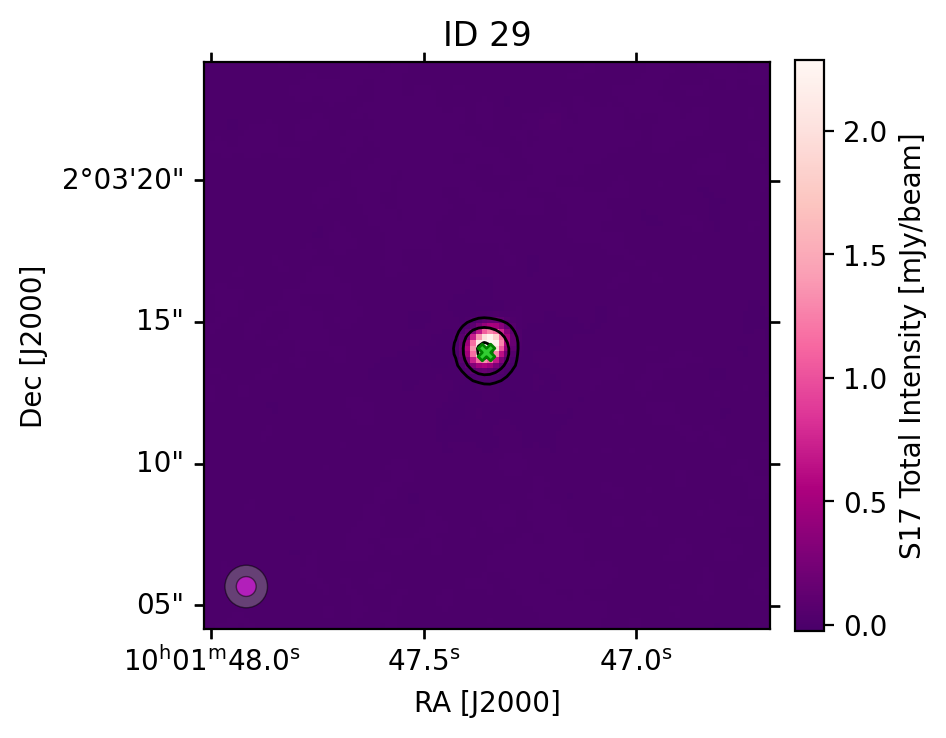} 
    \includegraphics[width=0.24\linewidth]{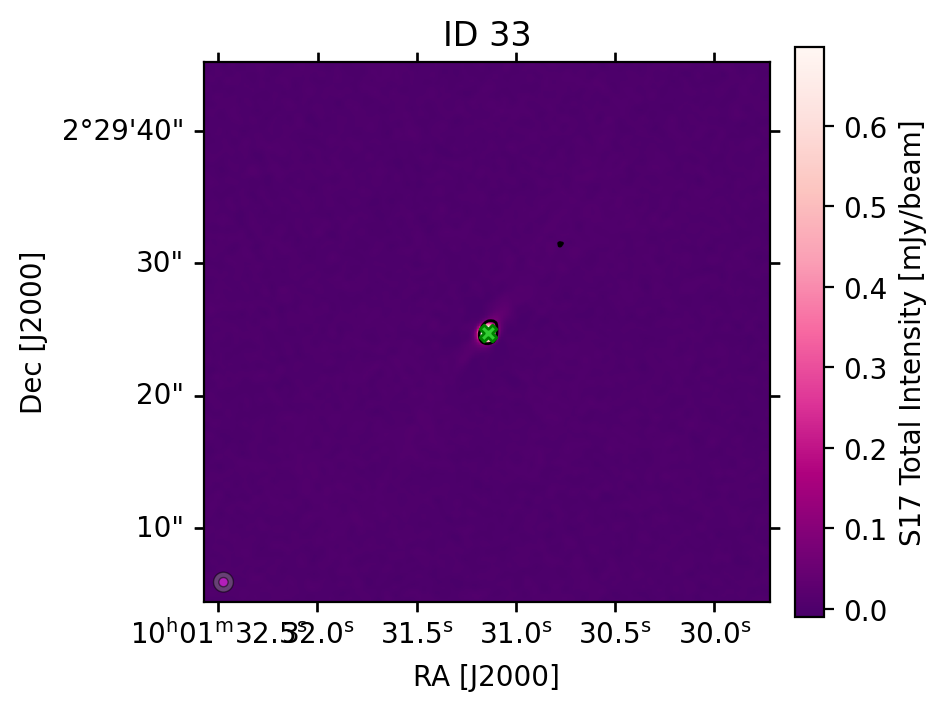} \\
    \includegraphics[width=0.24\linewidth]{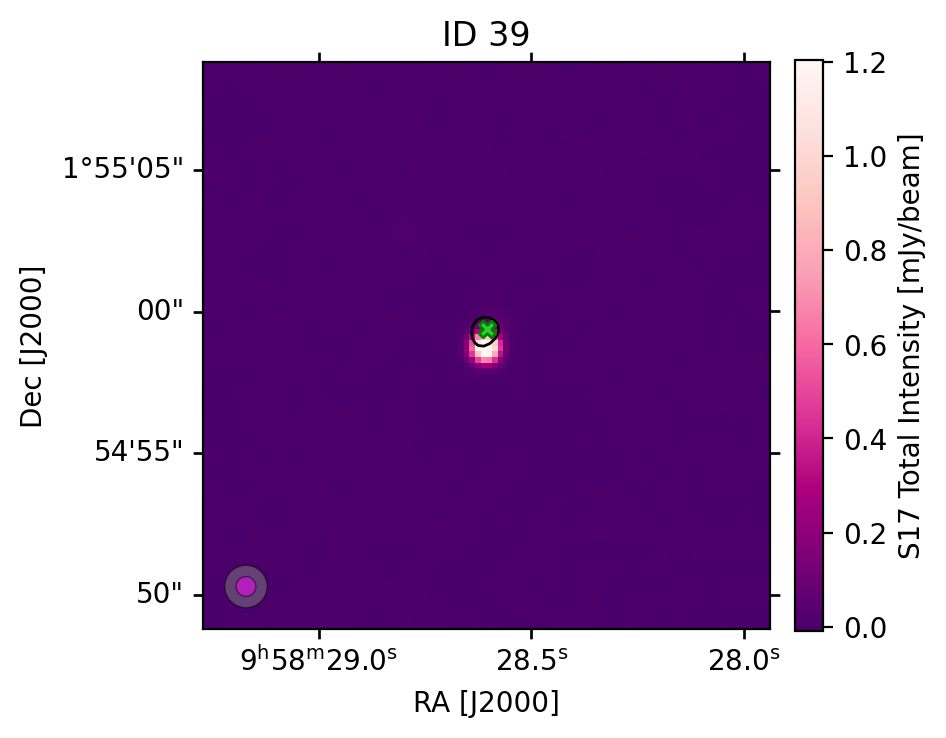}
    \includegraphics[width=0.24\linewidth]{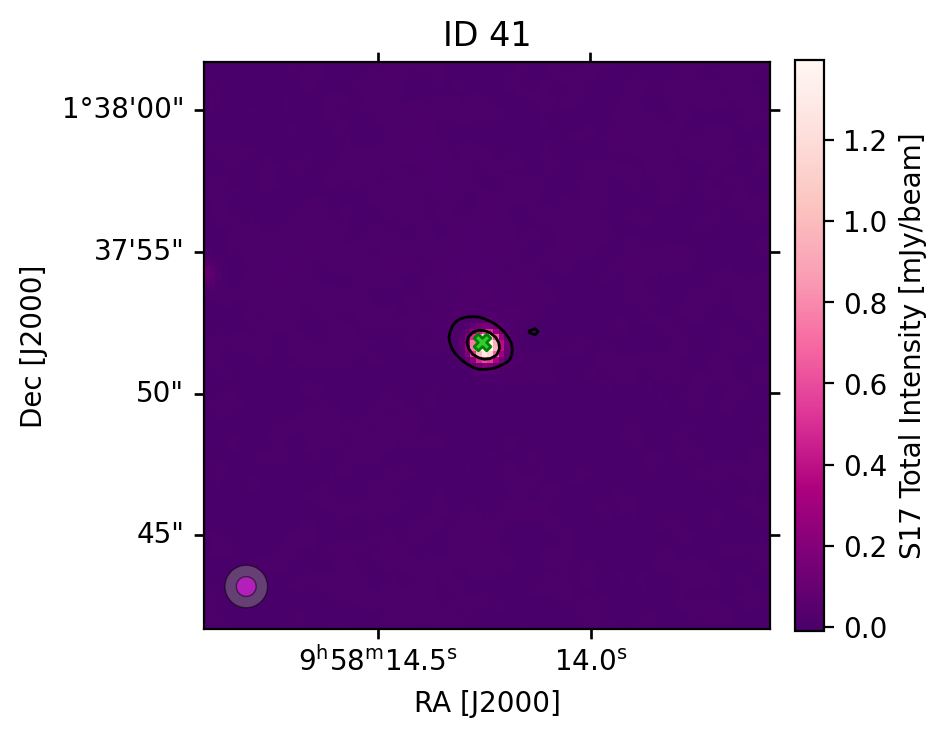} 
    \includegraphics[width=0.24\linewidth]{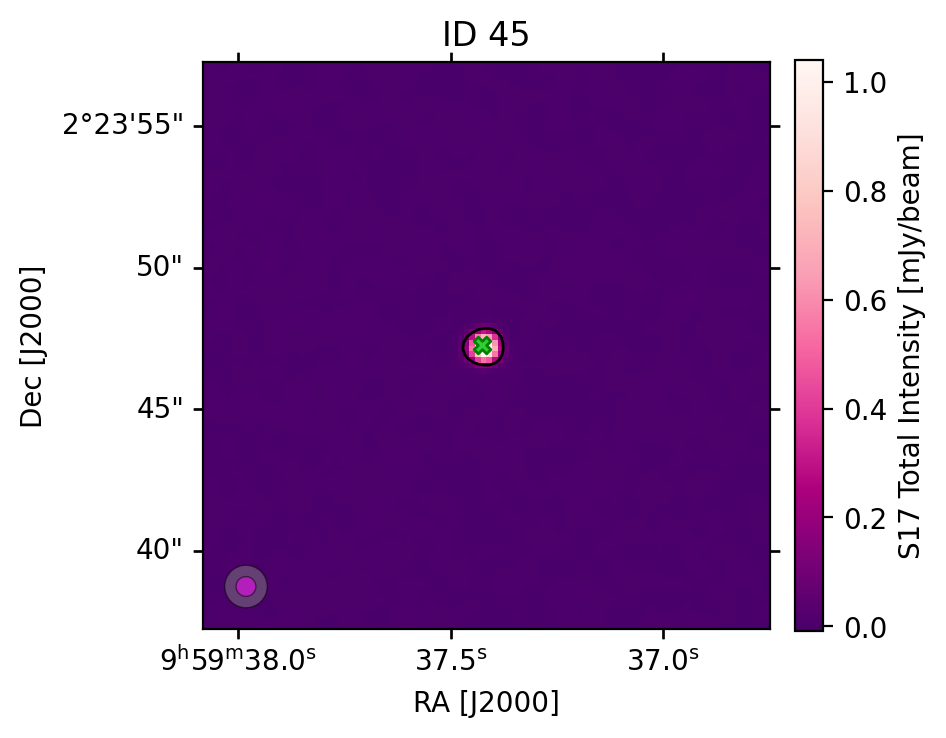}
    \includegraphics[width=0.24\linewidth]{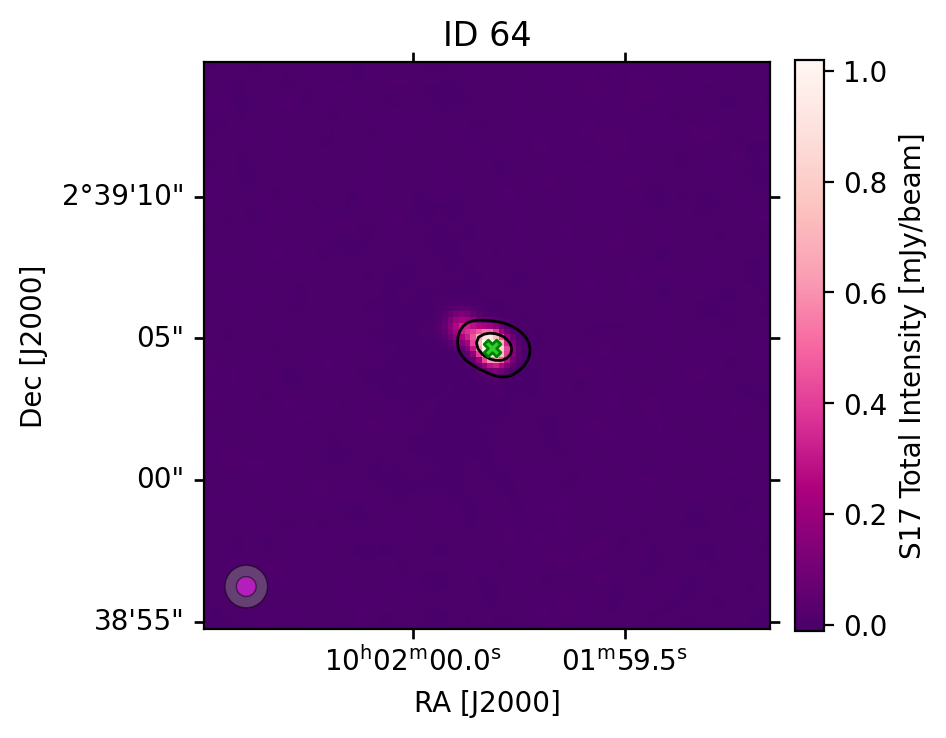} \\
    \includegraphics[width=0.24\linewidth]{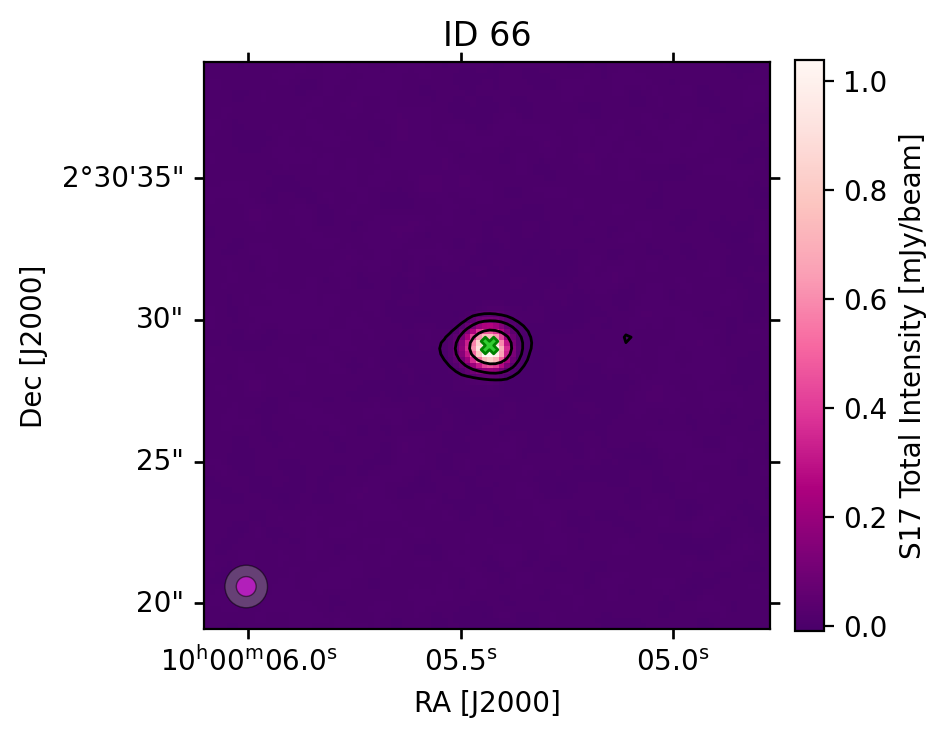}
    \includegraphics[width=0.24\linewidth]{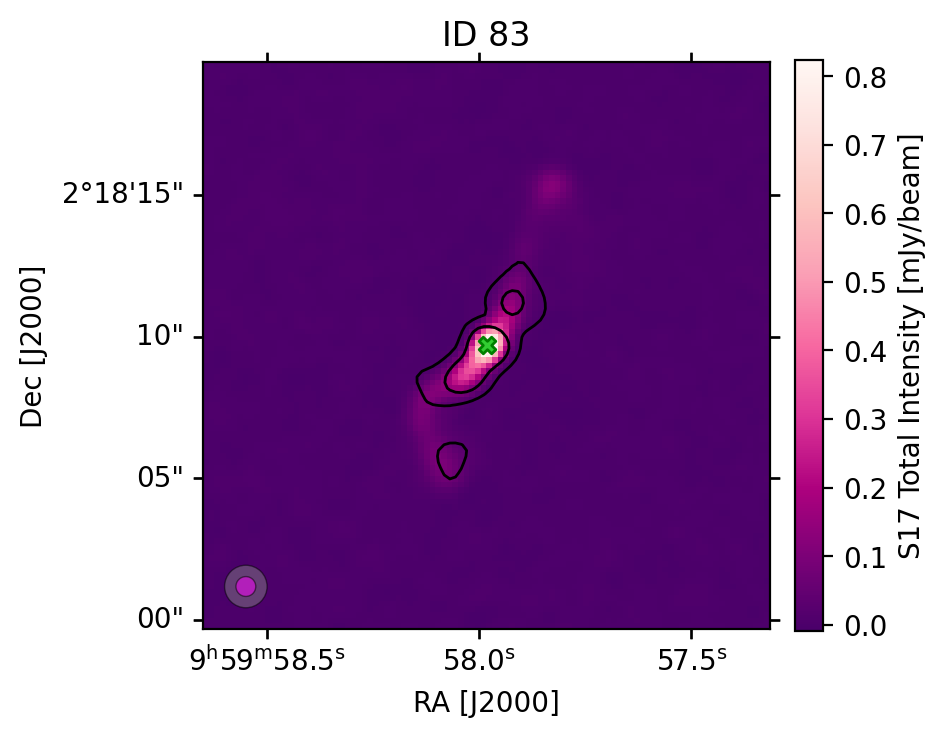}
    \includegraphics[width=0.24\linewidth]{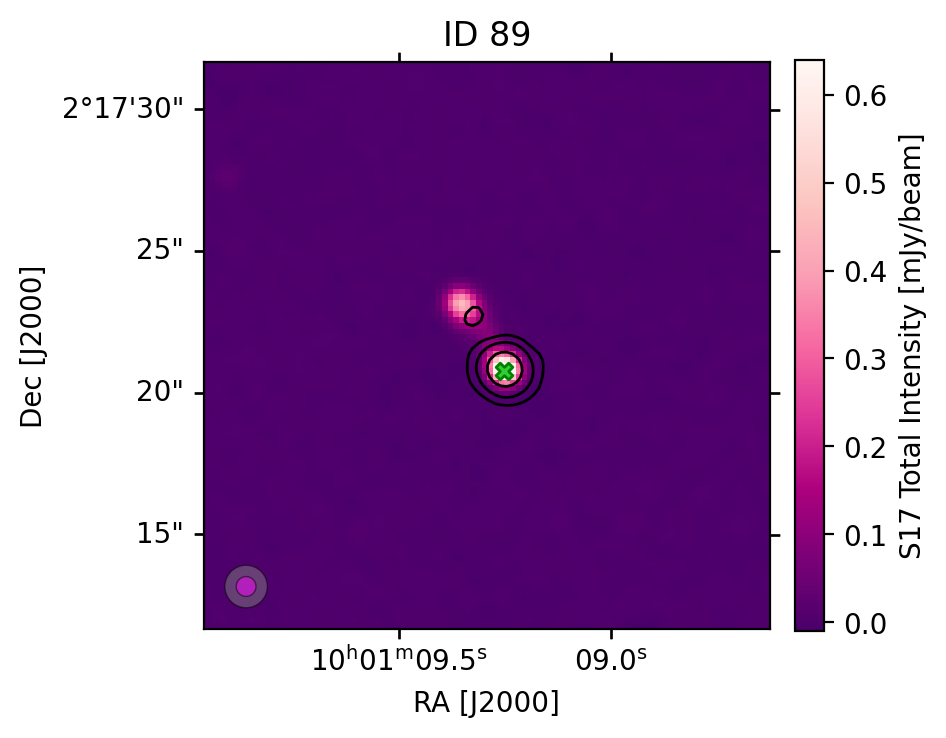}
    \includegraphics[width=0.24\linewidth]{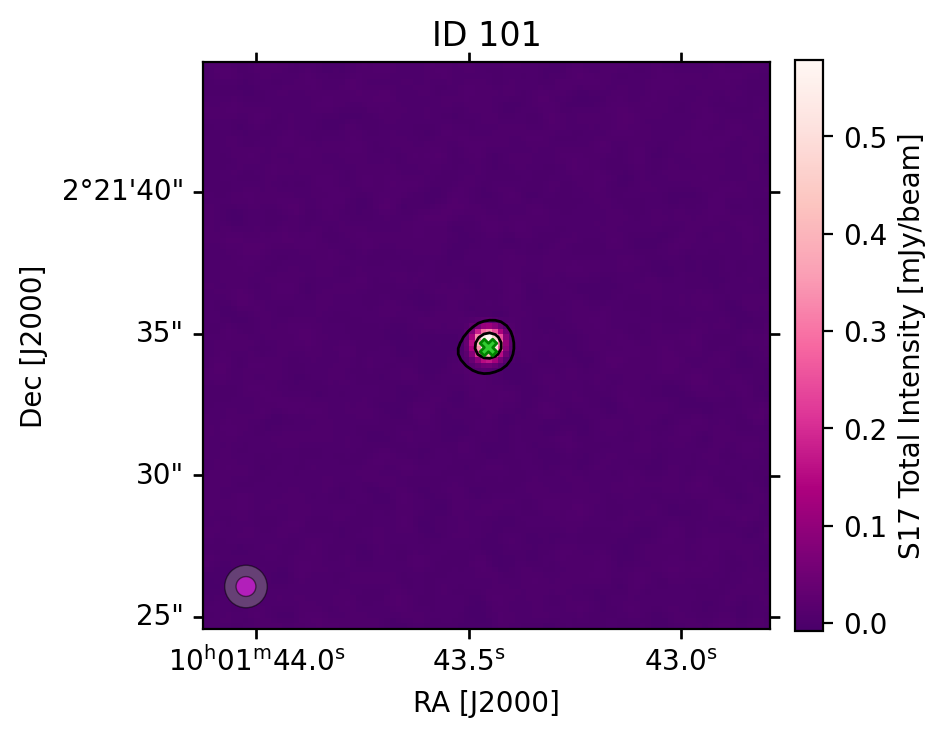} \\
    \includegraphics[width=0.24\linewidth]{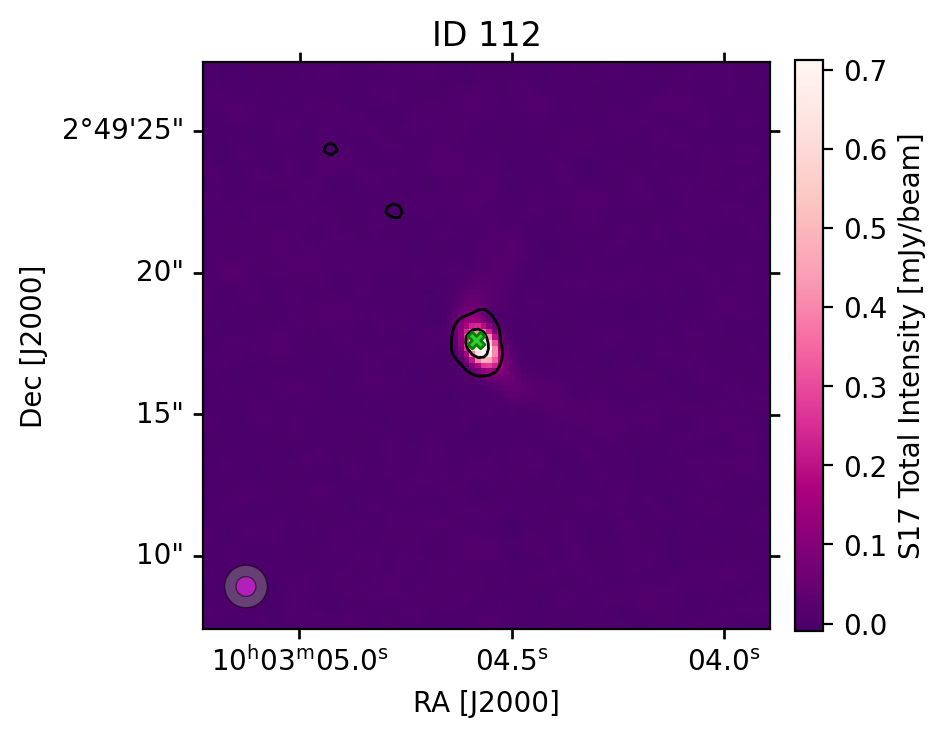}
    \includegraphics[width=0.24\linewidth]{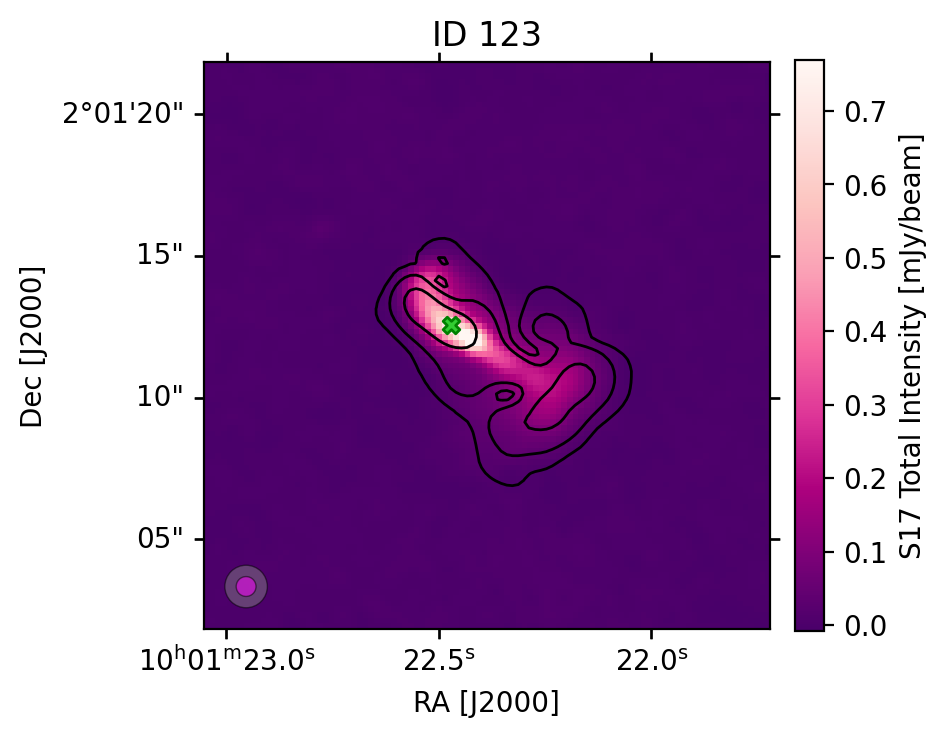}
    \includegraphics[width=0.24\linewidth]{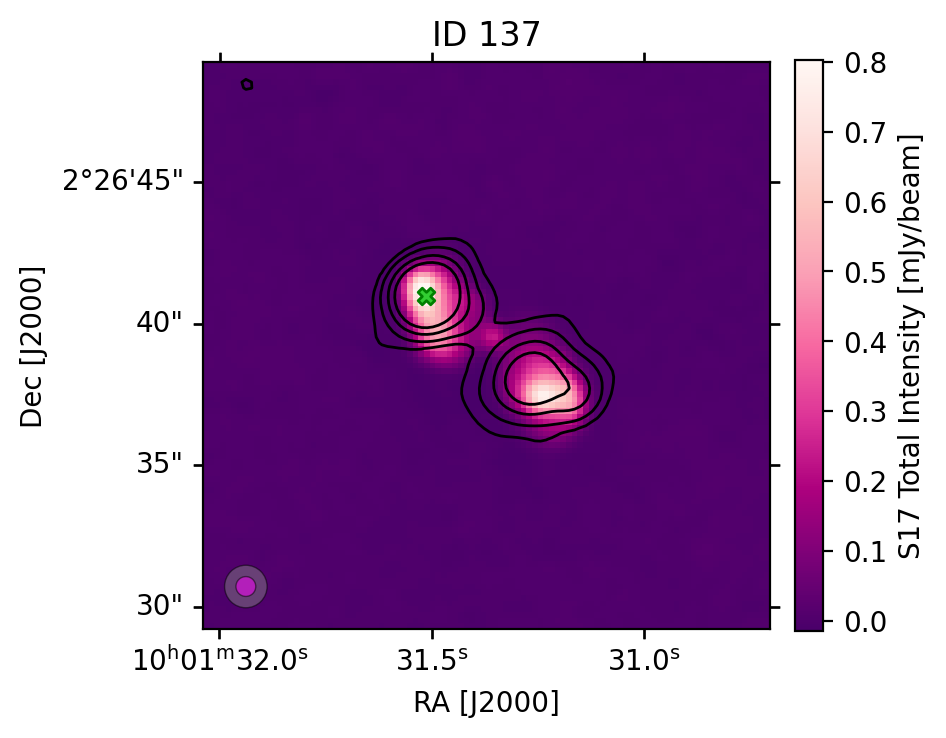}
    \includegraphics[width=0.24\linewidth]{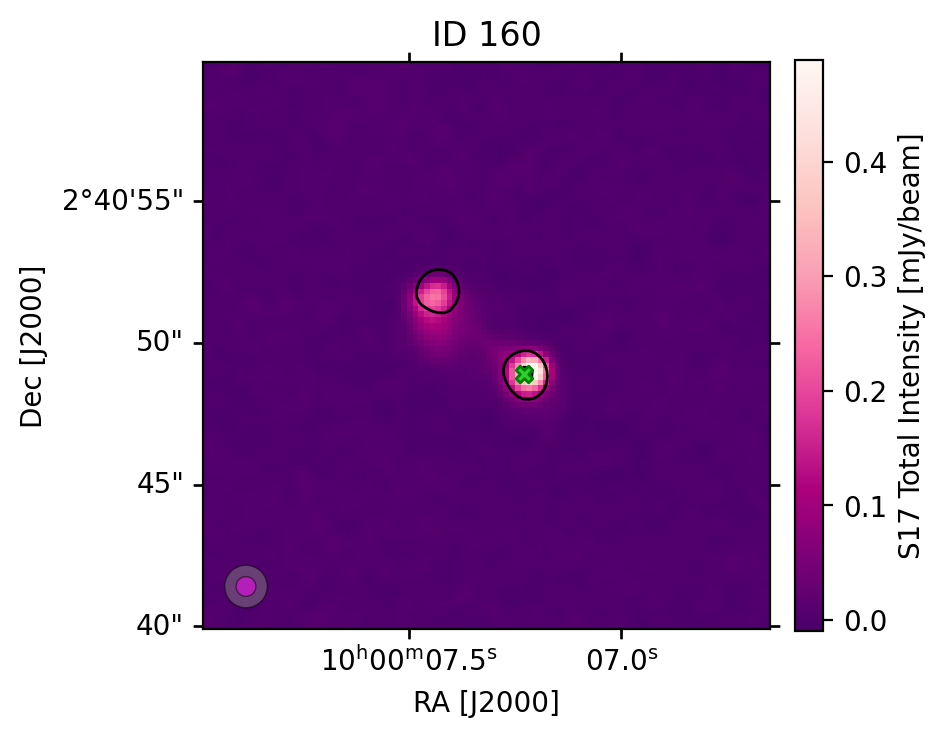} \\
    \includegraphics[width=0.24\linewidth]{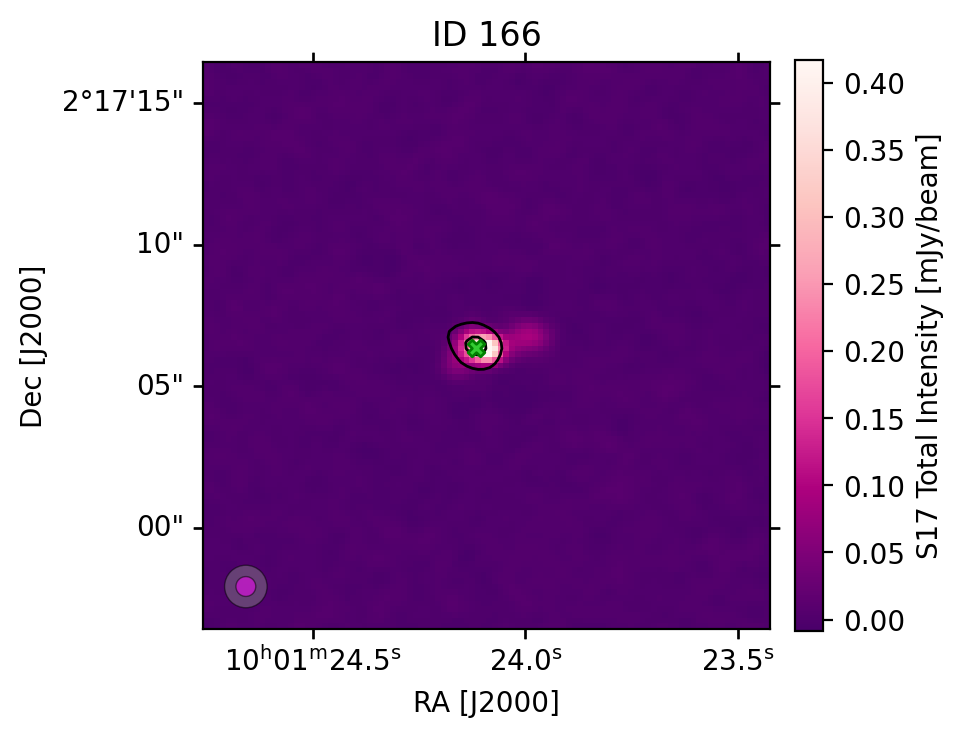}
    \includegraphics[width=0.24\linewidth]{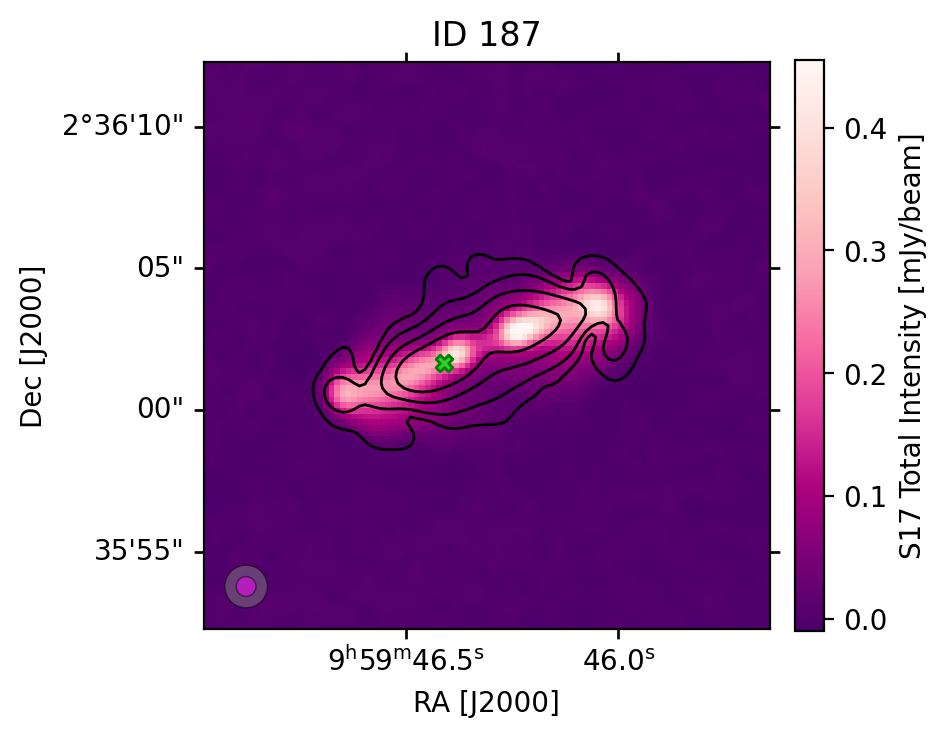}
    \includegraphics[width=0.24\linewidth]{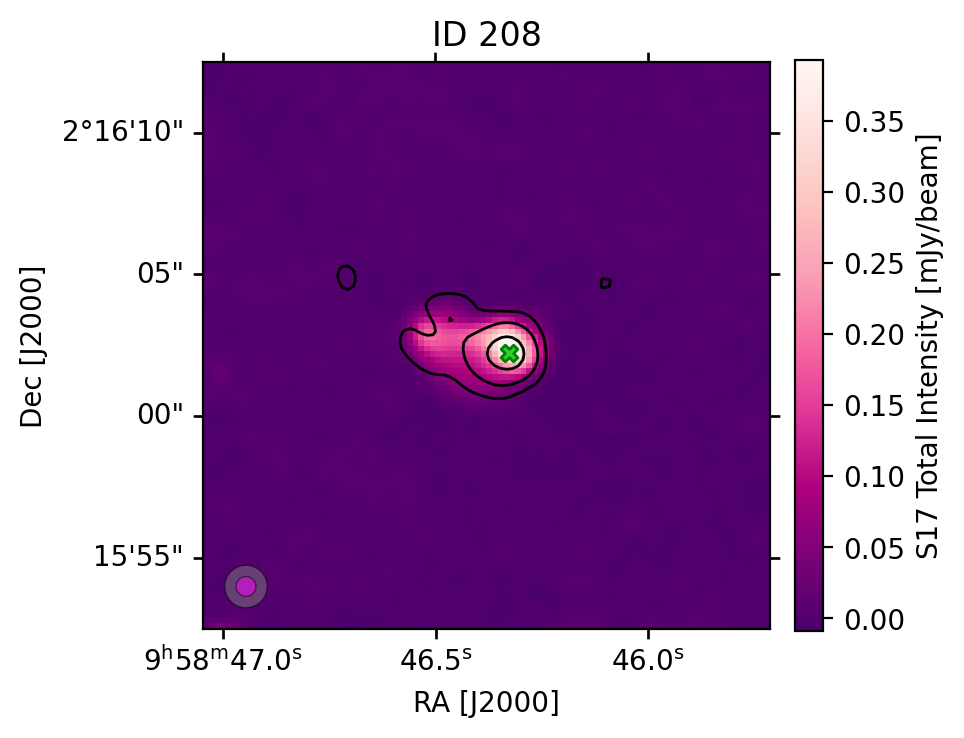}
    \includegraphics[width=0.24\linewidth]{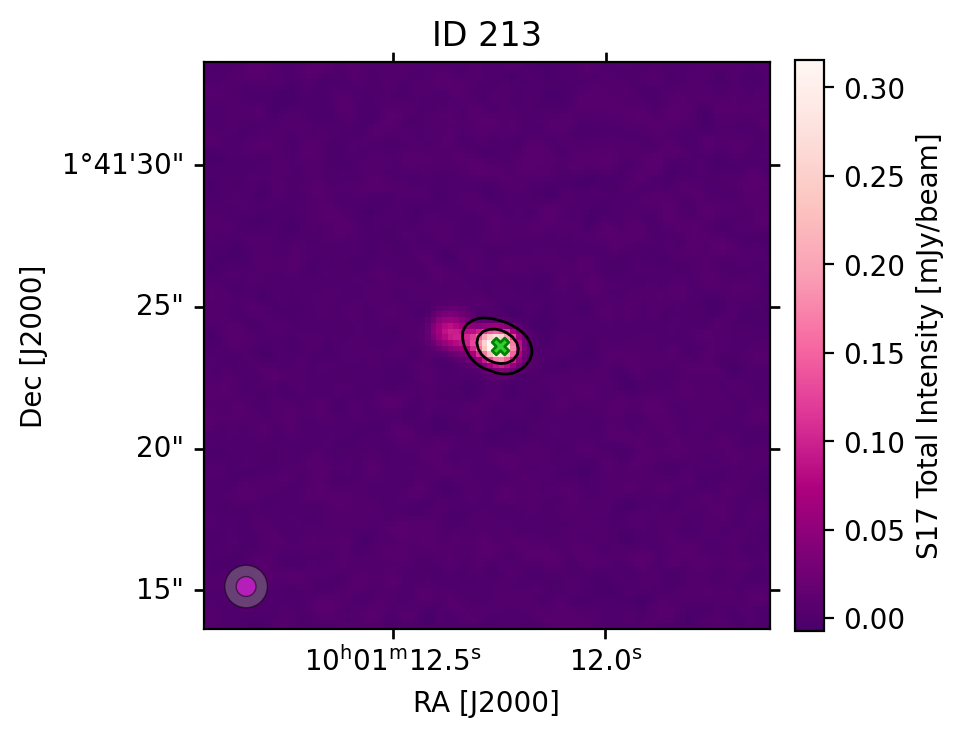} \\
    \includegraphics[width=0.24\linewidth]{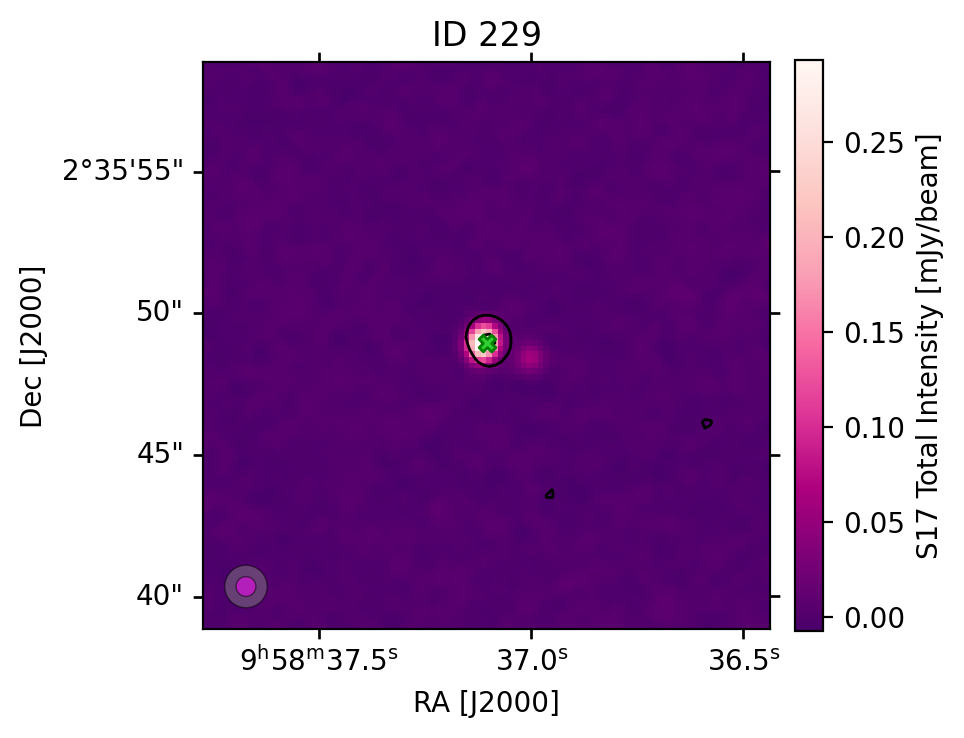}
    \includegraphics[width=0.24\linewidth]{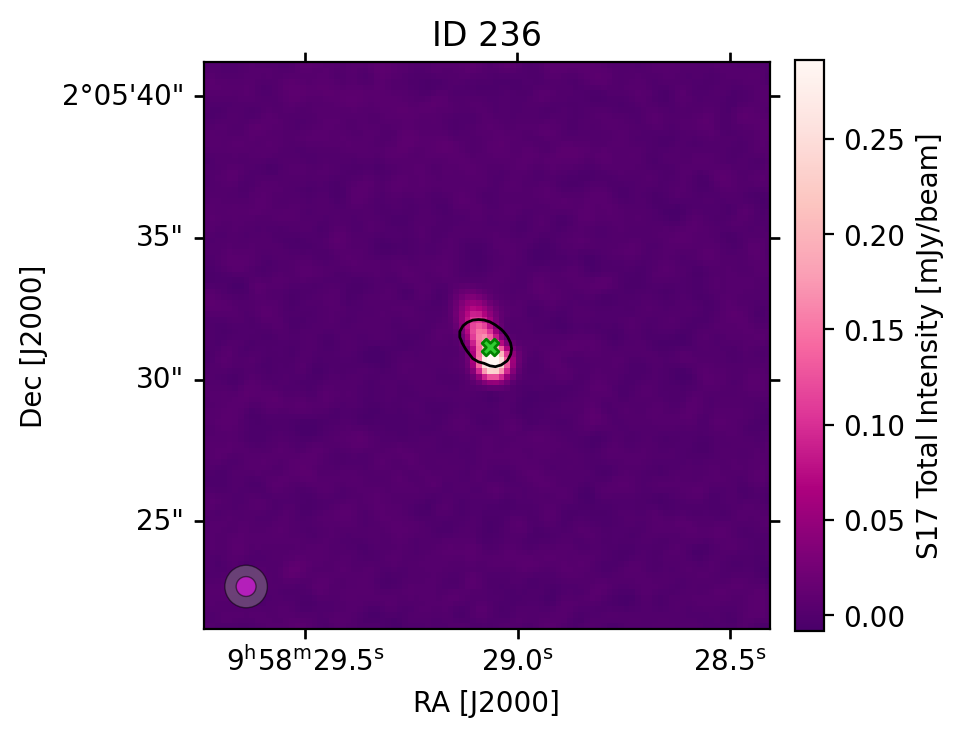}
    \includegraphics[width=0.24\linewidth]{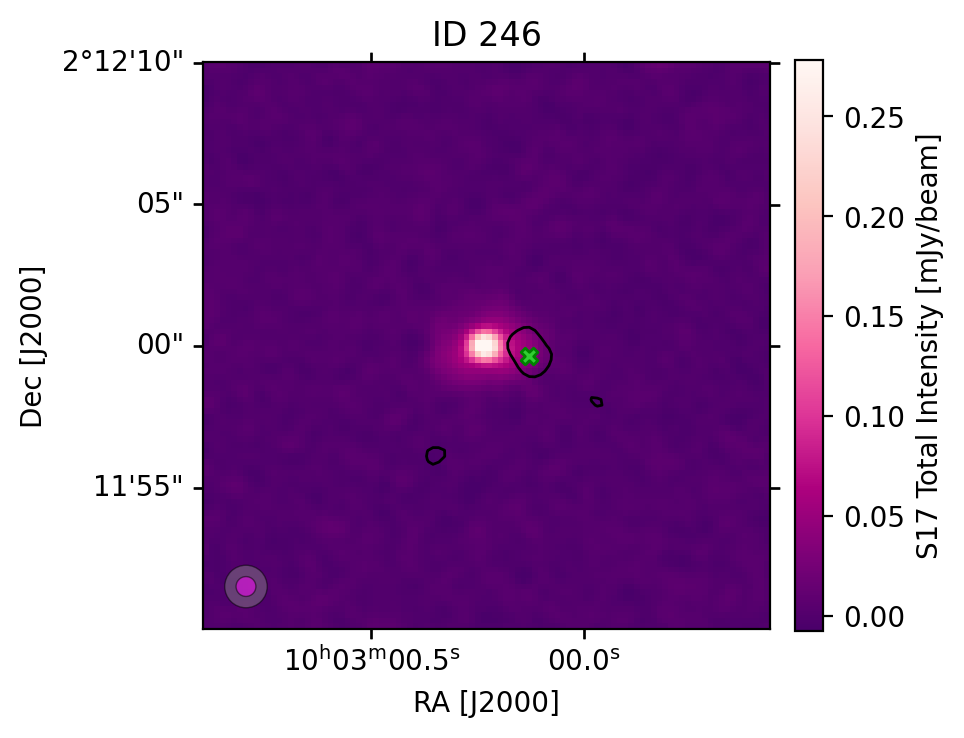} 
    \includegraphics[width=0.24\linewidth]{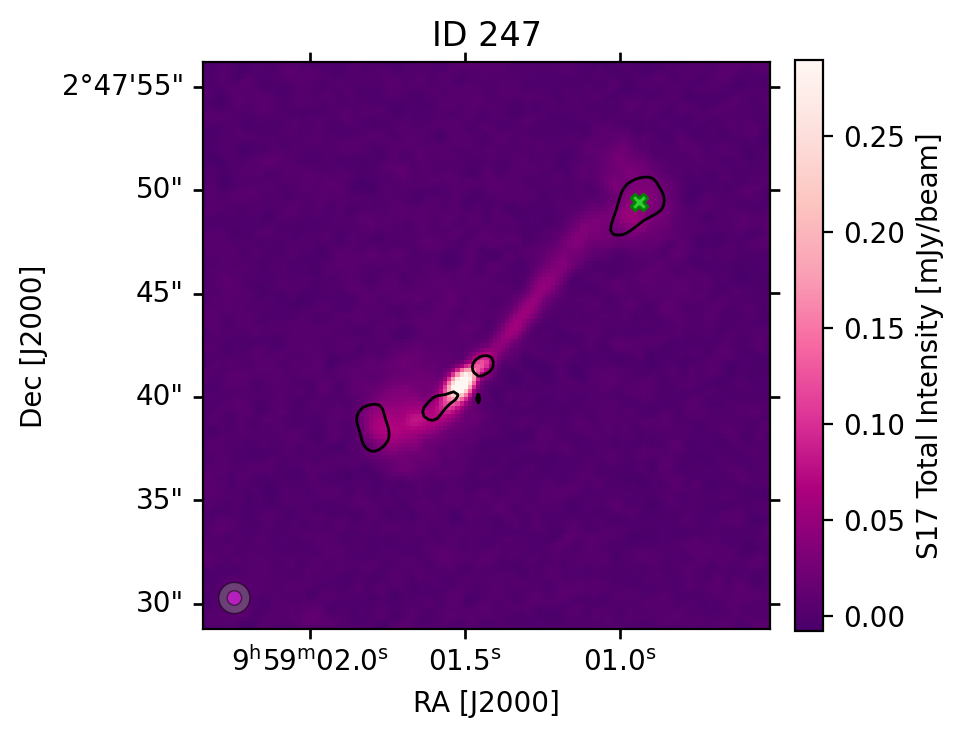} \\
    \caption[Total intensity images from \citetalias{Smolcic2017}, overlaid with the polarised intensity contours]{The total intensity images from \citetalias{Smolcic2017}, overlaid with the polarised intensity contours from this work. The contours correspond to $P =$~[22, 45, 90, 150]~$\mu$\jybeam. The restoring beams for the total intensity (magenta) and polarised intensity (grey) images are shown in the bottom left corner. The green cross indicates the position at which the Stokes I, Q, and U spectra are extracted for spectral analysis.}
    \label{appB:fig:image-cutouts}
\end{figure*}

\begin{figure*}\ContinuedFloat
    \centering
    \includegraphics[width=0.24\linewidth]{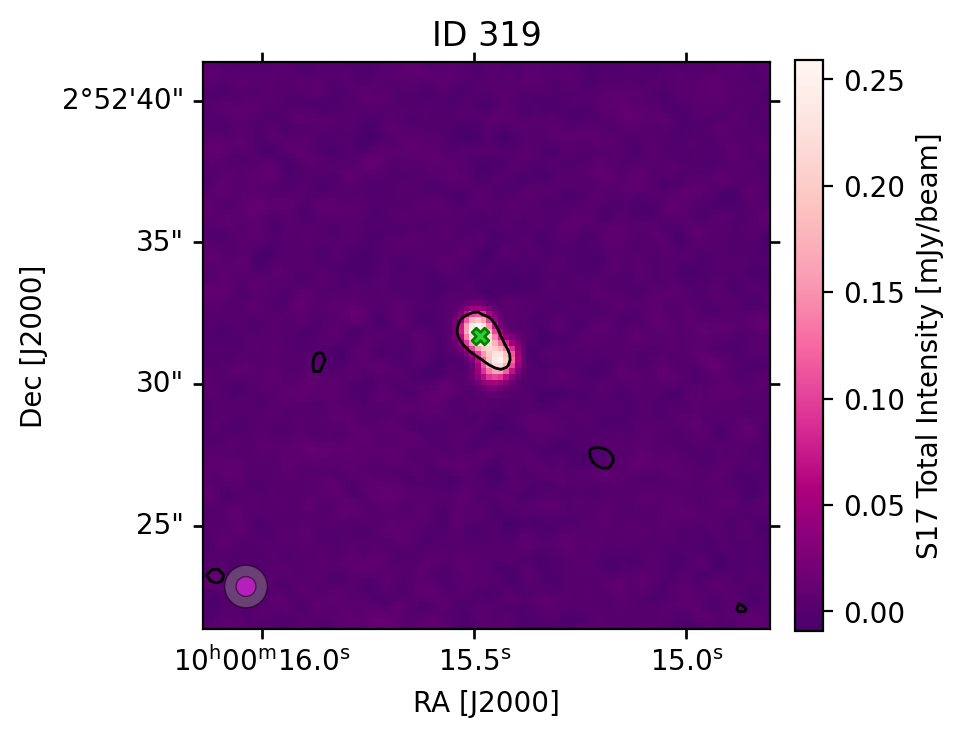}
    \includegraphics[width=0.24\linewidth]{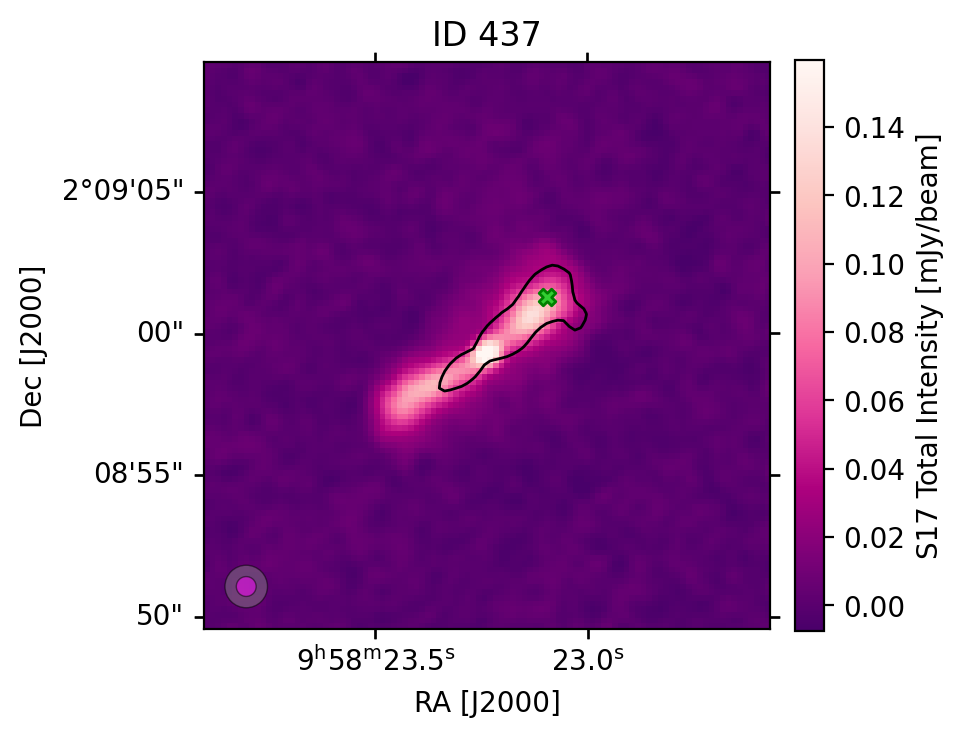}
    \includegraphics[width=0.24\linewidth]{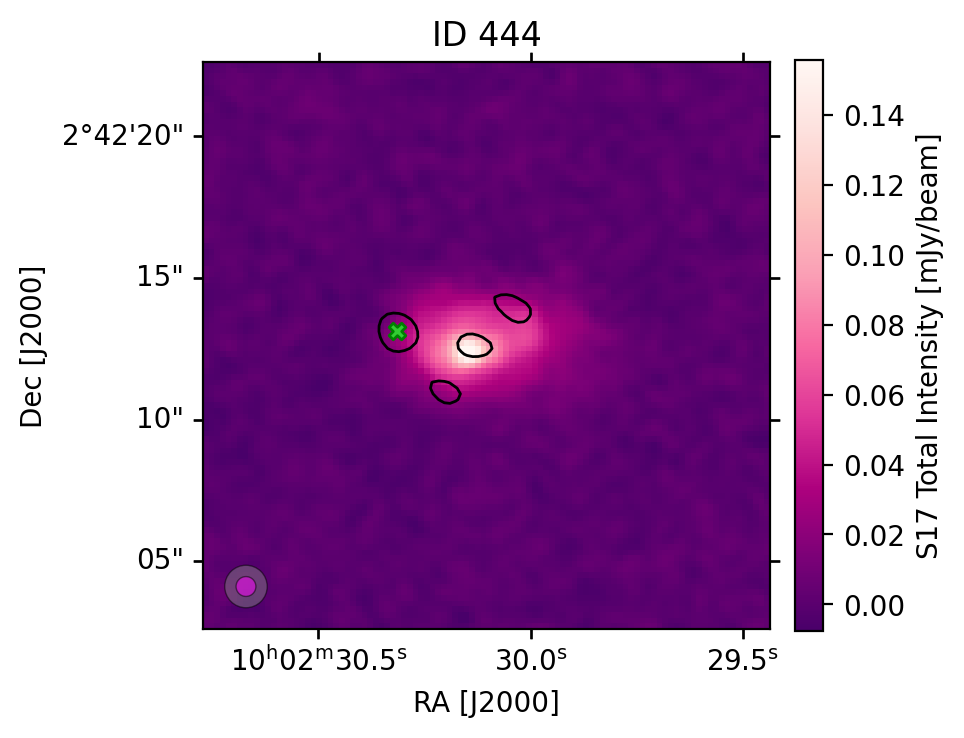}
    \includegraphics[width=0.24\linewidth]{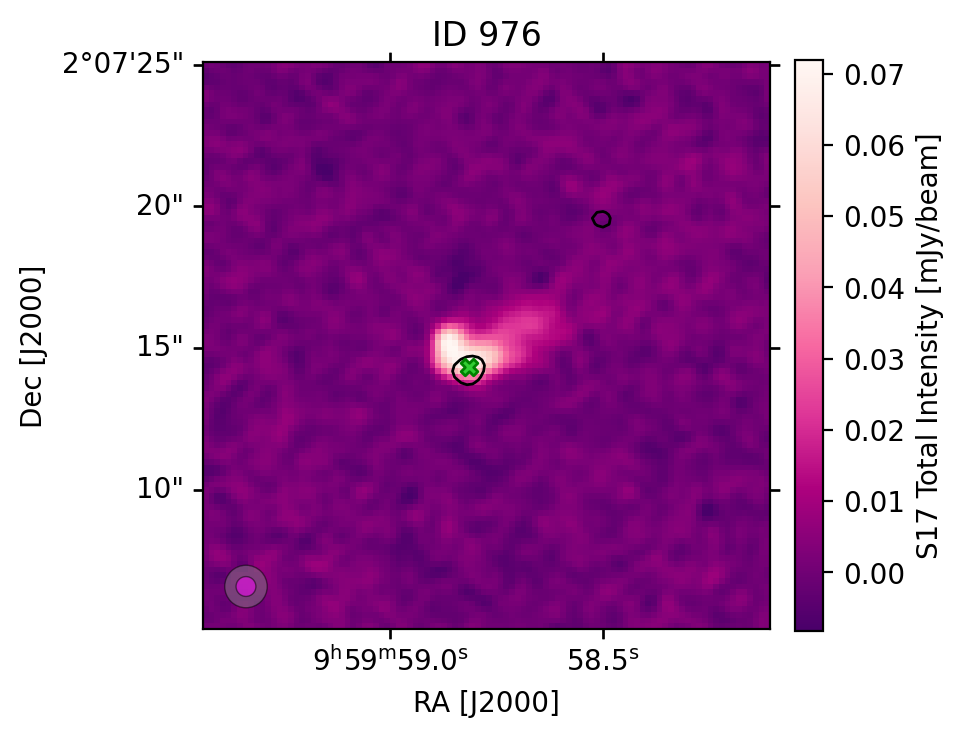} \\
    \includegraphics[height=0.25\linewidth]{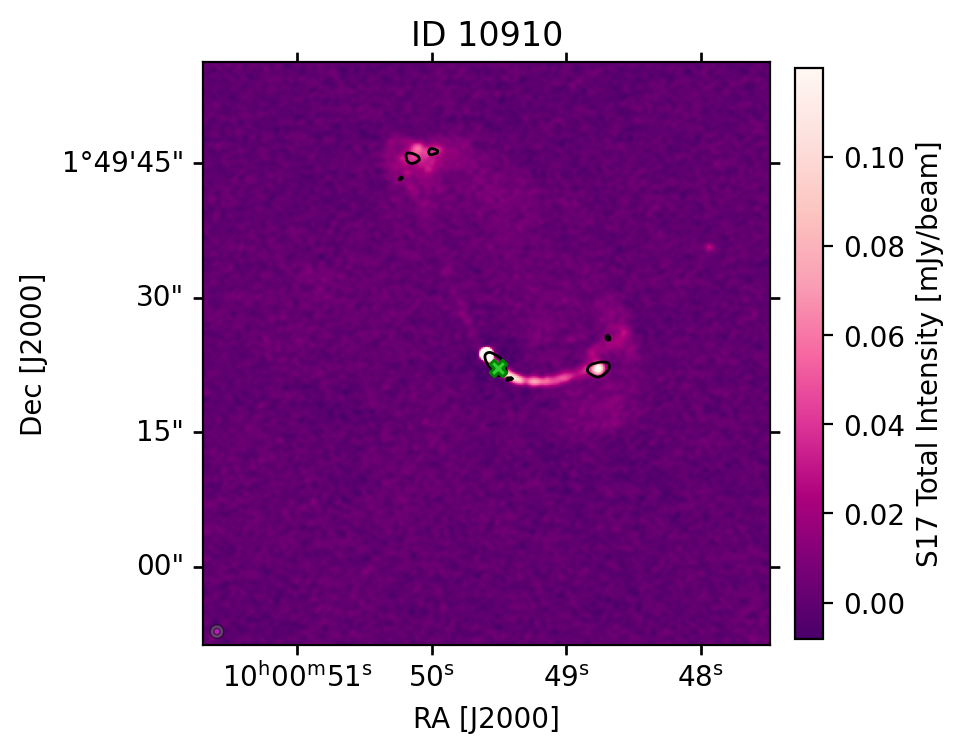}
    \includegraphics[height=0.25\linewidth]{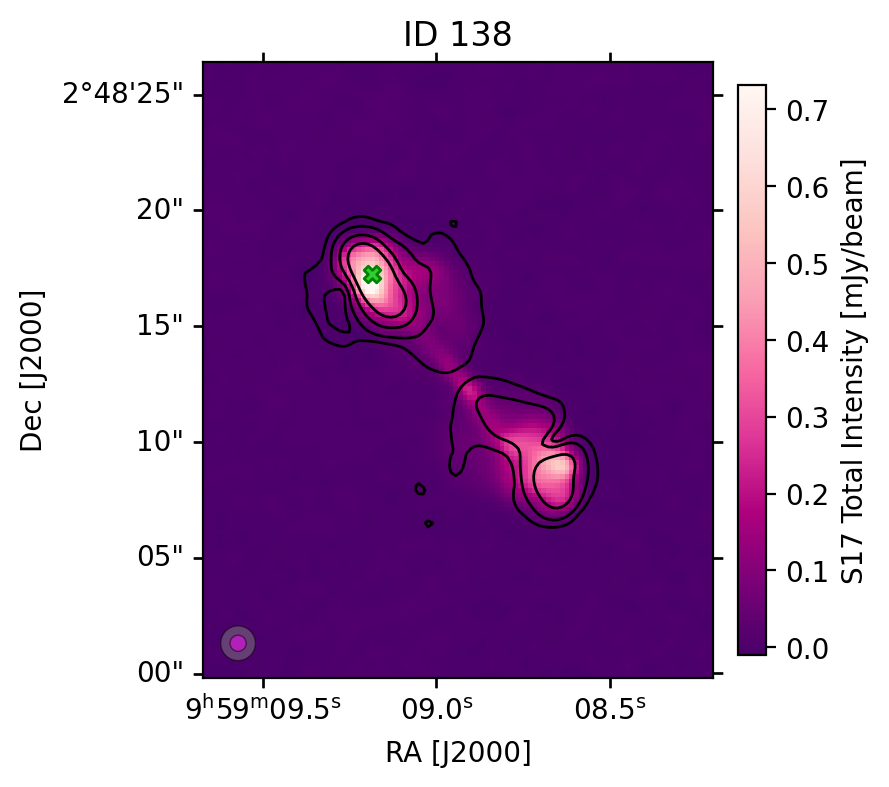}
    \includegraphics[height=0.25\linewidth]{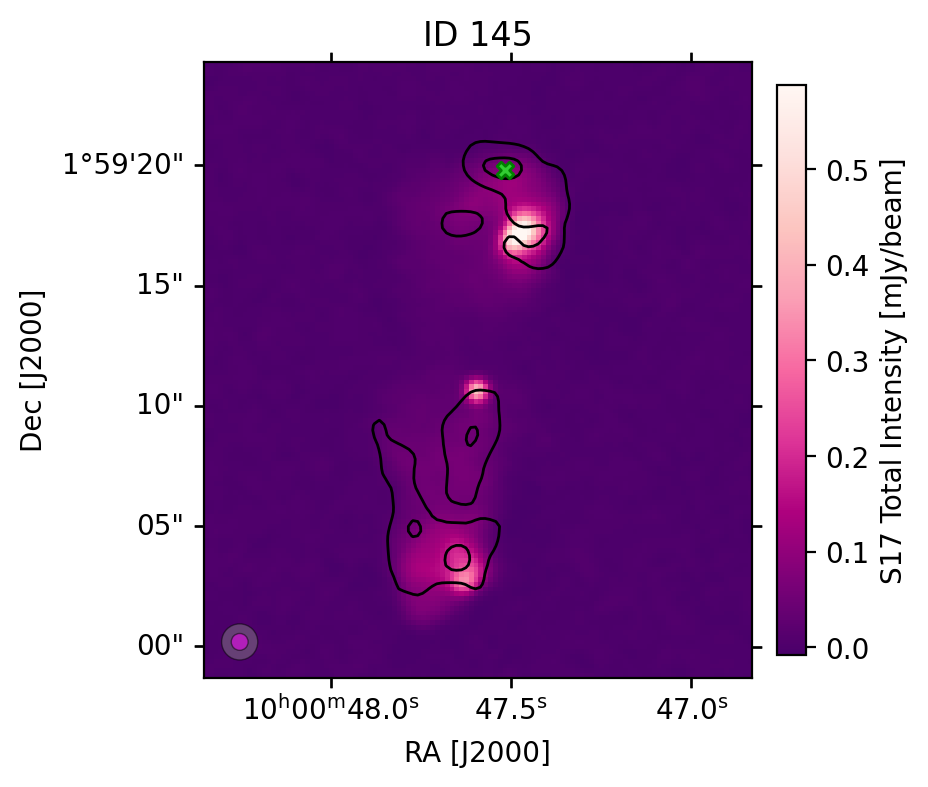} \\
    \includegraphics[height=0.3\linewidth]{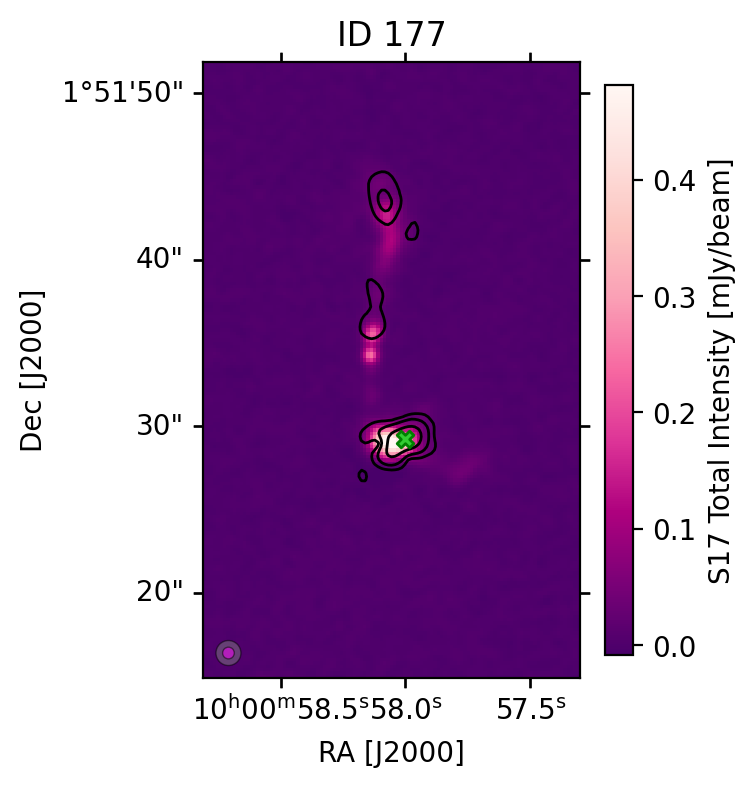} 
    \includegraphics[height=0.3\linewidth]{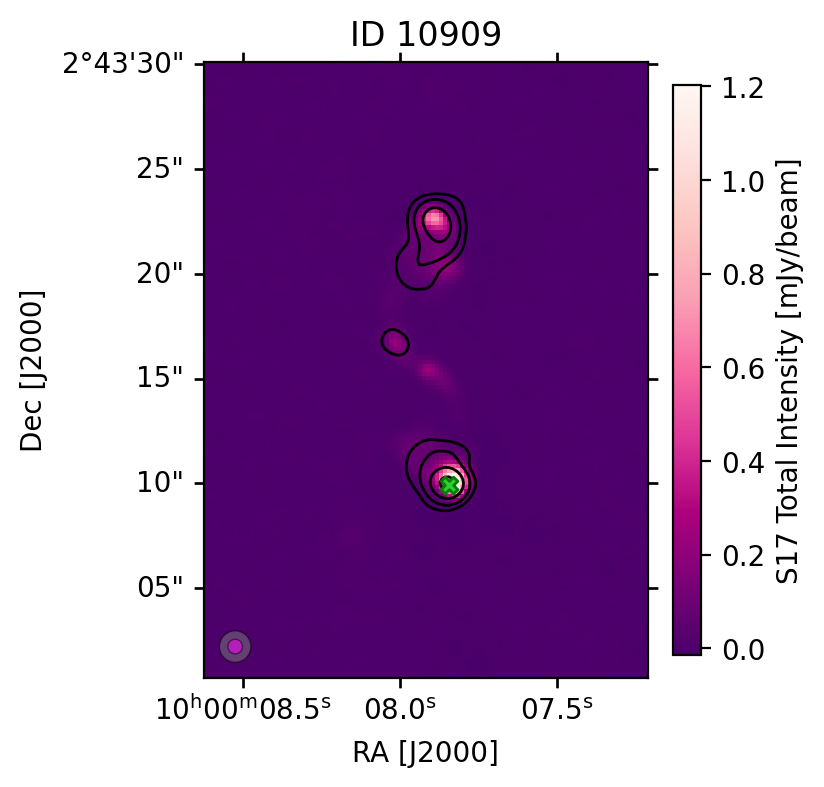}
    \includegraphics[height=0.3\linewidth]{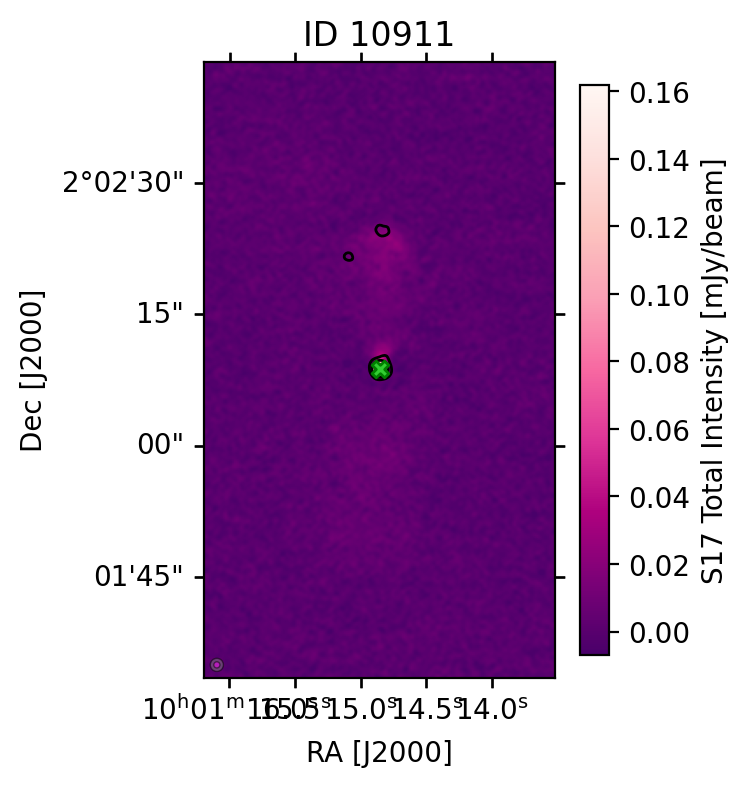} \\
    \includegraphics[height=0.22\linewidth]{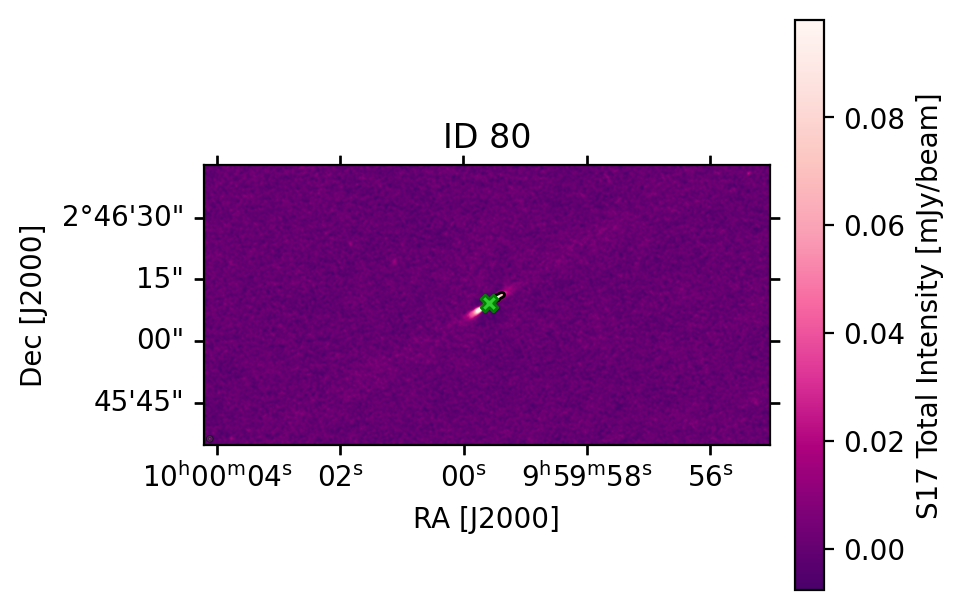}
    \includegraphics[height=0.22\linewidth]{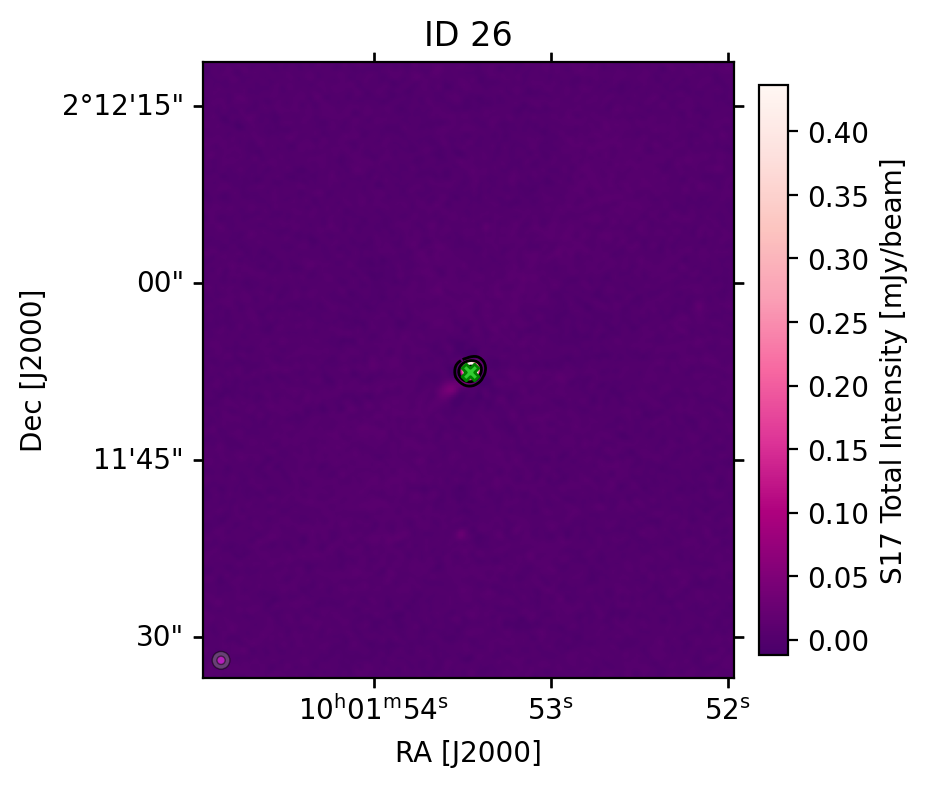}
    \includegraphics[height=0.22\linewidth]{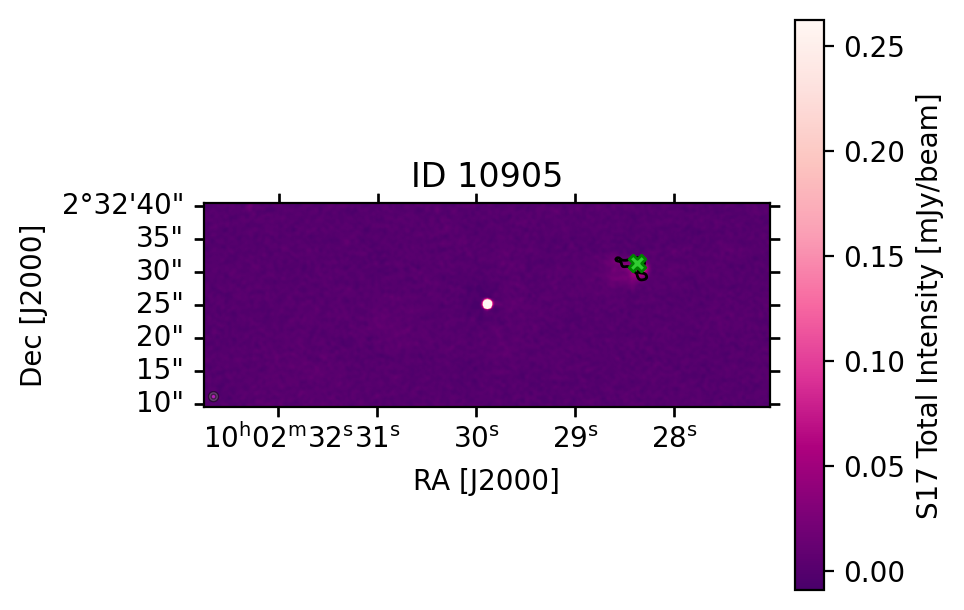} \\
    \includegraphics[height=0.25\linewidth]{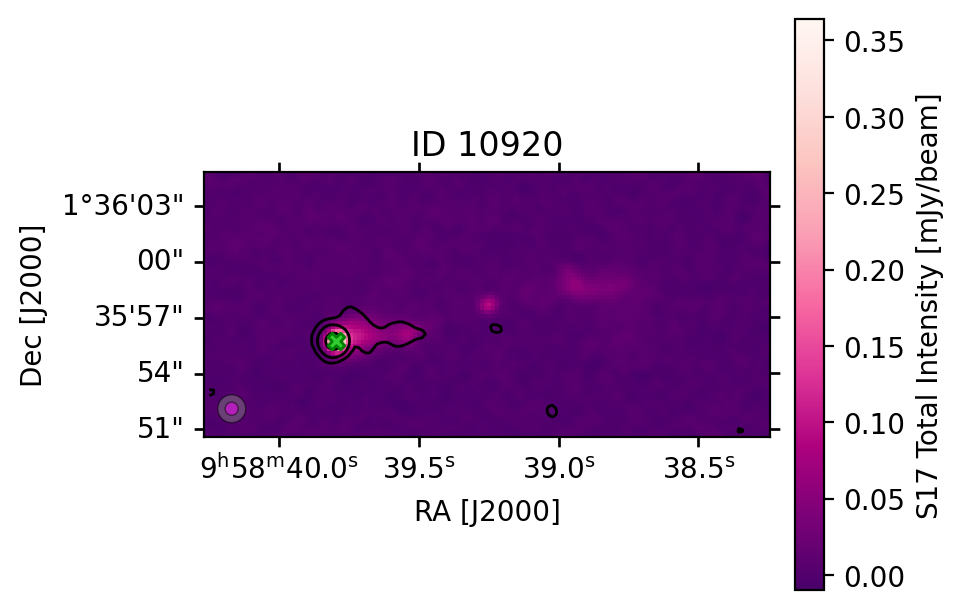}
    \includegraphics[height=0.25\linewidth]{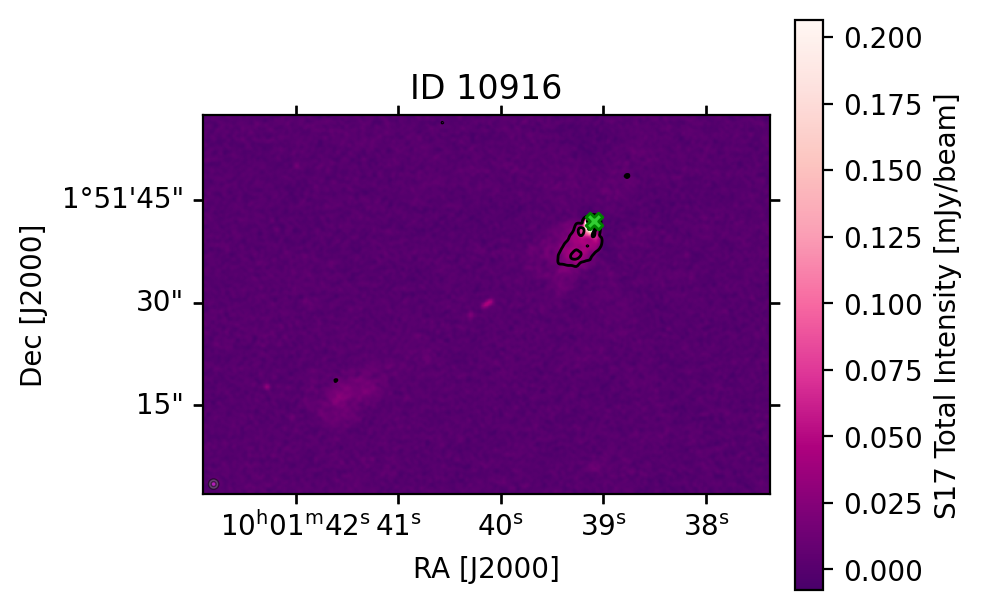} 
    \captionsetup{list=no}
    \caption{Total intensity images from \citetalias{Smolcic2017}, overlaid with the polarised intensity contours from this work, continued.}
\end{figure*}

\begin{figure*}\ContinuedFloat
    \centering
    \includegraphics[width=0.32\linewidth]{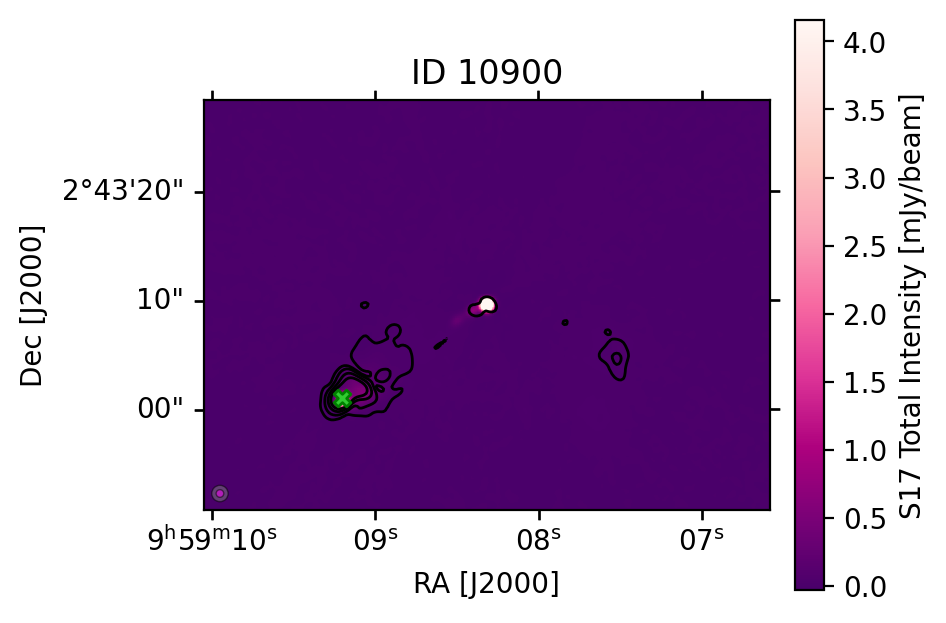} 
    \includegraphics[width=0.32\linewidth]{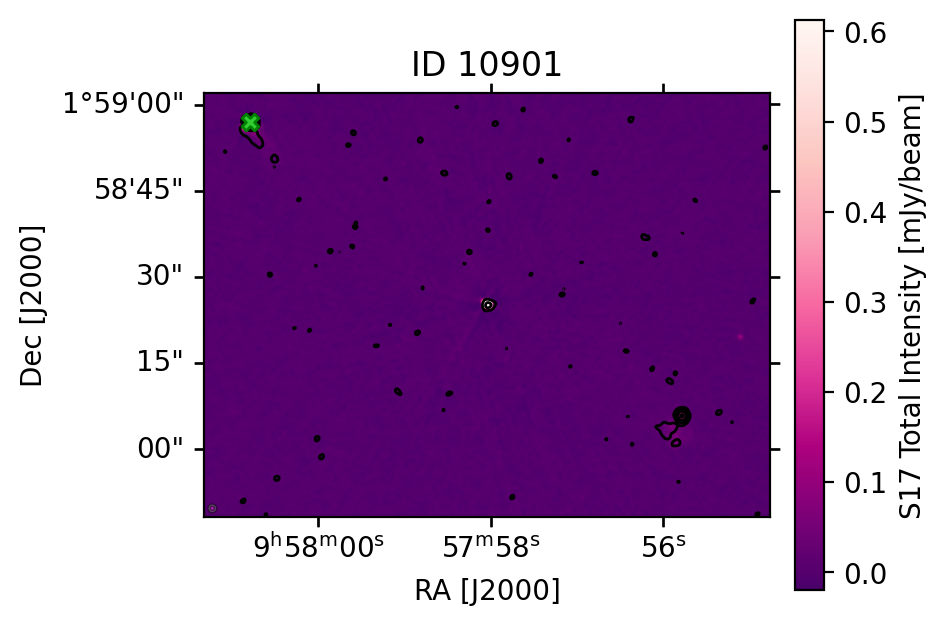}
    \includegraphics[width=0.32\linewidth]{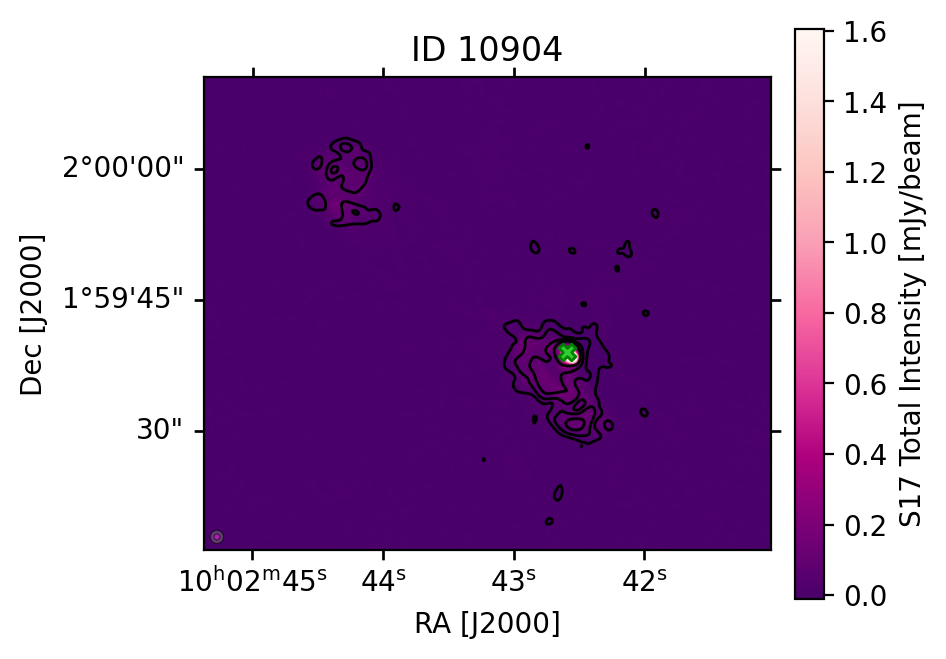} \\
    \includegraphics[width=0.32\linewidth]{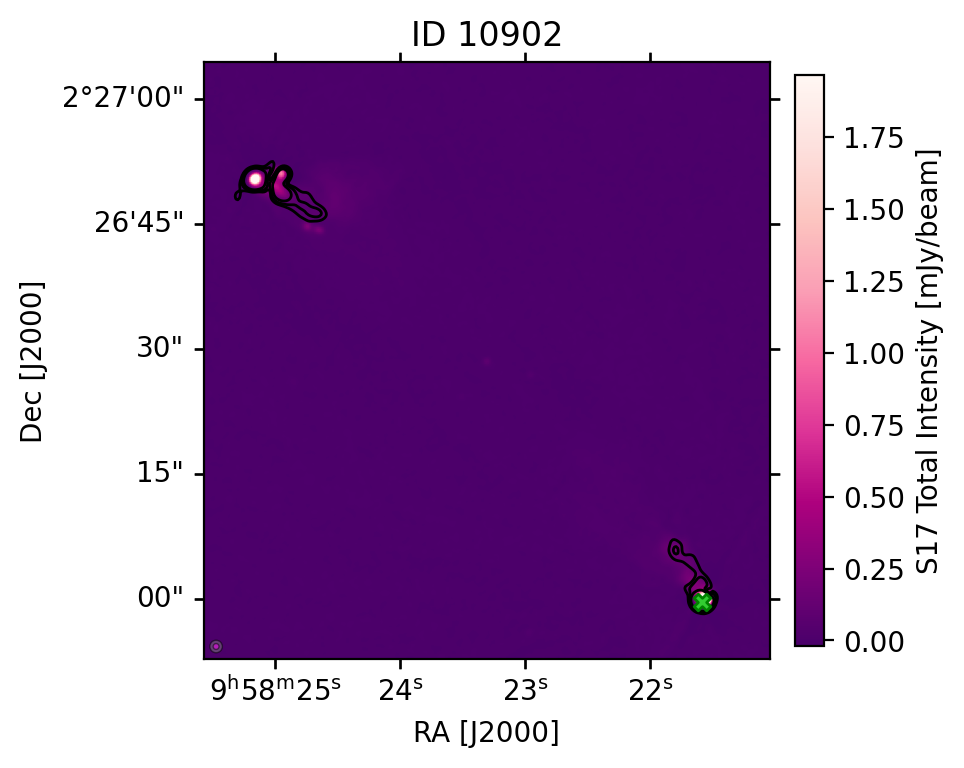}
    \includegraphics[width=0.32\linewidth]{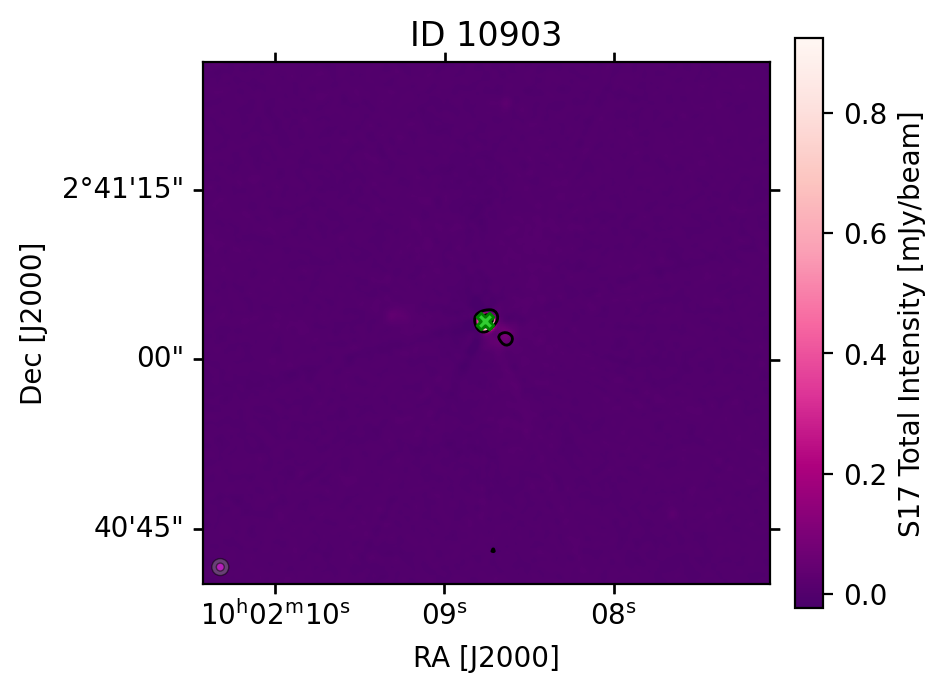} 
    \includegraphics[width=0.32\linewidth]{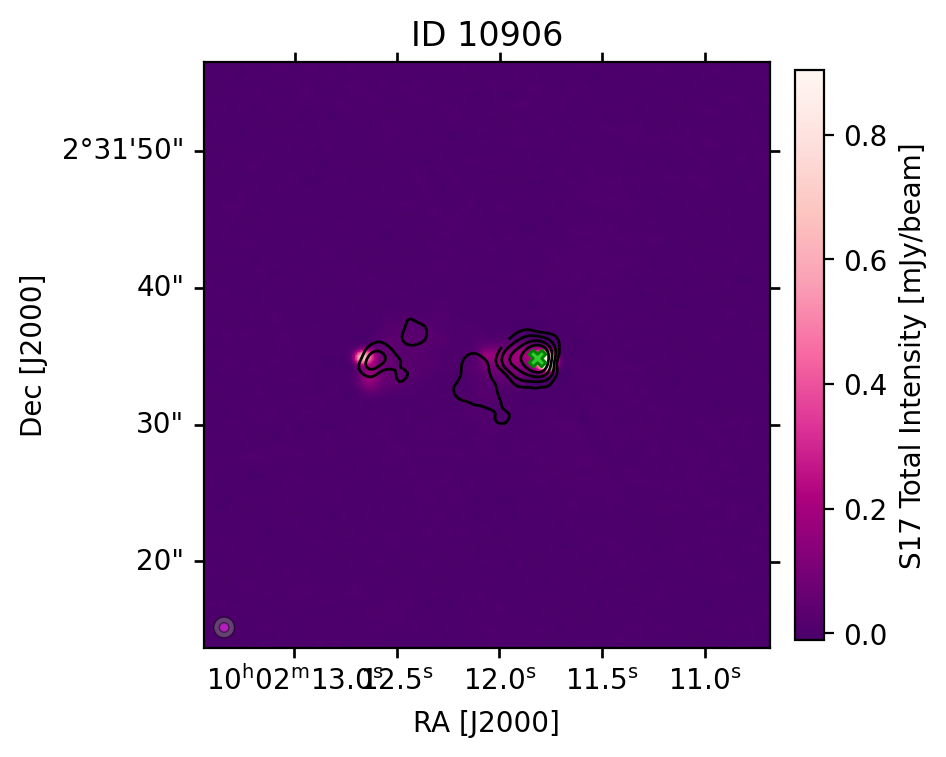}\\
    \includegraphics[width=0.32\linewidth]{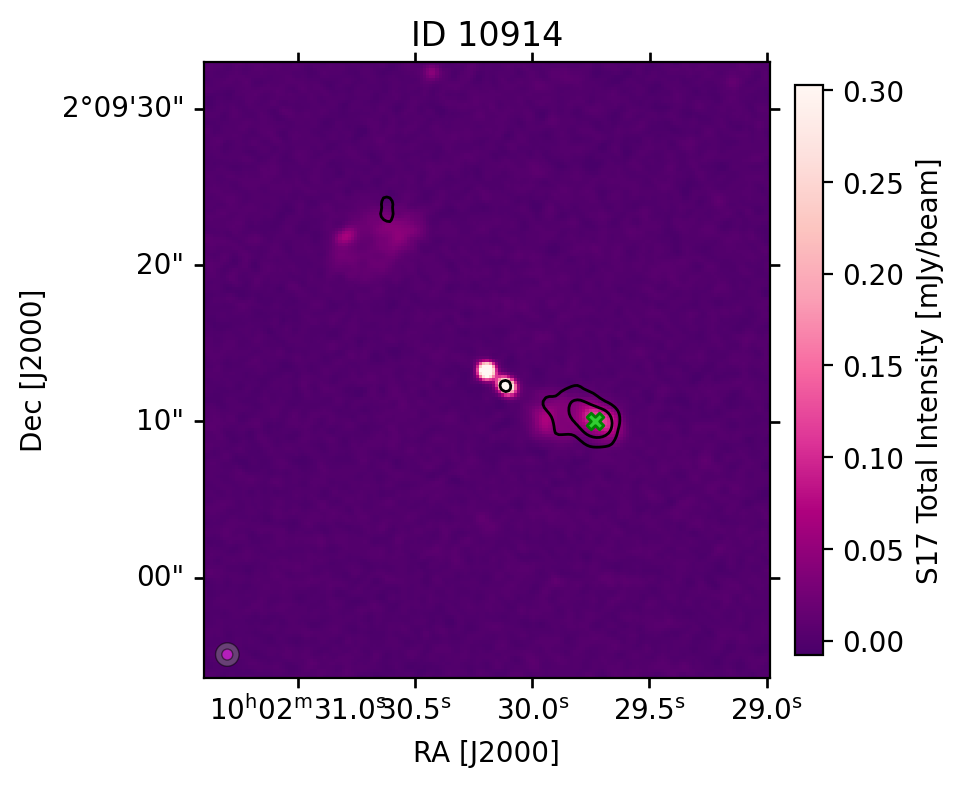} 
    \includegraphics[width=0.32\linewidth]{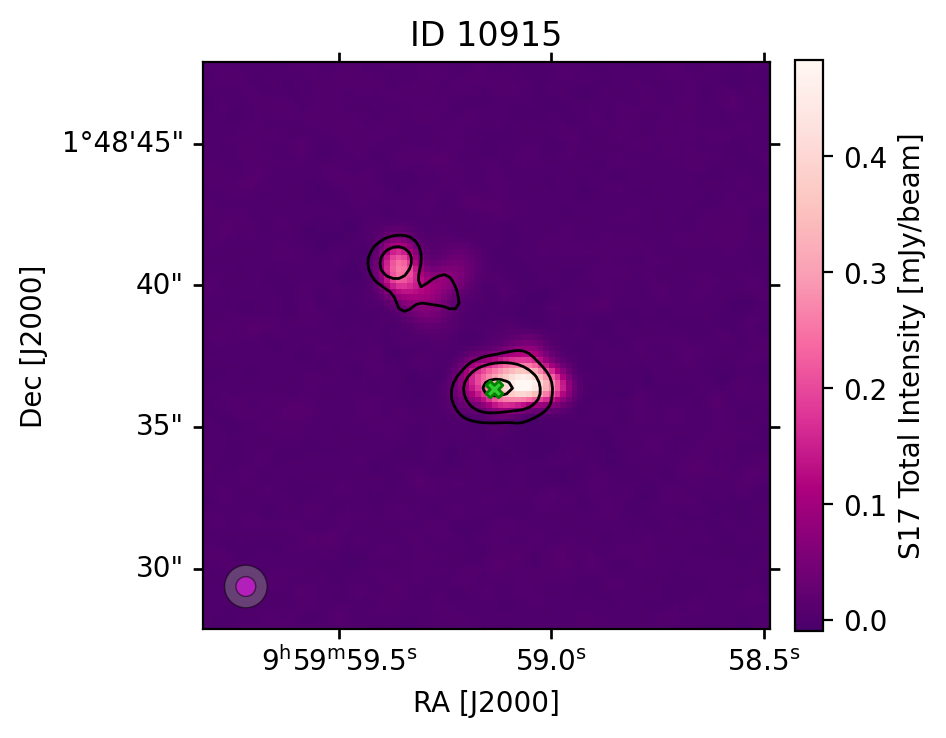}
    \includegraphics[width=0.32\linewidth]{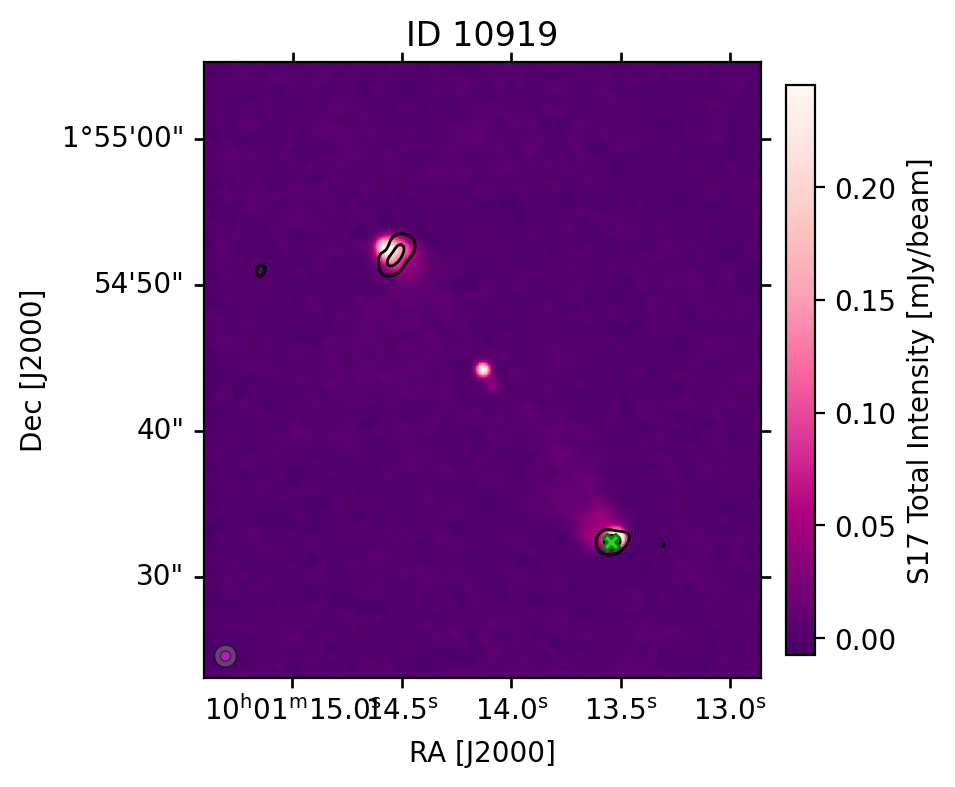} \\
    \includegraphics[width=0.32\linewidth]{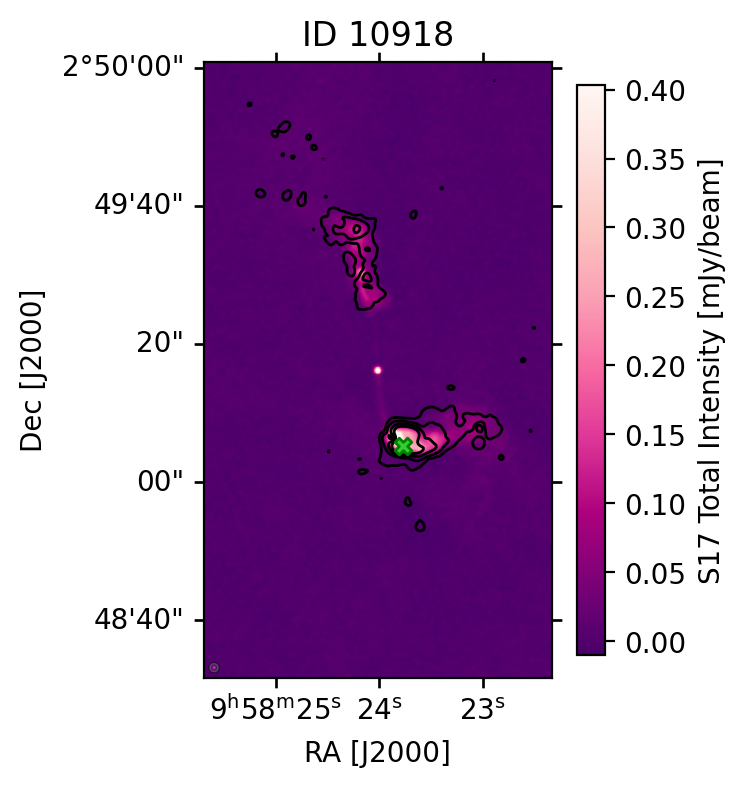}
    \includegraphics[width=0.32\linewidth]{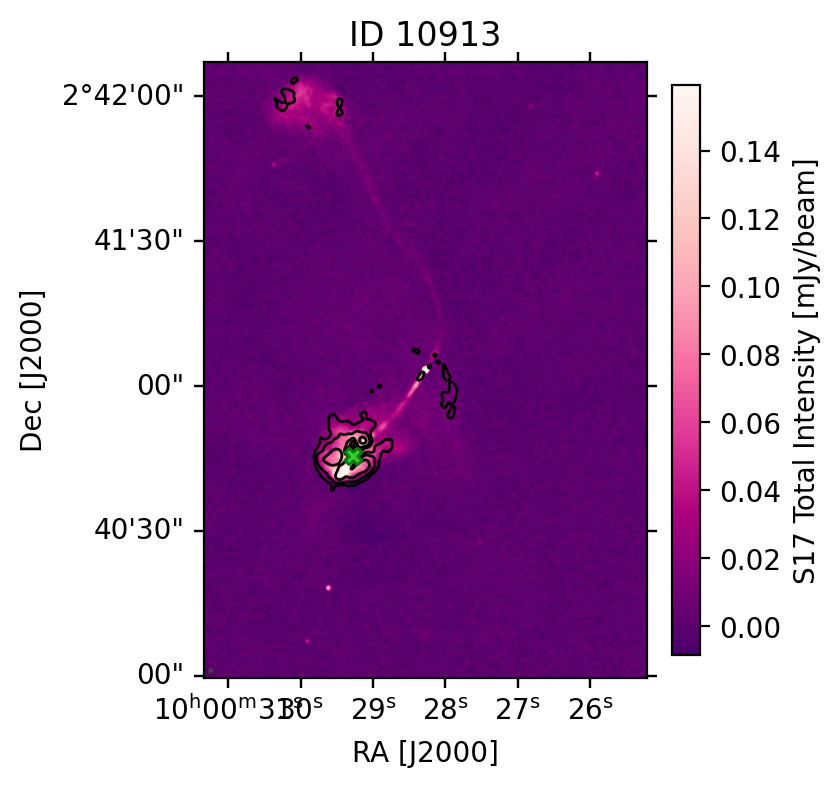}
    \includegraphics[width=0.32\linewidth]{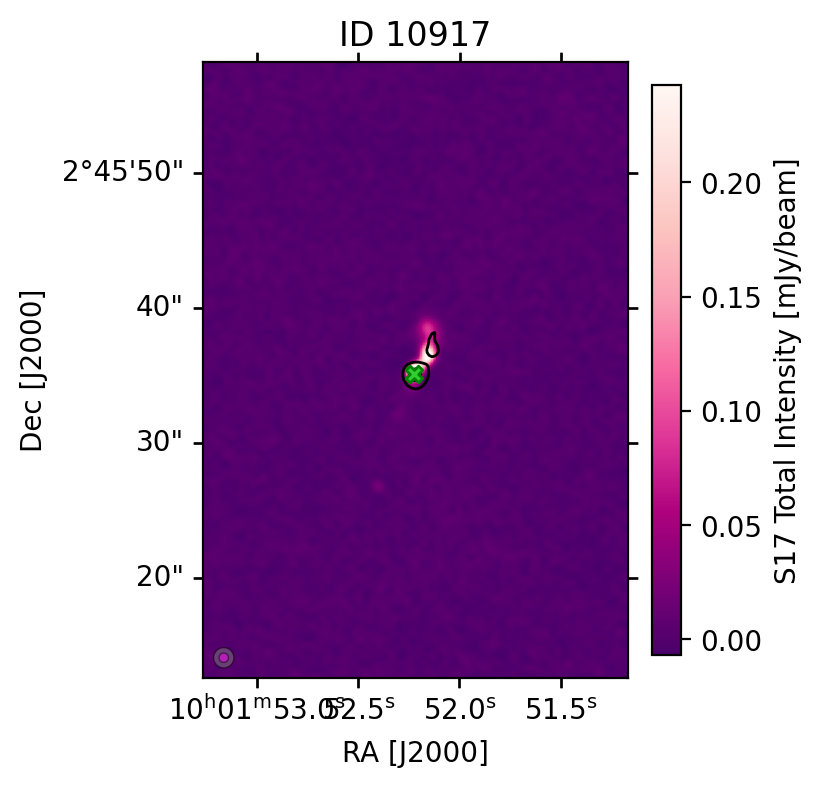} \\
    \includegraphics[width=0.32\linewidth]{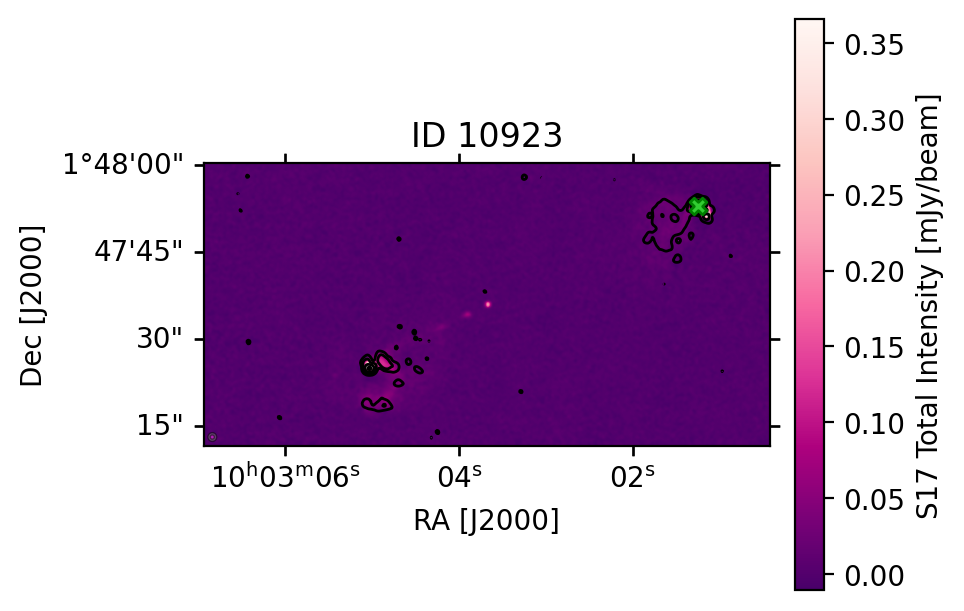}
    \includegraphics[width=0.32\linewidth]{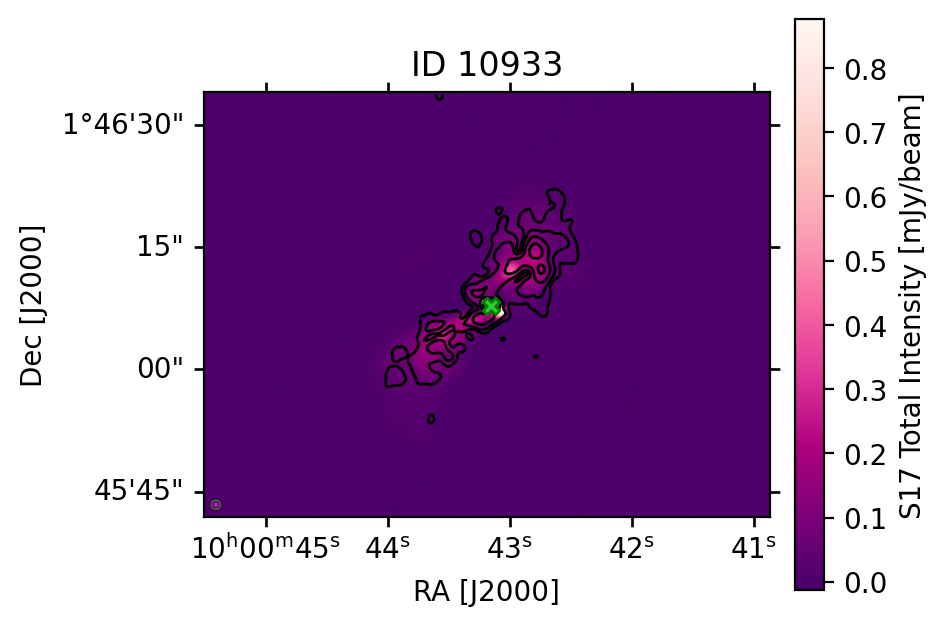}
    \includegraphics[width=0.32\linewidth]{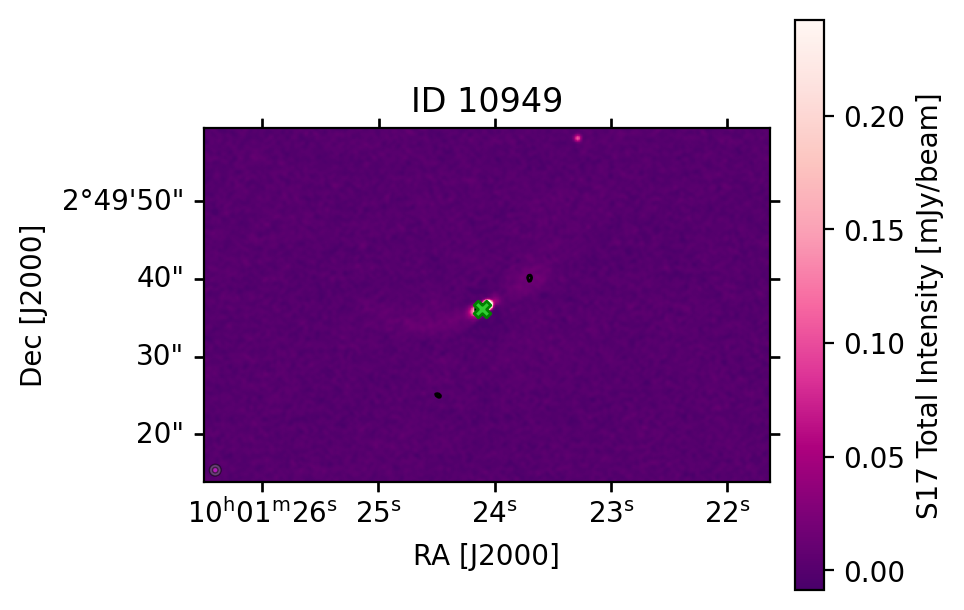}
    \captionsetup{list=no}
    \caption{Total intensity images from \citetalias{Smolcic2017}, overlaid with the polarised intensity contours from this work, continued.}
\end{figure*}

\begin{figure*}\ContinuedFloat
    \centering
    \includegraphics[width=0.32\linewidth]{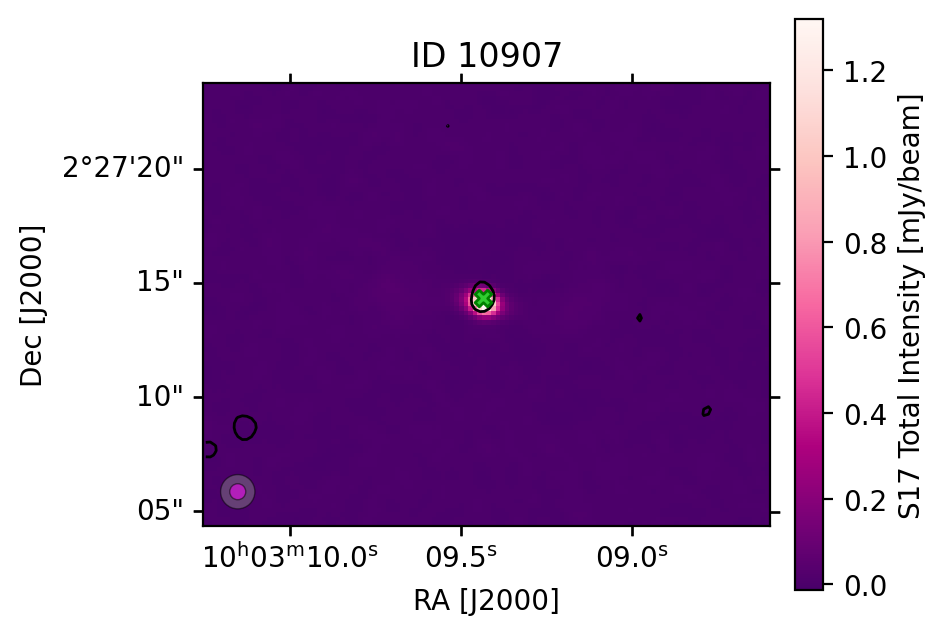}
    \includegraphics[width=0.32\linewidth]{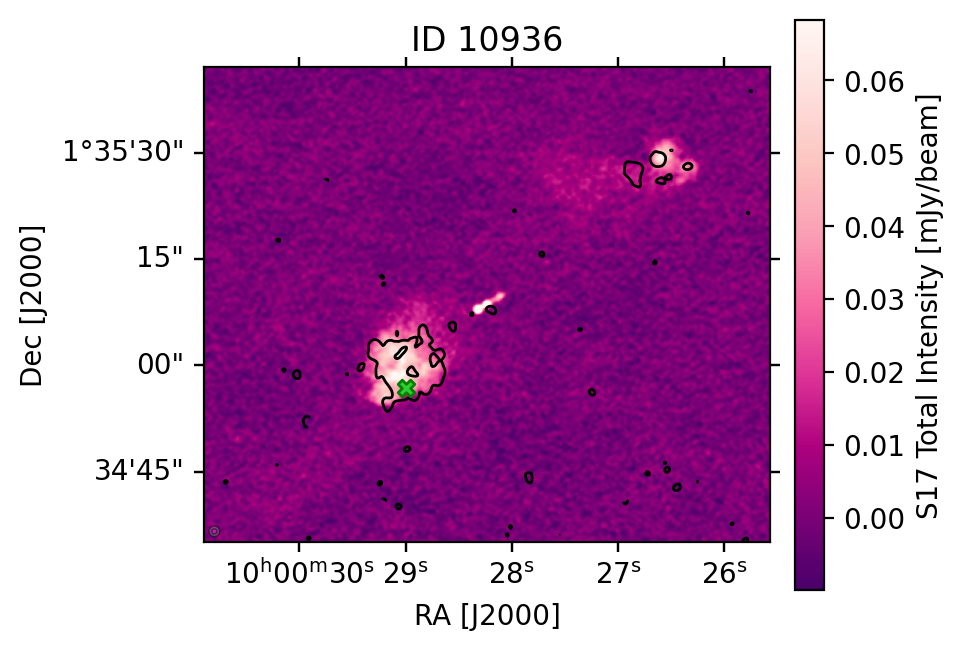}
    \includegraphics[width=0.32\linewidth]{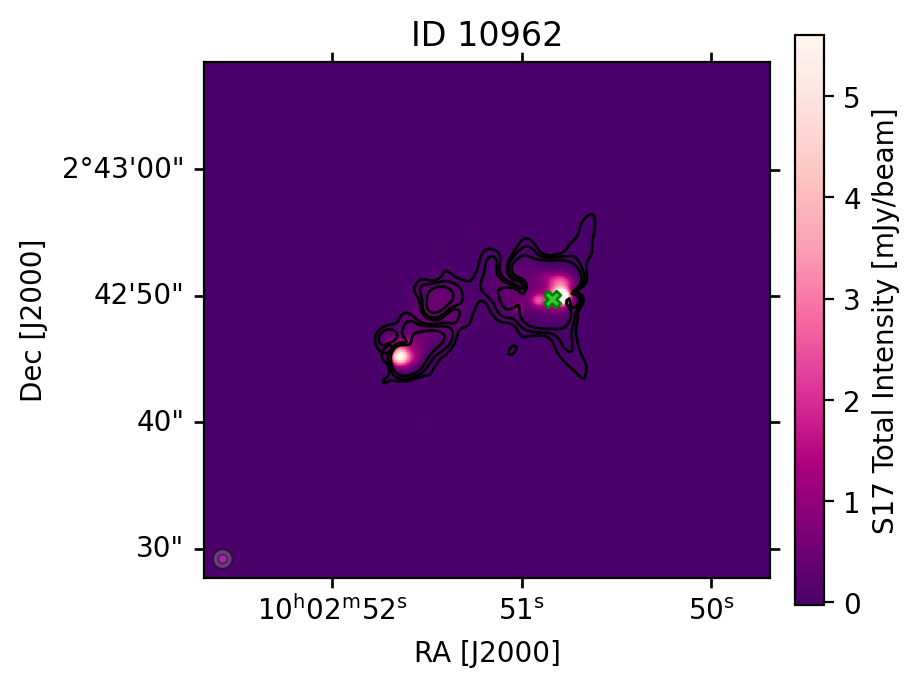} \\
    \includegraphics[width=0.32\linewidth]{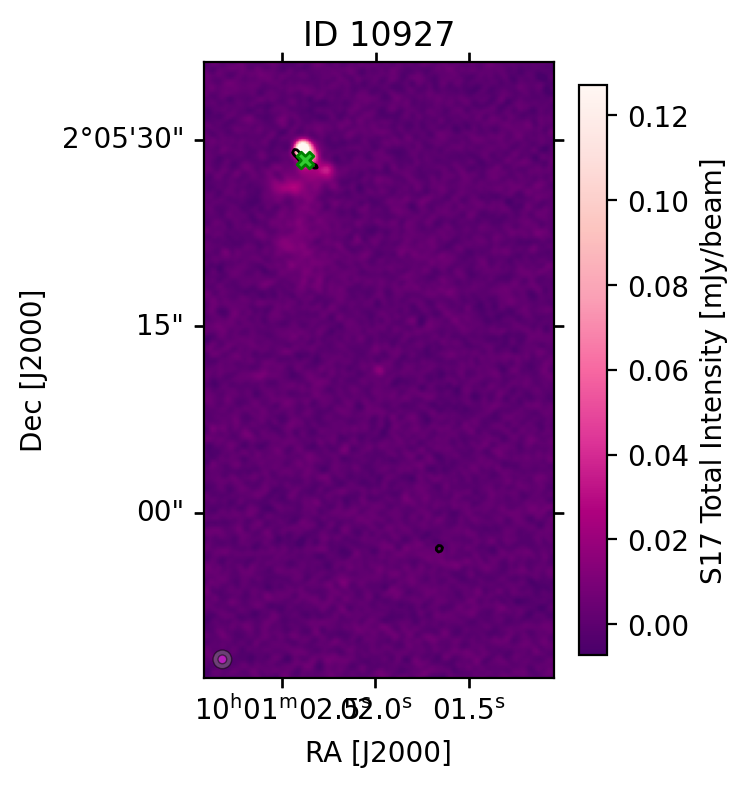}
    \includegraphics[width=0.32\linewidth]{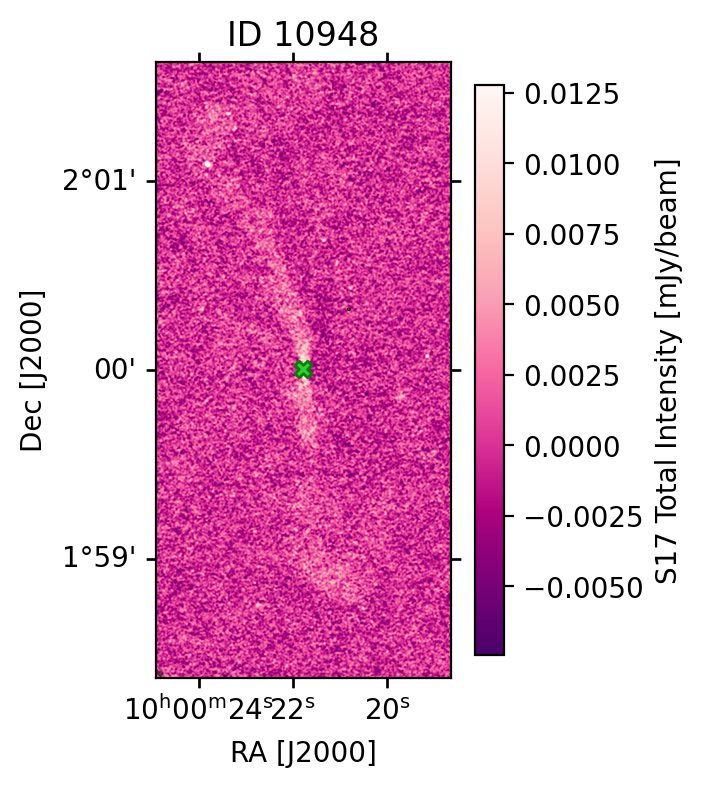}
    \includegraphics[width=0.32\linewidth]{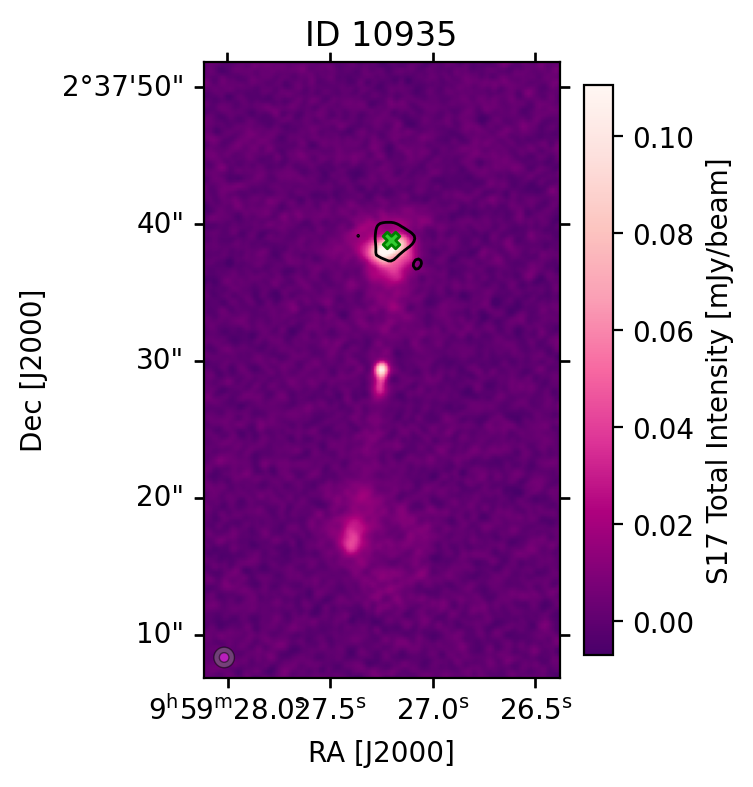} \\
    \includegraphics[width=0.32\linewidth]{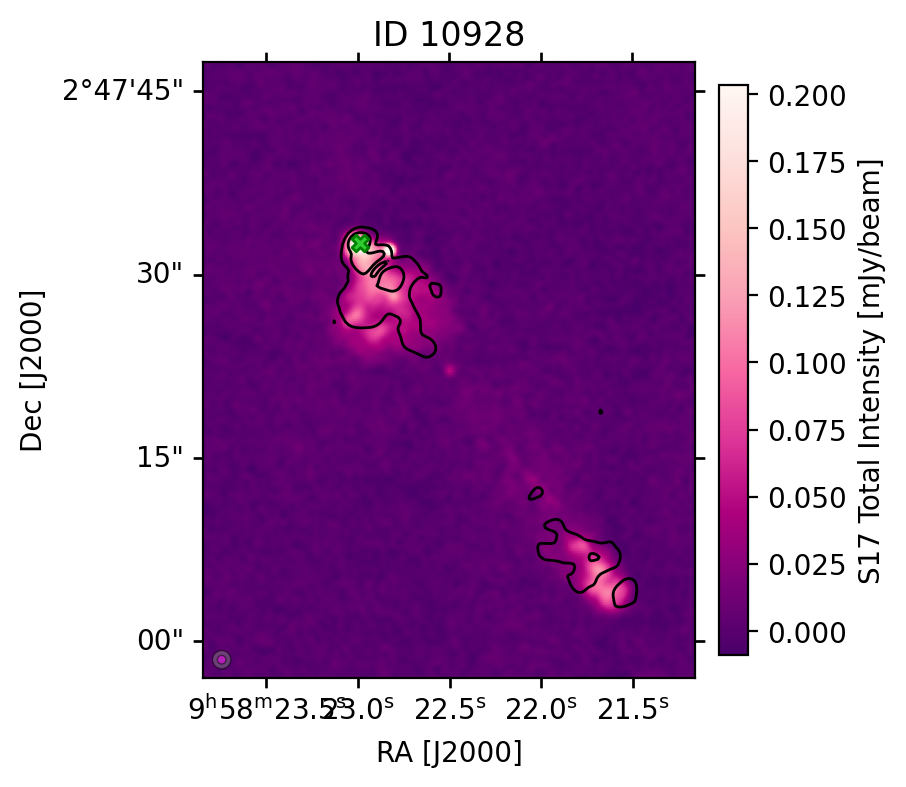}
    \includegraphics[width=0.32\linewidth]{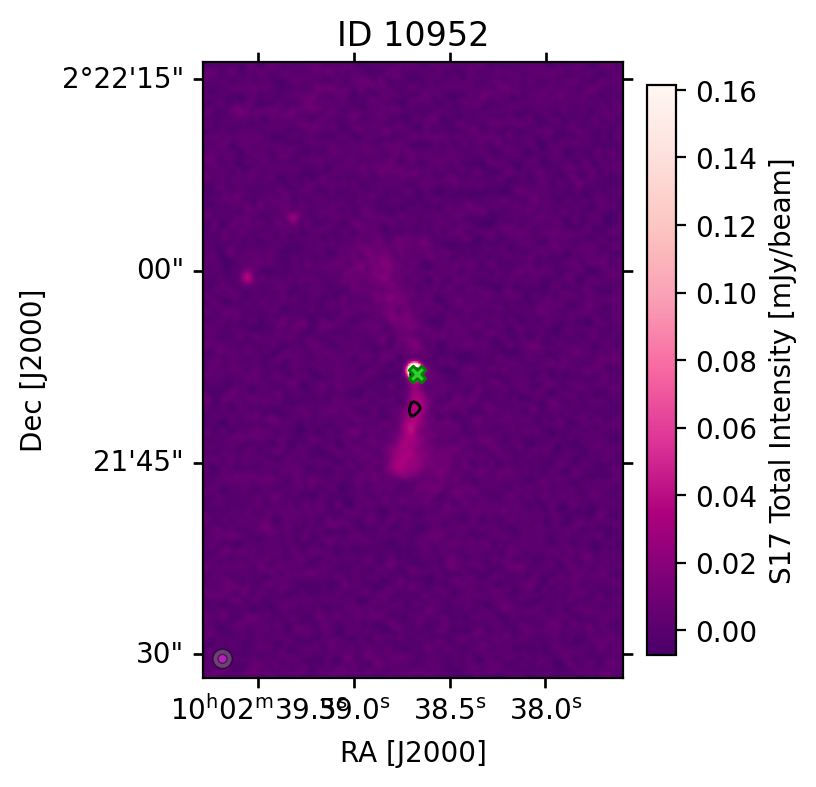}
    \includegraphics[width=0.32\linewidth]{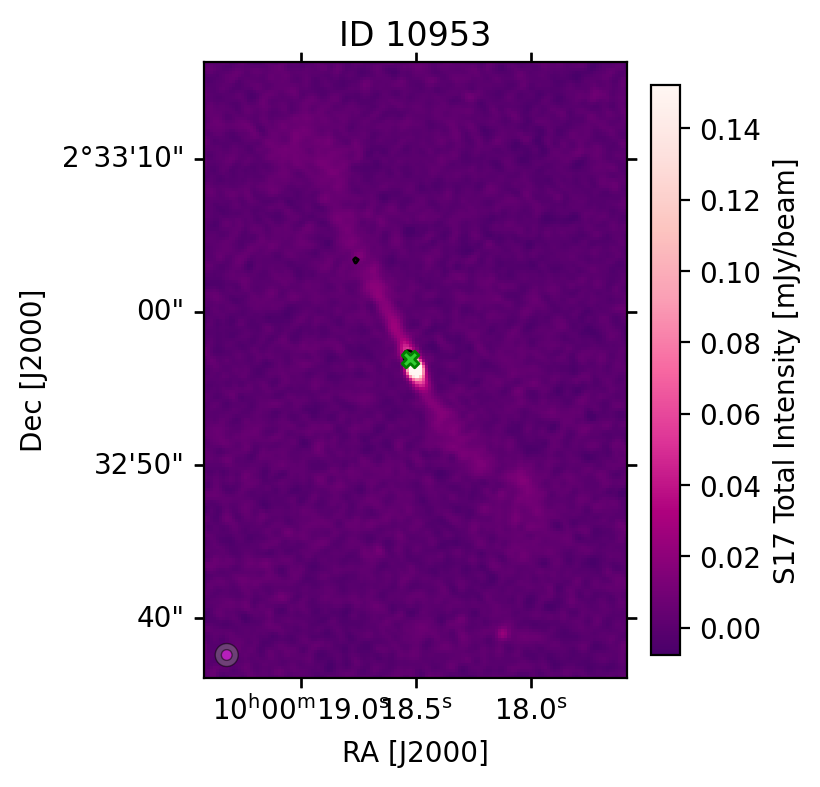} \\
    \includegraphics[width=0.32\linewidth]{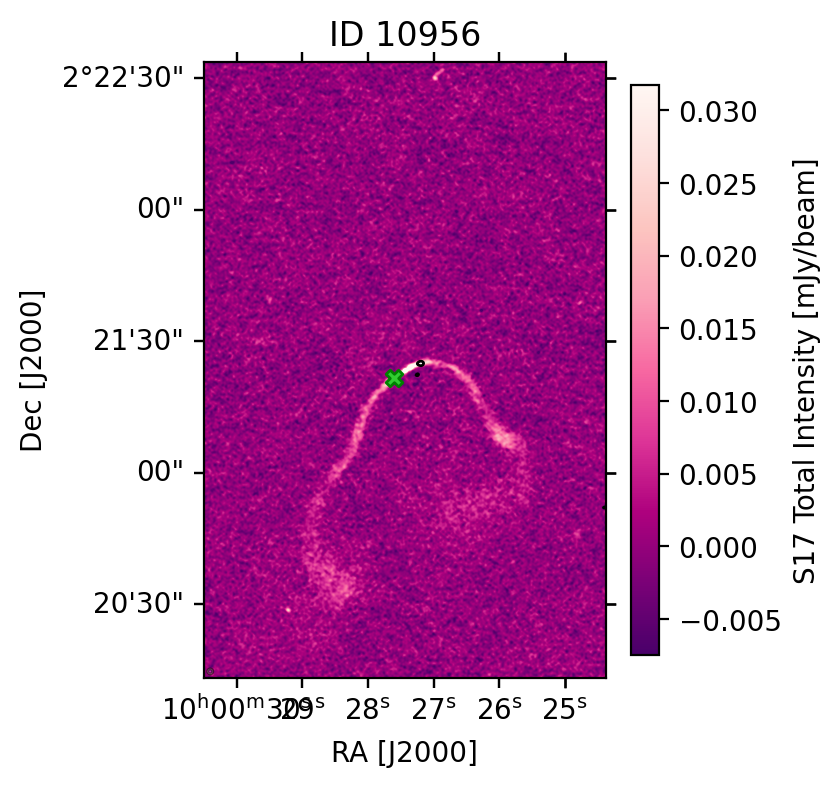}
    \includegraphics[width=0.32\linewidth]{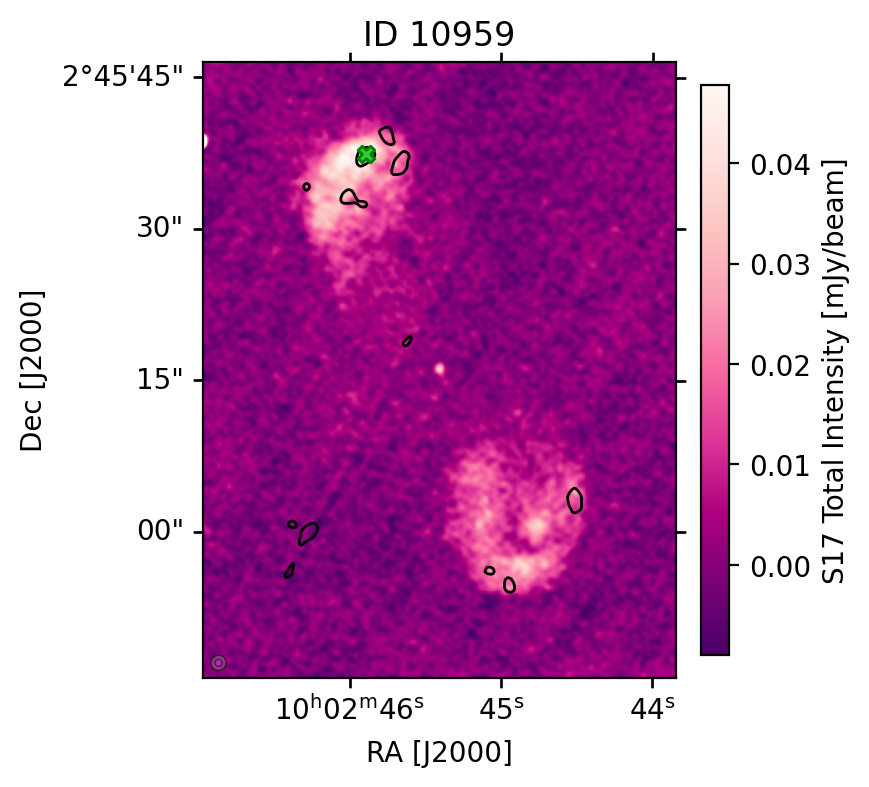}
    \captionsetup{list=no}
    \caption{Total intensity images from \citetalias{Smolcic2017}, overlaid with the polarised intensity contours from this work, continued.}
\end{figure*}
\end{appendix}

\end{document}